\documentclass[prl,twocolumn,10pt,groupedaddress,superscriptaddress,english,showpacs,aps]{revtex4-2}

\usepackage[normalem]{ulem}

\usepackage[utf8]{inputenc}
\usepackage{graphicx}
\usepackage{amsmath}
\usepackage{amssymb}
\usepackage{physics}
\usepackage{bbm}
\usepackage{bbold}
\usepackage{xspace}
\usepackage{color}
\usepackage{footnote}
\usepackage{comment}
\usepackage{placeins}
\usepackage{xr-hyper}
\usepackage{hyperref}
\usepackage{subfigure}

\usepackage{mathtools}
\usepackage{dsfont}

\definecolor{dgreen}{rgb}{0.0,0.5,0.0}
\definecolor{orange}{RGB}{252,77,6}
\definecolor{brown}{RGB}{200,127,50}
\definecolor{blue}{RGB}{00,000,100}
\definecolor{blue2}{RGB}{00,000,250}
\definecolor{green1}{RGB}{00,100,00}
\definecolor{green2}{RGB}{00,150,00}
\definecolor{green3}{RGB}{00,200,00}
\definecolor{green4}{RGB}{00,250,00}

\begin{document}

\title{Orbital Magnetic Field Driven Metal-Insulator Transition in Strongly Correlated Electron Systems}


\author{Georg Rohringer}
\affiliation{Theory and Simulation of Condensed Matter, Department of Physics, King's College London, The Strand, London WC2R 2LS, United Kingdom}
\affiliation{I. Institute of Theoretical Physics, University of Hamburg, 20355 Hamburg, Germany}
\author{Anton Markov}
\affiliation{Russian Quantum Center, Skolkovo Innovation City, 121205 Moscow, Russia}
\affiliation{Center for Nonlinear Phenomena and Complex Systems, Universit´e Libre de Bruxelles,
CP 231, Campus Plaine, B-1050 Brussels, Belgium}


\begin{abstract}
We study the effects of an orbital magnetic field on the Mott metal-insulator transition in the Hubbard-Hofstadter model.
We demonstrate that sufficiently large magnetic fields induce a Mott insulator-to-metal phase transition supporting our claim with dynamical mean field theory (DMFT) numerical results. 
For both competing phases (metal and insulator) we observe a magnetic-field-induced metallization reflected in an enhancement of kinetic and potential energy. 
The kinetic energy of the Mott insulator increases due to the Aharonov-Bohm effect experienced by electrons virtually tunneling around an elementary plaquette which is, however, suppressed by strong correlations.
The kinetic energy of the metallic phase, on the other hand, is more strongly affected by the magnetic field through a field-driven redistribution of spectral weight due to the formation of magnetic minibands. 
This leads to an increase of the kinetic energy which tends to stabilize the metallic state.
Our theoretical results might be relevant for recent experimental studies on magnetic field driven insulator-to-metal transitions in strongly correlated materials such as VO$_2$, $\lambda$-type organic conductors and moir\'e multilayers.

\end{abstract}

\maketitle

{\sl Introduction---}Transitions between metallic and insulating states in solids play a fundamental role in condensed matter physics. 
Apart from their theoretical importance they are potentially also technologically highly relevant as they allow controlling the flow of the electrical current. There exists a variety of distinct types of insulators that are driven by very different physical mechanisms:
Band insulators~\cite{Ashcroft1976} emerge due to the absence of spectral weight at the Fermi level, Slater insulators are induced by strong (antiferromagnetic) spin fluctuations~\cite{Schafer2015}, 
in Anderson insulators~\cite{Anderson1958,Evers2008} disorder leads to a localization of particles,
Peierls insulators~\cite{Peierls1996} originate from structural changes of the underlying lattice system, and
in charge transfer insulators~\cite{Zaanen1985} electrons or holes are transferred from the conduction to the valence band by correlation effects. A particularly interesting class are Mott insulators~\cite{Mott1968} where the insulating state is induced by correlation effects between the electrons due to their strong Coulomb repulsion ~\cite{Manzeli2017,Moon2021,Haule2017,Fei2022}.



A transition between a metallic and an insulating state can be triggered by different control parameters including temperature, strain~\cite{Gurunatha2020}, doping~\cite{Ling2019}, (chemical) pressure, or electric fields~\cite{Mazza2016}. Importantly for the context of this work, the transition between a metal and an insulator can be also controlled by the application of an external magnetic field~\cite{Kagawa2004,Sun2021}.
A good example for such a phenomenon in strongly correlated electron systems is the colossal magneto resistance observed in manganites~\cite{Ramirez1997,Dagotto2001,Salamon2001}, where an external magnetic field can change the resistivity by several orders of magnitude.

\begin{figure}[t!]
    \centering
    \includegraphics[width=0.5\textwidth]{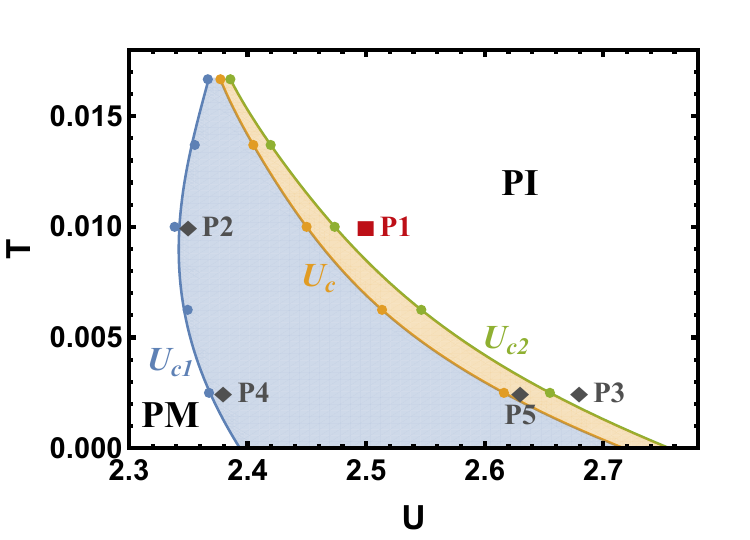}
    \caption{DMFT phase diagram of the $2d$ Hubbard model on a simple (bipartite) square lattice as a function of interaction strength $U$ and temperature $T$ at magnetic field $B\!=\!0$. $U_{c1}$ and $U_{c2}$ are the boundaries between the coexistence region (blue and orange shaded area) and the paramagnetic metallic (PM) and paramagnetic insulating (PI) phase, respectively. $U_c$ corresponds to the line where the thermodynamic phase transition occurs. P1 to P5 indicate the points for which calculations at finite orbital magnetic fields have been performed.}
    \label{fig:PhasediagramB0}
\end{figure}

\begin{figure*}[t!]
    \centering
    \includegraphics[width=1.0\textwidth]{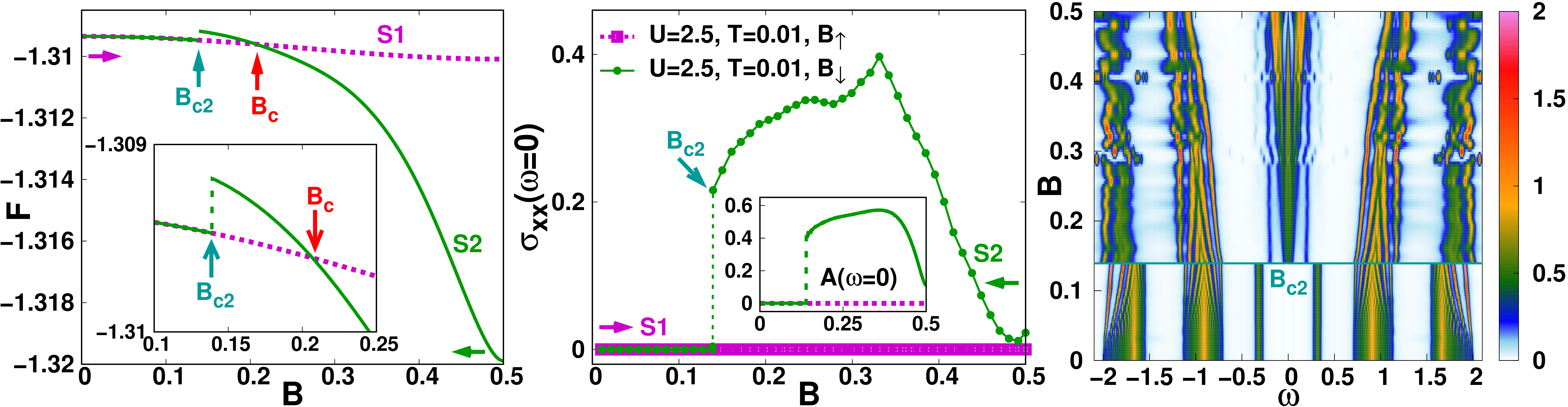}
    \caption{
    Free energy $F$ (left panel) and dc conductivity $\sigma_{xx}(\omega\!=\!0)$ (middle panel) as a function of the orbital magnetic field $B$ as well as local spectral function $A(\omega)$ (right panel) plotted as a heat map in the $(\omega,B)$ plane for the parameter set P1 in Fig.~\ref{fig:PhasediagramB0} corresponding to $U\!=\!2.5$ and $T=0.01$. $B_\uparrow$ and $B_\downarrow$ indicate that calculations have been started from $B\!=\!0$ (S1) or $B\!=\!0.5$ (S2), respectively. For $A(\omega)$ only results for S2 are shown.
    Inset left panel: blow up of the transition region.
    Inset middle panel: Spectral function at the Fermi level $A(\omega\!=\!0)$ as a function of $B$.   
    }
    \label{fig:InsulatorMetal}
\end{figure*}

Recently magnetic field driven insulator-to-metal transitions were experimentally observed in several strongly correlated materials including the transition metal oxide VO$_2$ ~\cite{Matsuda2020,Matsuda2022}, $\lambda$-type organic conductors~\cite{Fukoka2024}, and twisted bilayer graphene~\cite{Cao2018,Pizarro2019,Kwan2020}. 
The magnitudes of the critical magnetic field at which these transitions occur appear to be inversely proportional to the size of the unit cell when comparing this metallization effect across the different compounds.
Although a relevant contribution of the Zeeman term cannot be ruled out with absolute certainty, this observation rather suggests that the phase transitions are controlled by the magnetic flux through a unit cell, which has inspired us to study the {\sl orbital effects} of the magnetic field in the presence of strong electronic correlations~\cite{Czajka2006,Acheche2017,Markov2019,Vucicevic2021,Vucicevic2021a,Ding2022}. 

{\em Model and method}---We investigate the single-band Hubbard-Hofstadter model on a square lattice with hopping amplitudes $t_{ij}$ for electrons tunneling between sites $\mathbf{R}_i$ and $\mathbf{R}_j$, where a homogeneous magnetic field perpendicular to the two-dimensional lattice is coupled to the orbital degrees of freedom by multiplying $t_{ij}$ with a Peierls phase (for details see Sec.~\ref{Sec:TechnicalDetails} in Supplemental Material~\cite{Rohringer2024Supp} which includes Refs.~\cite{Wallerberger2019,Wang2001,Sherman1950,Hager1989,Harville1997,Niven1991,Wolfram2024}).
Correlation effects are introduced via a purely local Hubbard interaction $U$ within or close to the coexistence region between the metallic and the insulating state of the field-free model (see Fig.~\ref{fig:PhasediagramB0}).
To investigate these correlation effects, we exploit a recently developed extension of dynamical mean field theory (DMFT)~\cite{Vollhardt1992,Georges1992,Georges1992a,Georges1996} to finite orbital magnetic fields~\cite{Acheche2017,Markov2019,Vucicevic2021,Vucicevic2021a}.
Our analysis indeed predicts an orbital magnetic field driven metallization and a related metal-insulator transition that can be explained by a redistribution of spectral weight due to the formation of magnetic minibands and a delocalization effect induced by the magnetic field which drags the electrons on closed orbits around a plaquette of the underlying lattice.

We omit the Zeeman effect of the magnetic field to disentangle the relevance of an orbital magnetic field from its impact on the spin degrees of freedom.
The latter has been studied for the Hubbard model within DMFT~\cite{laloux1994effect,Bauer2007,van1997mott} and the Gutzwiller approach~\cite{vollhardt1984normal} predicting a crossover from a Mott to a fully polarized band insulator and an extended insulating region, respectively, rather than an enhancement of metallicity due to the magnetic field.
Let us, however note, that an investigation of the interplay between the response of orbital and spin degrees of freedom to a magnetic field is a very interesting future research direction.
Moreover, paramagnetic DMFT calculations neglect the strong antiferromagnetic fluctuations emerging for the bipartite lattice with nearest neighbor hopping $t\!=\!t_{i(i+1)}$, which we consider in the Letter just for simplicity.
To justify this restriction, we show that we obtain completely analogous results for frustrated lattices in Sec.~\ref{sec:FrustratedLattice} of Supplemental Material~\cite{Rohringer2024Supp}, where such fluctuations are strongly suppressed as is the case also for VO$_2$ or $\lambda$-type organic conductors.

In the following, we will use the half bandwidth $\frac{W}{2}\!=\!4t$ at $B\!=\!0$ as unit of energy while the magnetic field is given in terms of $\frac{\Phi_0}{a^2}$, where $a$ is the lattice constant. 
For convenience we set $\hbar\!=\!k_B\!=e\!=\!1$ allowing us to express temperature and inverse time in terms of our energy unit $\frac{W}{2}$ and the conductivity as dimensionless quantity.
{\sl Results}---We demonstrate the emergence of a magnetic-field-driven metal-insulator transition by performing DMFT calculations for the free energy $F$, the paramagnetic dc conductivity $\sigma_{xx} \equiv \sigma_{xx}(\omega = 0)$, and the local spectrum $A(\omega)$ as functions of the magnetic field. 
While the behavior of the free energy indicates the presence of a phase transition, the latter two quantities reveal its physical nature. In Fig.~\ref{fig:InsulatorMetal} we show our results for point P1 indicated by a red square in the phase diagram of Fig.~\ref{fig:PhasediagramB0}, corresponding to $U\!=\!2.5$ and $T\!=\!0.01$, where the system is a Mott insulator for $B\!=\!0$. 
Our data for points P2 to P5 as well as for systems where geometric frustration has been introduced by a next-nearest neighbor hopping $t'\!=\!t_{i(i+2)}$ are discussed in Secs.~\ref{Sec:InsulatorToMetal}, \ref{Sec:additionaldata}, and \ref{sec:FrustratedLattice} of Supplemental Material~\cite{Rohringer2024Supp}, respectively.
In the left panel, the free energy $F$ is shown as a function of the magnetic field $B$ for two different calculation sets, denoted by $B_\uparrow$ and $B_\downarrow$. 
For the first set, $B_\uparrow$, we started from the insulating state S1 at $B = 0$ and gradually increased the magnetic field (dashed purple line). The free energy evolves smoothly, and no signs of a phase transition are observed; i.e., the system remains in its initial thermodynamic state. 
For the second set, $B_\downarrow$, we started at $B = 0.5$ and gradually decreased the magnetic field (green solid line). 
Remarkably, at $B = 0.5$ we can stabilize not only state S1 but also a second solution, S2, which has a substantially lower free energy. 
As $B$ decreases, the difference in the free energy between the two solutions diminishes and eventually vanishes at $B_c = 0.2069$, indicating a thermodynamic phase transition from S2 to S1. 
We can stabilize S2 even for smaller values of $B < B_c$, where its free energy is already higher than that of S1, until S2 becomes unstable at $B_{c2} = 0.1388$ and collapses onto S1. 
This behavior is a distinctive hallmark of a first-order phase transition, where $B_{c2}$ marks the lower boundary of the coexistence region between states S1 and S2. 

The dc conductivity~\cite{Kubo1957} $\sigma_{xx}\!\equiv\!\sigma_{xx}(\omega\!=\!0)$ and spectral function $A(\omega)$ in the middle and right panels allow us to unambiguously identify the states S1 and S2 as insulator and metal, respectively.
In fact, we can see that $\sigma_{xx}$ [as well as $A(\omega\!=\!0)$ in the inset of the middle panel] is finite for S2 (filled green circles) for $B\!>\!B_{c2}$ and it vanishes for S1 (empty violet squares) on which S2 collapses at $B\!\le\!B_{c2}$.
Similarly, the spectral density $A(\omega)$ which is shown for state S2 as a heat map in the $(\omega,B)$ plane in the right panel features a charge gap around $\omega\!=\!0$ for $B\!<\!B_{c2}$ whose size $U\!-\!W\!\approx\!0.5$ is in excellent agreement with predictions for the gap size in the Mott insulating state in the literature~\cite{Sangiovanni2006,Wang2009}.
For $B\!>\!B_{c2}$ we observe final spectral weight around $\omega\!=\!0$ signaling a metal-insulator transition at $B_{c2}$.
More details about the spectral function are provided in Sec.~\ref{Sec:spectralfunction} of Supplemental Material~\cite{Rohringer2024Supp} (see also Refs.~\cite{Fei2021,Nogaki2023,Nogaki2023a,Jarrell1996,Kaufmann2023} therein).


Let us note that the upper boundary $B_{c1}$, beyond which the system always remains in the metallic state S2, is not reached for the largest magnetic field $B = 0.5$~\footnote{Larger magnetic fields ($B > 0.5$) are not meaningful in our model, as it is symmetric under the transformation $B\rightarrow 1-B$.} in the bipartite case considered here. Instead, $B_{c1}$ is found at lower interaction strengths (albeit without $B_{c2}$), as discussed in Sec.\ref{Sec:InsulatorToMetal} of Supplemental Material~\cite{Rohringer2024Supp}. For systems with strong geometrical frustration~\cite{Hatsugai1990,Hyttrek2024}, both boundaries $B_{c1}$ and $B_{c2}$ of the coexistence region are located within the interval $B \in [0, 0.5]$ (see Fig.\ref{fig:motttransitionfrustratedt1.0} in Sec.\ref{sec:FrustratedLattice} of Supplemental Material~\cite{Rohringer2024Supp}).

\begin{figure}[t!]
    \centering
    \includegraphics[width=0.5\textwidth]{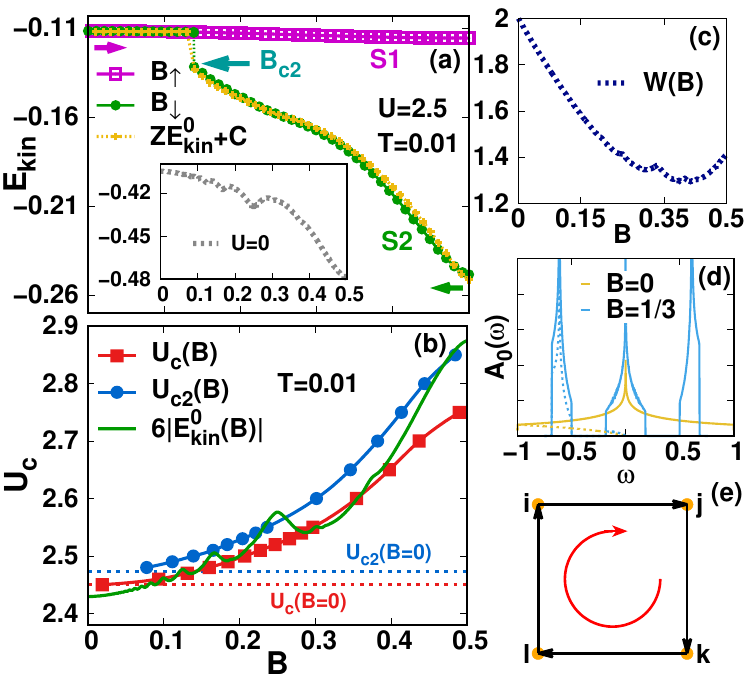}
    \caption{(a) Kinetic energy for $U\!=\!2.5$ and $T\!=\!0.01$ (point P1 in Fig.~\ref{fig:PhasediagramB0}) as a function of the orbital magnetic field $B$ obtained by the two sets of calculations $B_\uparrow$ and $B_\downarrow$ discussed in Fig.~\ref{fig:InsulatorMetal}. 
    Yellow crosses correspond to the noninteracting kinetic energy $E_\text{kin}^0$ (shown in the inset) rescaled by the quasiparticle weight $Z$ and shifted by a magnetic field independent constant $C$.
    (b) Critical interactions $U_c$ and $U_{c2}$ (see Fig.~\ref{fig:PhasediagramB0}) as a function of $B$ compared to the kinetic energy of the noninteracting system for $T\!=\!0.01$. (c) Spectral width $W(B)$ of the noninteracting Hofstadter model as function of $B$. (d) Noninteracting density of states $A_0(\omega)$ for $B\!=\!0$ and $B\!=\!\frac{1}{3}$. Dashed lines indicate the (negative) contribution $-\omega A_0(\omega)$ to the kinetic energy. (e) Illustration of an electron hopping around a plaquette due to the magnetic field.}
    \label{fig:KineticEnergyBeta400}
\end{figure}

{\sl Discussion}---Let us now address the question which physical mechanism is at work to drive the transition between a metal and an insulator upon tuning the orbital magnetic field $B$. 
A first naive attempt to understand this phenomenon is to investigate the magnetic field dependence of the spectral width $W(B)$ of the noninteracting system which is defined as the difference between the maximum and minimum eigenvalue $E_{l\mathbf{k}}$ of the dispersion matrix $\varepsilon_{ll'}(\mathbf{k})$ (see Eq.~(\ref{equ:HarperMatrix}) in Sec.~\ref{Sec:TechnicalDetails} of Supplemental Material~\cite{Rohringer2024Supp}), i.e., $W(B)=\max_{l\mathbf{k}}E_{l\mathbf{k}}-\min_{l\mathbf{k}}E_{l\mathbf{k}}$.
The corresponding results are shown in panel~(c) of Fig.~\ref{fig:KineticEnergyBeta400}.
We observe that $W(B)$ {\em decreases} with $B$ which enhances the interaction-spectral width ratio $\frac{U}{W(B)}$ and should, hence, lead to a more insulating state rather than the observed metallization.
Next, we consider the second moment of the noninteracting density of states $M_2\!=\!\frac{1}{q}\sum_\mathbf{k}\Tr\varepsilon_\mathbf{k}^2$ which is approximately related to the critical interaction $U_c\!\propto\!\sqrt{M_1}$~\cite{Bulla2000}.
However, quite remarkably for the Hofstadter model $M_1$ does not depend on the magnetic field $B$ and consequently no effect of the orbital magnetic field on the Mott transition can be expected from this consideration.
Hence, a more thorough analysis of the observed phenomenon is required.

From a general perspective the Mott transition is governed by the balance between kinetic and potential energy. 
The Mott insulating phase is characterized by a small absolute value of kinetic and potential energy compared to the metallic state due to the localization of particles.
An increase in the (absolute) value of these thermodynamic variables thus signals a tendency toward metallization. 
We indeed observe such an enhancement of $E_\text{kin}$ with increasing magnetic field for both solutions S1 (filled green circles) and S2 (empty purple squares) in panel~(a) of Fig.~\ref{fig:KineticEnergyBeta400} (analogous results for the potential energy $E_\text{pot}$ are shown in Sec.~\ref{sec:potentialenergyU2.5beta100} of Supplemental Material~\cite{Rohringer2024Supp}). 
The effect is more pronounced in the metallic regime $B\!>\!B_{c2}$ of S2, where $E_{\text{kin}}$ becomes substantially larger and grows rapidly with $B$. 
A similar trend was reported in determinant quantum Monte Carlo~\cite{Ding2022} and ED~\cite{Czajka2006} studies, albeit without addressing the metal-insulator transition.

On the metallic side the increase of $E_\text{kin}$ with $B$ is in excellent quantitative agreement with the evolution of the noninteracting kinetic energy $E_\text{kin}^0$ [see inset of panel (a)] renormalized by the quasiparticle weight $Z$, corresponding to the enhancement of the effective mass of a quasiparticle due to correlations $\frac{1}{Z}\!=\!\frac{m_\text{eff}}{m}$ (for the rigorous definition see Sec.~\ref{sec:potentialenergyU2.5beta100} in Supplemental Material~\cite{Rohringer2024Supp}), and shifted by a field-independent constant $C$ which accounts for the contribution from the incoherent portion of the spectrum (yellow crosses).
This behavior of $E_\text{kin}$ is consistent with predictions of the Gutzwiller approach~\cite{gutzwiller1965correlation,Brinkman1970,Vollhardt1984} which allows for a physical transparent understanding of the Mott transition in Hubbard-type models at low temperature.
This method predicts a critical interaction strength, at which the metal-insulator transition occurs, of $U_c\!=\!8|E^0_\text{kin}|$.
Hence, as $\lvert E_\text{kin}^0\rvert$ increases with $B$ we can expect a corresponding increase of $U_c$ or, for a fixed value of $U$, a transition from an insulator to a metal at the value of $B$, where $8\lvert E_\text{kin}^0\rvert$ reaches this value of $U$.
To quantify these expectations and determine the energy scale that is responsible for the magnetic-field driven metal-insulator transition we show $U_c$ (filled red squares) and $U_{c2}$ (filled blue circles) as a function of $B$ in Fig.~\ref{fig:KineticEnergyBeta400}(b) [which are obtained by inverting the monotonous functions $B_c(U)$ and $B_{c2}(U)$] and compared it to $\lvert E_\text{kin}^0(B)\rvert$ (green line).
We find a good agreement for $U_c(B)$ with $6\lvert E_\text{kin}^0(B)\rvert$ where the reduction of the factor $8$ in the original version of the Gutzwiller approach~\cite{Brinkman1970} to $6$ is indeed in excellent agreement with the predictions obtained from an improved treatment of fluctuations within the Gutzwiller method in Ref.~\cite{Fabrizio2017} and the reduced value of $U_c$ at finite temperatures with respect to $T\!=\!0$~\footnote{The improved treatment of fluctuations in Ref.~\cite{Fabrizio2017} reduces $U_c$ by a factor of $\sim0.822$. Moreover, according to our phase diagram the critical $U_c\!=\!2.71$ at $T\!=\!0$ is reduced to $U_c\!=\!2.45$ at $T\!=\!0.01$. Hence, we have $U_c=(2.45/2.71)\times0.822\times8|E_\text{kin}^0|=5.945|E_\text{kin}^0| \sim 6|E_\text{kin}^0|$.}.

We, hence, conclude that the magnetic-field-driven metal-insulator transition is governed by the increase of $\lvert E_\text{kin}^0\rvert$ with $B$.
The increase can be explained by a redistribution of spectral weight in the noninteracting system due to the formation of magnetic minibands that generically feature van Hove singularities at finite energies, which are associated with the topological properties of the system~\cite{naumis2016topological}.
This is illustrated exemplary in Fig.~\ref{fig:KineticEnergyBeta400}(d) for $B\!=\!\frac{1}{3}$. 
The enhanced contribution to $\lvert E_\text{kin}^0\rvert$ arises from the van Hove singularity at $\omega\!\approx\!-0.55$ for $B\!=\!\frac{1}{3}$, while the singularity at $\omega\!=\!0$ for $B\!=\!0$ has minimal impact on $E_\text{kin}$. 


In the Mott insulating regime, the magnetic-field-induced increase of kinetic (and potential) energy  can be explained in a simplified way by a strong coupling expansion~\cite{Pairault1998,Pairault2000,sen1995large} of these thermodynamic variables in terms of $\frac{t}{U}$ which is carried out explicitly in Sec.~\ref{sec:largeUexpansion} of Supplemental Material~\cite{Rohringer2024Supp}.
The leading order $B$-dependent term for both $E_\text{pot}$ and $\lvert E_\text{kin}\rvert$ is proportional to $-\frac{t^4}{U^3}\cos(2\pi B)$ and indeed leads to an increase of these energies when $B$ increases from $0$ to $0.5$.
This behavior can be attributed to the Aharonov-Bohm~\cite{aharonov1959significance} effect experienced by the electrons hopping around a plaquette [Fig.~\ref{fig:KineticEnergyBeta400}(e)] which leads to a delocalization of these particles.

{\sl Relation to experimental studies.} At the end let us briefly address three experimental studies for which our theoretical analysis of the orbital-magnetic-field-driven insulator-to-metal transition can be potentially relevant. In all these strongly correlated materials an insulator-to-metal transition was observed for magnetic fields corresponding to a substantial part of the flux quantum which indicates the importance of the orbital effects of the magnetic field. Moreover, the relevance of the predicted mechanism is supported by the fact that the value of the critical magnetic field strength is approximately inversely proportional to the unit cell area. Let us, however, stress that the exact magnetic field at which the transition occurs in these complicated materials can obviously not be predicted by a simple single-band Hubbard model calculation.


(i) In recent experiments~\cite{Matsuda2020,Matsuda2022} a magnetic-field-driven insulator-to-metal transition was observed in tungsten-doped vanadium dioxide V$_{1-x}$W$_x$O$_2$ at doping $x\!=\!0.06$, and ultrahigh magnetic fields $B\!=\!500$T. 
Let us note that no metal-insulator transition has been observed in the experiment for lower W doping in the considered range of the magnetic field, which indeed fits our predictions for the role of the orbital effects.
Since the lattice constant in V$_{1-x}$W$_x$O$_2$, which can be estimated by $a\!\approx\!0.6$nm, decreases with decreasing doping~\cite{Choi2020,Lu2019} a gradually increasing magnetic field is required to generate the same total flux through a unit cell at which the phase transition can be observed. 

(ii) In $\lambda$-type organic conductors~\cite{Fukoka2024} a magnetic field induced transition has been found at $B\!\sim\!50$T, which is much lower than the corresponding field strength at which the transition occurs in VO$_2$.
Within the scenario of the orbital magnetic field driven Mott transition this can be attributed to the much larger unit cell in $\lambda$-type organic conductors (lattice constant $a\!\sim\!1.2$ nm), which facilitates large fluxes also for smaller magnetic fields.

(iii)Finally, a magnetic field driven phase transition was observed in magic angle twisted bilayer graphene~\cite{Cao2018} at moderate magnetic fields of $B\approx 5T$ for the half-filled band close to the first magic angle. 
Within the picture of the orbital-magnetic-field-driven Mott transition this low value of $B_c$ can be explained by the large lattice constant at this twist angle of $\approx 13.6$ nm.  
Let us note that Zeeman effect of the magnetic field could not easily explain the appearance of the metallic phase as discussed in Refs.~\cite{Pizarro2019,Kwan2020}

{\sl Conclusions and Outlook}---We have demonstrated the emergence of a metal-insulator transition in the Hubbard-Hofstadter model which is driven by the orbital effect of an external magnetic field on the electrons.
As for the standard interaction-driven MIT we observe a first order behavior with a coexistence region extending to a magnetic field $B_{c2}$ while the actual thermodynamic transition occurs at $B_c\!>\!B_{c2}$.
We have argued that this magnetic-field-driven change of state can be explained by a metallization effect due to the orbital magnetic field which is indicated by the increase of the kinetic and potential energy with increasing $B$. We have identified the redistribution of spectral weight due to the formation of magnetic minibands and the magnetic field induced virtual motion of an electron around a plaquette as theoretical explanations for this effect from a metallic and insulating perspective, respectively.
In the final part of the Letter, we have discussed the relevance of our results for recent experiments on VO$_2$, $\lambda$-type organic conductors and twisted bilayer graphene and argue that orbital effects of an external magnetic field might play an important role for the magnetic field induced insulator-to-metal transition in these materials. 

In this study we have focused on the orbital effect of the magnetic field and neglected its impact on the spin degrees of freedom via a Zeeman term.
In general, the importance of the orbital effects of a magnetic field with respect to the spin Zeeman term can be investigated experimentally by changing the direction of the magnetic field. 
While the Zeeman term should be unaffected, the orbital effects should change quite significantly.
Another possibility to disentangle the spin and orbital effects of magnetic fields are synthetic gauge fields~\cite{Dalibard2015} generated by lasers with periodic driving which couple only to orbital degrees of freedom. 
Hence, a comprehensive  investigation of the interplay between the effects of a magnetic field on orbital and spin degrees of freedom (Zeeman term), which will allow us to work out a unified picture of the response of correlated electron systems to magnetic fields, is a highly interesting novel research direction.

Our prediction could be potentially of technological use, as it provides a new switching mechanism between an insulator and a metal. In particular, the discussion is relevant for the moir\'e materials with large unit cells, such that the orbital effects of the magnetic become significant at moderate magnetic fields of a few Tesla. In these systems~\cite{kang2021cascades}, the orbital effects of the magnetic field are under a current focus of both experimental and theoretical effort ~\cite{Cao2018,kang2021cascades,Vucicevic2021,kometter2023hofstadter}.  Let us also note that the mechanism we discussed is not specific to the Mott insulators and could be extended to other types of insulators.   


{\sl Acknowledgements}---We thank T. Sch\"afer, A. Toschi, and G. Sangiovanni for useful discussions. G. R. acknowledges financial support from the Deutsche Forschungsgemeinschaft (DFG) through
Projects No. 407372336 and No. 449872909. A. A. M. acknowledges support from the Russian Quantum Center in the framework of the Russian Quantum Technologies Roadmap and from the  European Research Council (LATIS project). The authors gratefully acknowledge the computing time granted by the Resource Allocation Board and provided on the supercomputers Lise and Emmy/Grete at NHR@ZIB and NHR@Göttingen as part of the NHR infrastructure. These centers are jointly supported by the Federal Ministry of Education and Research and the state governments participating in the NHR (www.nhr-verein.de). The calculations for this research were conducted with computing resources under the project hhp00048.

\clearpage

\hypersetup{pageanchor=false}
\setcounter{page}{1}

\renewcommand{\theHsection}{S\arabic{section}}
\renewcommand{\theHfigure}{S\arabic{figure}}
\renewcommand{\theHequation}{S\arabic{equation}}
\renewcommand{\theHtable}{S\arabic{table}}

\setcounter{section}{0}
\setcounter{figure}{0}
\setcounter{table}{0}
\setcounter{equation}{0}

\renewcommand{\thesection}{S\arabic{section}}
\renewcommand{\thefigure}{S\arabic{figure}}
\renewcommand{\thetable}{S\arabic{table}}
\renewcommand{\theequation}{S\arabic{equation}}

\thispagestyle{empty}

\onecolumngrid
\includegraphics[width=1.0\linewidth]{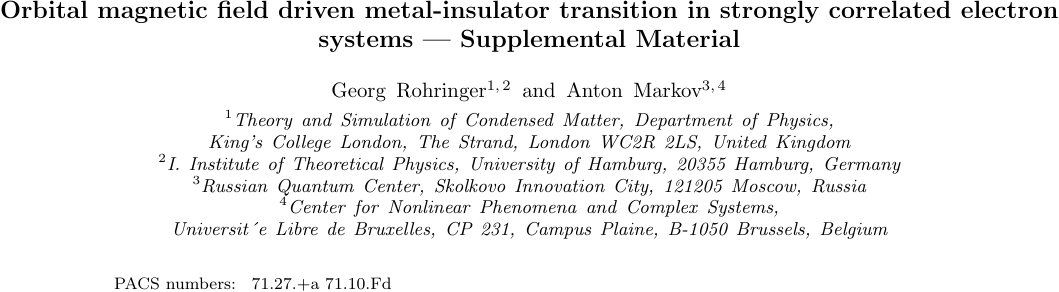}
\vspace{0.05cm}
\twocolumngrid

\section{Calculation of the free energy within DMFT}
\label{Sec:TechnicalDetails}

In this section, we provide a number of technical details for our DMFT calculations.
More specifically, in Sec.~\ref{Sec:DMFTHofstadter} we discuss the Hubbard-Hofstadter Hamiltonian and the DMFT approach to treat it.
In Sec.~\ref{Sec:FreeEnergy} we present the calculation of the free energy and in Sec.~\ref{Sec:NumericalImprovements} we outline specific numerical improvements which have allowed us to perform calculations on a very fine $B$ grid.

\subsection{DMFT for the Hubbard-Hofstadter model}
\label{Sec:DMFTHofstadter}

The Hamiltonian of the Hubbard-Hofstadter model is given by
\begin{equation}
	\label{equ:Hamiltonian}
	\hat{H}=-\sum_{\langle ij \rangle,\sigma}t_{ij}e^{i\phi_{ij}}\hat{c}^\dagger_{i\sigma}\hat{c}_{j\sigma}+U\sum_i\hat{n}_{i\uparrow}\hat{n}_{i\downarrow},
\end{equation}
where $\hat{c}^{(\dagger)}_{i\sigma}$ annihilates (creates) an electron with spin $\sigma\!=\!\uparrow,\downarrow$ at lattice site $\mathbf{R}_i$, $\hat{n}_{i\sigma}\!=\!\hat{c}^\dagger_{i\sigma}\hat{c}_{i\sigma}$ denotes the number of particles at lattice site $\mathbf{R}_i$, $t_{ij}e^{i\phi_{ij}}$ is the hoping amplitude between sites $\mathbf{R}_i$ and $\mathbf{R}_j$ of the square lattice and $U$ corresponds to the local Coulomb interaction between two particles at the same lattice site $\mathbf{R}_i$.
The magnetic field dependence is encoded in the phase factor $\phi_{ij}$ as
\begin{equation}
	\label{equ:PeierlsPhase}
	\phi_{ij}=\frac{2\pi}{\Phi_0}\int_{\mathbf{R}_i}^{\mathbf{R}_j}\mathbf{A}(\mathbf{r})\cdot d\mathbf{r},
\end{equation}
where $\phi_{ij}$ is the so-called Peierl's factor which guarantees local gauge invariance of the model.
The vector potential $\mathbf{A}(\mathbf{r})$ for the uniform magnetic field in $z$-direction $\mathbf{B}\!=\!(0,0,B)$ in the Landau gauge is given by:
\begin{equation}
	\label{equ:VectorPotential}
	\mathbf{A}(\mathbf{r})=-B\begin{pmatrix}y\\0\\0\end{pmatrix}.
\end{equation}

The additional phase factor is position-dependent and therefore formally breaks the translational invariance of the system in $y$-direction, although from a physical perspective the system should remain lattice translational invariant for a homogeneous magnetic field.
However, for magnetic fields which are rational multiples of the flux quantum $\Phi_0$ divided by the volume of the unit cell $a^2$, i.e., $B\!=\!\frac{\Phi_0}{a^2}\frac{p}{q}$, where $p,q\in\mathds{N}$ are coprime, the phase factor becomes periodic with a period of $q$ in $y$-direction.
We can, hence, define a translational invariant lattice of magnetic unit cells with $q$ lattice sites in $y$-direction.
The non-interacting part of the Hamiltonian can now be transformed to momentum space yielding the $q\times q$ dispersion (or Harper) matrix of the system\cite{S-Markov2019,S-Vucicevic2021a}

\begin{widetext}
	\begin{equation}  
		\label{equ:HarperMatrix}
		\varepsilon_{ll'}(\mathbf{k})=
		\begin{pmatrix}
			- 2 t \cos\left(k_x\right) & -t  & 0 & \ldots & -t e^{iqk_y}\\
			-t & - 2 t \cos\left(k_x+ \frac{2 \pi p}{q}\right)& - t   & 0  &\ldots \\
			& & & \ddots&  & &\\
			-t e^{-iqk_y}&0 &\ldots& -t &- 2 t \cos\left(k_x+ \frac{2 \pi p (q-1)}{q}\right)
		\end{pmatrix},
	\end{equation}
\end{widetext}
where we have set the lattice constant $a\!=\!1$ and considered only the case of nearest neighbor hopping $t\!=\!t_{i(i+1)}$. The corresponding matrix for the case with a finite next-nearest neighbor hopping parameter $t'\!=\!t_{i(i+2)}$ can be found in Refs.~\cite{S-Hatsugai1990,S-Hyttrek2024}.
The Green's function $G(i\nu,\mathbf{k})\!=\![(i\nu+\mu)\mathds{1}_{q\times q}-\varepsilon(\mathbf{k})-\Sigma(i\nu,\mathbf{k})]^{-1}$ and the self-energy $\Sigma(i\nu,\mathbf{k})$ of the system are $q\times q$ matrices where the matrix indices correspond to the position of the lattice site within the magnetic unit cell.
Within the DMFT approximation a purely local self-energy is assumed which implies that the matrix $\Sigma(i\nu,\mathbf{k})$ does not depend on the momentum $\mathbf{k}$, has no off-diagonal elements (which would correspond to a nonlocal self-energy between different sites within the magnetic unit cell), and is the same at all lattice sites due to the homogeneity of the system\cite{S-Acheche2017,S-Markov2019,S-Vucicevic2021,S-Vucicevic2021a}.
Hence, the self-energy matrix is proportional to a local scalar self-energy $\Sigma_\text{loc}(i\nu)$ times the $q\times q$ unit matrix.
Within this approximation the corresponding DMFT lattice Green's function reads
\begin{equation}
	\label{equ:GreensFunctionDMFT}
	G_\text{DMFT}(i\nu,\mathbf{k})=\frac{1}{[i\nu+\mu-\Sigma_\text{loc}(i\nu)]\mathds{1}_{q\times q}-\varepsilon(\mathbf{k})}.
\end{equation}
$\Sigma_\text{loc}(i\nu)$ can be obtained from an auxiliary Anderson impurity model (AIM)
\begin{equation}
	\label{equ:AIM}
	\hat{H}_\text{AIM}=\underset{\hat{H}_\text{bath}}{\underbrace{\sum_{n\sigma}\epsilon_n \hat{a}^\dagger_{n\sigma}\hat{a}_{n\sigma}}}+\underset{\hat{H}_\text{hyb}}{\underbrace{\sum_{n\sigma}V_n \hat{a}^\dagger_{n\sigma}\hat{c}_{\sigma}+\hat{c}^\dagger_\sigma\hat{a}_{n\sigma}}}+\underset{\hat{H}_\text{imp}}{\underbrace{U\hat{n}_\uparrow\hat{n}_\downarrow}},
\end{equation}
where $\hat{a}^{(\dagger)}_{n\sigma}$ annihilates (creates) an electron with spin $\sigma$ at the (non-interacting) bath site $n$, $\epsilon_n$ is the local onsite energy of this bath site, $\hat{c}^{(\dagger)}_\sigma$ is the annihilation (creation) operator for an electron at the (interacting) impurity, $V_n$ corresponds to the hybridization between the impurity and the bath and $U$ denotes the interaction energy between two particles at the impurity.
Model (\ref{equ:AIM}) can be solved numerically exactly by exact diagonalization (ED) if a finite number $N$ of bath sites is considered.
The bath parameters $\epsilon_n$ and $V_n$ are determined by the condition that the impurity Green's function is equivalent to the local part of the DMFT Green's function in Eq.~(\ref{equ:GreensFunctionDMFT}).
This assumption is formally expressed by the selfconsistency condition
\begin{equation}
	\label{equ:DMFTselfconsistency}
	G_\text{imp}(i\nu)=\sum_\mathbf{k}[G_\text{DMFT}(i\nu,\mathbf{k})]_{ll},
\end{equation}
where $\sum_\mathbf{k}$ denotes the normalized integral over the magnetic Brillouin zone defined by $k_x\!\in\![-\pi,\pi)$ and $k_y\!\in\![-\frac{\pi}{q},\frac{\pi}{q})$.
Note that all diagonal elements of the momentum-summed DMFT Green's function are equivalent due to the homogeneity of the problem.
Let us also remark that for testing and for the frustrated case with a finite next-nearest neighbor hopping parameter $t'\!=\!t_{i(i+2)}$ we have also exploited a continuous time quantum Monte Carlo (CTQMC) algorithm in its hybridization expansion as provided in the \texttt{w2dynamics} package \cite{S-Wallerberger2019} as impurity solver.

\subsection{Calculation of free energy}
\label{Sec:FreeEnergy}

Within DMFT the lattice free energy $F$ can be obtained from the impurity free energy $F_\text{imp}$ via the following equation\cite{S-Georges1996}
\begin{equation}
	\label{equ:FreeEnergy}
	F=F_\text{imp}-\frac{1}{q}\frac{2}{\beta}\sum_{\nu\mathbf{k}}\ln\left(\det\left[\frac{G_\text{imp}(i\nu)\mathds{1}_{q\times q}}{G_\text{DMFT}(i\nu,\mathbf{k})}\right]\right),
\end{equation}
where a factor of two in front of the frequency and momentum sum originates from the two spin projections of an electron and $\beta\!=\!\frac{1}{T}$ denotes the inverse temperature.
The impurity free energy $F_\text{imp}$ is given by the difference between the free energy $F_\text{AIM}$ of the AIM in Eq.~(\ref{equ:AIM}) and the free energy $F_\text{bath}$ of the bath, i.e., $F_\text{imp}\!=\!F_\text{AIM}\!-\!F_\text{bath}$, where $F_\text{AIM}\!=\!-\frac{1}{\beta}\ln Z_\text{AIM}$ and $F_\text{bath}\!=\!-\frac{1}{\beta}\ln Z_\text{bath}$.
Within ED the partition functions $Z_\text{AIM}$ and $Z_\text{bath}$ of the entire AIM and the bath, respectively, can be straightforwardly evaluated as $Z_\text{AIM}\!=\!\Tr e^{-\beta \hat{H}_\text{AIM}}\!=\!\sum_{j=1}^{4^{N+1}}e^{-\beta E_j}$ and $Z_\text{bath}\!=\!\Tr e^{-\beta \hat{H}_\text{bath}}\!=\!\prod_{n=1}^N(1\!+\!e^{-\beta\epsilon_n})^2$, where $E_j$ denotes the numerically calculated eigenvalues of $\hat{H}_\text{AIM}$.
Let us mention that within QMC the determination of $F_\text{imp}$ is more difficult as a Wang-Landau algorithm\cite{S-Wang2001} would be required to directly measure the partition function $Z_\text{imp}$ of the impurity.

\subsection{Numerical improvements}
\label{Sec:NumericalImprovements}

In our study we have performed calculations for various values of $B\!\sim\!\frac{p}{q}$ on a fine grid in the interval [0,0.5].
This, however, requires rather large values of $q$ corresponding to large sizes of the $q\times q$ matrix 
\begin{equation}
	\label{equ:DefineM}
	M(i\nu,\mathbf{k})\!=\![i\nu+\mu-\Sigma_\text{loc}(i\nu)]\mathds{1}_{q\times q}\!-\!\varepsilon(\mathbf{k})
\end{equation}
in the denominator of Eq.~(\ref{equ:GreensFunctionDMFT}).
The calculation of the DMFT lattice Green's function $G_\text{DMFT}(i\nu,\mathbf{k})$ in Eq.~(\ref{equ:GreensFunctionDMFT}) requires an inversion of this potentially large matrix while for the evaluation of the lattice free energy $F$ in Eq.~(\ref{equ:FreeEnergy}) the determinant of this matrix has to be calculated.
These matrix operations typically scale as $q^3$ with the matrix size $q$ and have to be carried out for all values of the fermionic Matsubara frequency $\nu$ and the momentum $\mathbf{k}$. Hence, this task can indeed become challenging for matrix sizes of the order of $q\!\sim\!1000$.
In the following, we will outline three numerical improvements based on (i) the specific structure of the matrix $M(i\nu,\mathbf{k})$, (ii) a method which allows to find the fraction with the smallest denominator in a given interval and (iii) the appropriate consideration of the high-frequency tail of the summand in Eq.~\ref{equ:FreeEnergy}.
The calculations are presented for the bipartite case $t'\!=\!0$ but the structure of the dispersion matrix is the same for finite next-nearest neighbor hopping and, hence, the generalization to this case is straightforward.

\subsubsection{Sherman-Morrison-Woodburry identities}
\label{Sec:ShermanMorrisonWoodburry}

The specific structure of the dispersion Matrix $\varepsilon(\mathbf{k})$ allows for a decomposition of the matrix $M(i\nu,\mathbf{k})$ as
\begin{widetext}
	\begin{equation}
		\label{equ:MatrixDecompose}
		M(i\nu,\mathbf{k})=
		\underset{M_\text{t}(i\nu,k_x)}{\underbrace{
				\begin{pmatrix}
					C_1^{\nu k_x}-t & t & 0 & \cdots & 0 \\
					t & C_2^{\nu k_x} & t & \ddots & \vdots \\
					0 & \ddots & \ddots & \ddots & 0 \\
					\vdots & \ddots & t & C_{q-1}^{\nu k_x} & t \\
					0 & \cdots & 0 & t & C_q^{\nu k_x}-t
		\end{pmatrix}}}+
		t
		\underset{M_\text{r1}(k_y)=\mathbf{v}(k_y)\otimes \mathbf{v}^\dagger(k_y)}{\underbrace{
				\begin{pmatrix}
					1 \\
					0 \\
					\vdots \\
					0 \\
					e^{-iqk_y}
				\end{pmatrix}
				\otimes
				\begin{pmatrix}
					1 & 0 & \cdots & 0 & e^{iqk_y}
		\end{pmatrix}}}
	\end{equation}
\end{widetext}
where we have used the short cut $C_j^{\nu k_x}\!=\!i\nu+\mu-\Sigma_\text{loc}(i\nu)+2t\cos[k_x+2\pi\frac{p(j-1)}{q}]$.
We can now see that $M_\text{t}(i\nu,k_x)$ is a tridiagonal matrix depending on $\nu$ and $k_x$ while $M_\text{r1}(k_y)$ depends only on $k_y$ and is a matrix of rank one as it is built from the outer product of a complex vector $\mathbf{v}(k_y)$ and its complex conjugate.
According to the Sherman–Morrison–Woodbury identities\cite{S-Sherman1950,S-Hager1989,S-Harville1997} we can now calculate the inverse and the determinant of the full (complex) matrix $M(i\nu,\mathbf{k})$ by evaluating the corresponding inverse and determinant of the (real) tridiagonal matrix $M_\text{t}(i\nu,k_x)$ plus (times) a specific correction term
\begin{subequations}
	\label{equ:ShermanMorrisonWoodburry}
	\begin{align}
		&M^{-1}=M_\text{t}^{-1}-t\frac{M_\text{t}^{-1}\mathbf{v}\mathbf{v}^\dagger M_\text{t}^{-1}}{1+\mathbf{v}^\dagger M_\text{t}^{-1}\mathbf{v}}\label{equ:ShermanMorrisonWoodburryInverse},\\
		&\det M=\det M_\text{t}\left(1+t\mathbf{v}^\dagger M_\text{t}^{-1}\mathbf{v}\right).\label{equ:ShermanMorrisonWoodburryDeterminant}
	\end{align}
\end{subequations}
For better readability we have suppressed the dependencies of the matrices $M$ and $M_\text{t}$ as well as of the vector $\mathbf{v}$ on $\nu$, $k_x$ and $k_y$.
Eqs.~(\ref{equ:ShermanMorrisonWoodburry}) speed up our numerical calculations in two ways: (i) the evaluation of the inverse or the determinant of a tridiagonal matrix scales linearly with the matrix size $q$ (in contrast to the cubic scaling $q^3$ for the full matrix); (ii) the actual calculation of the inverse or the determinant has to be performed independently only for all values of $\nu$ and $k_x$ but not for $k_y$ since the tridiagonal matrix $M_\text{t}$ does not depend on $k_y$.
Let us also note that due to the sparse structure of the vector $\mathbf{v}$ (only two elements are non-zero) the numerical complexity of the matrix-vector multiplication $M_\text{t}^{-1}\mathbf{v}$ is independent of the matrix size $q$.

\subsubsection{Minimal denominator in a given interval}
\label{Sec:MinimalDenominator}

To perform a comprehensive analysis of the discussed observables (free energy, DC conductivity, spectral function, kinetic and potential energies and quasiparticle lifetimes) as a function of the magnetic field we have used a very fine grid of roughly $N\!\sim\!1500$ magnetic field points in the interval $B\!=\!\frac{p}{q}\!\in\![0,0.5]$. 
This corresponds to a set of magnetic fields $B\!=\!\frac{i}{2N}, i=1,\ldots,N$.
Hence, we would have to perform $N$ calculations with matrix size of $q\!=\!2N$ which represents a considerable numerical effort in spite of the simplifications due to the specific structure of the dispersion matrix discussed in Sec.~\ref{Sec:ShermanMorrisonWoodburry}.
We, hence, did not conduct our calculations for magnetic fields corresponding to the edges of each interval $\left[\frac{i-1}{2N},\frac{i}{2N}\right], i=1,\ldots,N$ but rather for the fraction $\frac{p}{q}\!\in\!\left[\frac{i-1}{2N},\frac{i}{2N}\right]$ with the {\em smallest} denominator in this interval.
This fraction $\frac{p}{q}\!\in\!\left[\frac{i-1}{2N},\frac{i}{2N}\right]$ can be determined by constructing a so-called Farey sequence\cite{S-Niven1991} of intervals
\begin{equation}
	\left[\frac{a_{n+1}}{b_{n+1}},\frac{c_{n+1}}{d_{n+1}}\right]=
	\begin{cases} 
		\left[\frac{a_{n}+c_n}{b_{n}+d_n},\frac{c_{n}}{d_{n}}\right],\;\text{if}\;\frac{a_{n}+c_n}{b_{n}+d_n}<\frac{i-1}{2N} \\
		\left[\frac{a_{n}}{b_{n}},\frac{a_{n}+c_n}{b_{n}+d_n}\right],\;\text{if}\;\frac{a_{n}+c_n}{b_{n}+d_n}>\frac{i}{2N},
	\end{cases}
\end{equation}
where the starting values for our sequence are $a_0\!=\!0$ and $b_0\!=\!c_0\!=d_0\!=\!1$.
We terminate the sequence at step $n$ when either the fraction $\frac{a_n}{b_n}$ or the fraction $\frac{c_n}{d_n}$ is inside the interval $\left[\frac{i-1}{2N},\frac{i}{2N}\right]$.
This resulting fraction has then the smallest possible denominator in the given interval.

\subsubsection{Improved frequency sums}
\label{Sec:ImprovedFrequencySums}

Equation~(\ref{equ:FreeEnergy}) for the evaluation of the DMFT lattice free energy $F$ requires to sum the logarithm of the determinant of the impurity and DMFT lattice Green's function over all Matsubara frequencies.
In practice, this sum has obviously to be truncated at a given finite frequency which gives rise to a truncation error.
To mitigate this error, we have expanded the addend in Eq.~(\ref{equ:FreeEnergy}) in terms of $\frac{1}{i\nu}$
\begin{align}
	\frac{1}{q}\sum_\mathbf{k}\ln\left(\det\left[\frac{G_\text{imp}(i\nu)\mathds{1}_{q\times q}}{G_\text{DMFT}(i\nu,\mathbf{k})}\right]\right)=&\frac{-2t^2+\sum_n V_n^2}{(i\nu)^2}\nonumber\\&+\mathcal{O}\left[\frac{1}{(i\nu)^3}\right].
\end{align}
We can subtract the first term on the right-hand side of this equation in the truncated numerical frequency sum and sum it up analytically for all frequencies which yields $-\frac{\beta}{4}\left[-2t^2+\sum_n V_n^2\right]$.
This procedure improves considerably the accuracy of our numerical results as our truncation error depends cubically (instead of linearly) on the inverse cutoff frequency.

\subsection{Strong coupling expansion of \texorpdfstring{$E_\text{kin}$}{Ekin} and \texorpdfstring{$E_\text{pot}$}{Epot}}
\label{sec:largeUexpansion}

In this section, we provide a derivation for the strong coupling expansion which has been exploited in the main text to estimate the magnetic field dependence of $E_\text{kin}$ and $E_\text{pot}$ in the Mott insulating regime.
Within this approach the interaction term of the Hubbard model in Eq.(\ref{equ:Hamiltonian}) corresponds to the unperturbed part of the Hamiltonian while the hopping part represents the perturbation.
Hence, this method, which has been introduced in Refs.~\cite{S-Pairault1998,S-Pairault2000}, can be interpreted as an expansion around the atomic limit which leads to an expansion of the correlation functions in powers of the hopping $t$.
In its lowest order $t^0$, we can simply replace the exact self-energy of the system by its atomic limit counterpart $\Sigma_\text{AL}(i\nu)$ which (for the half-filled case) is given by
\begin{equation}
	\label{equ:sigmaAL}
	\Sigma_\text{AL}(i\nu)=\frac{U}{2}+\frac{U^2}{4i\nu}.
\end{equation}
This approximation can now be exploited in the standard equations for $E_\text{kin}$ and $E_\text{pot}$ which gives
\begin{subequations}
	\label{equ:ekinepotapprox}
	\begin{align}
		\label{equ:ekinapprox}&E_\text{kin}=2\frac{1}{\beta}\frac{1}{q}\sum_{\nu\mathbf{k}}\Tr\left[\frac{\varepsilon_\mathbf{k}}{\left[i\nu+\mu-\Sigma_\text{AL}(i\nu)\right]\mathds{1}_{q\times q}-\varepsilon_\mathbf{k}}\right]\\
		\label{equ:epotapprox}&E_\text{pot}=\frac{1}{\beta}\frac{1}{q}\sum_{\nu\mathbf{k}}\Tr\left[\frac{\Sigma_\text{AL}(i\nu)\mathds{1}_{q\times q}}{\left[i\nu+\mu-\Sigma_\text{AL}(i\nu)\right]\mathds{1}_{q\times q}-\varepsilon_\mathbf{k}}\right],
	\end{align}
\end{subequations}
where the $q\!\times\!q$ dispersion matrix $\varepsilon_\mathbf{k}$ is given in Eq.~(\ref{equ:HarperMatrix}).
We now want to expand the above expressions for $E_\text{kin}$ and $E_\text{pot}$ in powers of $t$ which obviously correspond to an expansion in powers of $\varepsilon_\mathbf{k}$.
This is achieved by the following identity
\begin{align}
	\label{equ:highUexpansionG}
	\frac{1}{\left[i\nu+\mu-\Sigma_\text{AL}(i\nu)\right]\mathds{1}_{q\times q}-\varepsilon_\mathbf{k}}=\sum_{n=0}^\infty \frac{\varepsilon_\mathbf{k}^{n}}{\left[i\nu+\mu-\Sigma_\text{AL}(i\nu)\right]^{n+1}}.
\end{align}
Inserting this series into Eqs.~(\ref{equ:ekinepotapprox}) and considering that momentum sums over odd powers of $\varepsilon_\mathbf{k}$ vanish we obtain the following expressions for $E_\text{kin}$ and $E_\text{pot}$ up to the fourth power in $\varepsilon_\mathbf{k}$
\begin{subequations}
	\label{equ:expandekinepot}
	\begin{align}
		\label{equ:expandekin}E_\text{kin}=&2\frac{1}{\beta}\sum_\nu\frac{1}{\left[i\nu+\mu-\Sigma_\text{AL}(i\nu)\right]^2}\frac{1}{q}\sum_\mathbf{k}\Tr\left[\varepsilon_\mathbf{k}^2\right]\nonumber\\+&2\frac{1}{\beta}\sum_\nu\frac{1}{\left[i\nu+\mu-\Sigma_\text{AL}(i\nu)\right]^4}\frac{1}{q}\sum_\mathbf{k}\Tr\left[\varepsilon_\mathbf{k}^4\right]+\mathcal{O}\left(\varepsilon_\mathbf{k}^6\right)\\
		\label{equ:expandepot}E_\text{pot}=&\frac{U}{4}+\frac{1}{\beta}\sum_\nu\frac{\frac{U^2}{4i\nu}}{\left[i\nu+\mu-\Sigma_\text{AL}(i\nu)\right]}\nonumber\\+&\frac{1}{\beta}\sum_\nu\frac{\frac{U^2}{4i\nu}}{\left[i\nu+\mu-\Sigma_\text{AL}(i\nu)\right]^3}\frac{1}{q}\sum_\mathbf{k}\Tr\left[\varepsilon_\mathbf{k}^2\right]\nonumber\\+&\frac{1}{\beta}\sum_\nu\frac{\frac{U^2}{4i\nu}}{\left[i\nu+\mu-\Sigma_\text{AL}(i\nu)\right]^5}\frac{1}{q}\sum_\mathbf{k}\Tr\left[\varepsilon_\mathbf{k}^4\right]+\mathcal{O}\left(\varepsilon_\mathbf{k}^6\right).
	\end{align}
\end{subequations}
Considering Eq.~(\ref{equ:sigmaAL}) for the atomic limit approximation of $\Sigma_\text{AL}(i\nu)$ and taking into account that $\mu\!=\!\frac{U}{2}$ in the half filled bipartite system we can perform the Matsubara sums analytically using \texttt{Mathematica}\cite{S-Wolfram2024}.
While this leads to rather complicated expression we can achieve a substantial simplification by considering the limit of zero temperature $T\!\rightarrow\!0$. Taking furthermore into account that $\frac{1}{q}\sum_\mathbf{k}\Tr\left[\varepsilon_\mathbf{k}^2\right]\!=\!4t^2$ we obtain the following expressions for the kinetic and potential energies
\begin{subequations}
	\label{equ:epotekinT0}
	\begin{align}
		\label{equ:ekinT0}E_\text{kin}&=-\frac{4t^2}{U}+\frac{1}{2U^3}\frac{1}{q}\sum_\mathbf{k}\Tr\left[\varepsilon_\mathbf{k}^4\right]+\mathcal{O}\left(\varepsilon_\mathbf{k}^6\right)\\
		\label{equ:epotT0}E_\text{pot}&=\frac{t^2}{2U}-\frac{3}{32U^3}\frac{1}{q}\sum_\mathbf{k}\Tr\left[\varepsilon_\mathbf{k}^4\right]+\mathcal{O}\left(\varepsilon_\mathbf{k}^6\right).
	\end{align}
\end{subequations}
The sum over the fourth power of the dispersion relation evaluates to
\begin{equation}
	\label{equ:eps4}
	\frac{1}{q}\sum_\mathbf{k}\Tr\left[\varepsilon_\mathbf{k}^4\right]=4t^4\left[7+\cos(2\pi B)\right].
\end{equation}
Let us stress again that, quite remarkably, the magnetic field appears only in the fourth moment $\frac{1}{q}\sum_\mathbf{k}\Tr\left[\varepsilon_\mathbf{k}^4\right]$ and higher moments of the non-interacting density of states while the second moment $\frac{1}{q}\sum_\mathbf{k}\Tr\left[\varepsilon_\mathbf{k}^2\right]$ is field independent.
Hence, the effect of the orbital magnetic field sets in only at the order $\frac{t^4}{U^3}$.
Inserting Eq.~(\ref{equ:eps4}) into Eqs.~(\ref{equ:epotekinT0}) we obtain
\begin{subequations}
	\label{equ:ekinepotfinal}
	\begin{align}
		&E_\text{kin}=-\frac{4t^2}{U}+2\frac{t^4}{U^3}\left[7+\cos\left(2\pi B\right)\right]+\mathcal{O}\left(t^6\right)\label{equ:ekinfinal}\\
		&E_\text{pot}=\frac{t^2}{2U}-\frac{3}{8}\frac{t^4}{U^3}\left[7+\cos\left(2\pi B\right)\right]+\mathcal{O}\left(t^6\right).\label{equ:epotfinal}
	\end{align}    
\end{subequations}
The corresponding magnetic field dependent parts of $E_\text{kin}$ and $E_\text{pot}$ in the order $\frac{t^4}{U^3}$ are then given by $\pm\frac{t^4}{U^3}\cos(2\pi B)$.
In real space, the diagram which generates this contribution can be interpreted as an electron hopping around a plaquette as it is illustrated in panel~(e) of Fig.~\ref{fig:KineticEnergyBeta400} of the main text.
The increase of the absolute values of kinetic and potential energy with $B$ can be then attributed to the Aharonov-Bohm\cite{S-aharonov1959significance} effect which the electron experiences during its virtual hopping process around the plaquette.

\begin{figure*}
	\centering
	\includegraphics[width=0.49\textwidth]{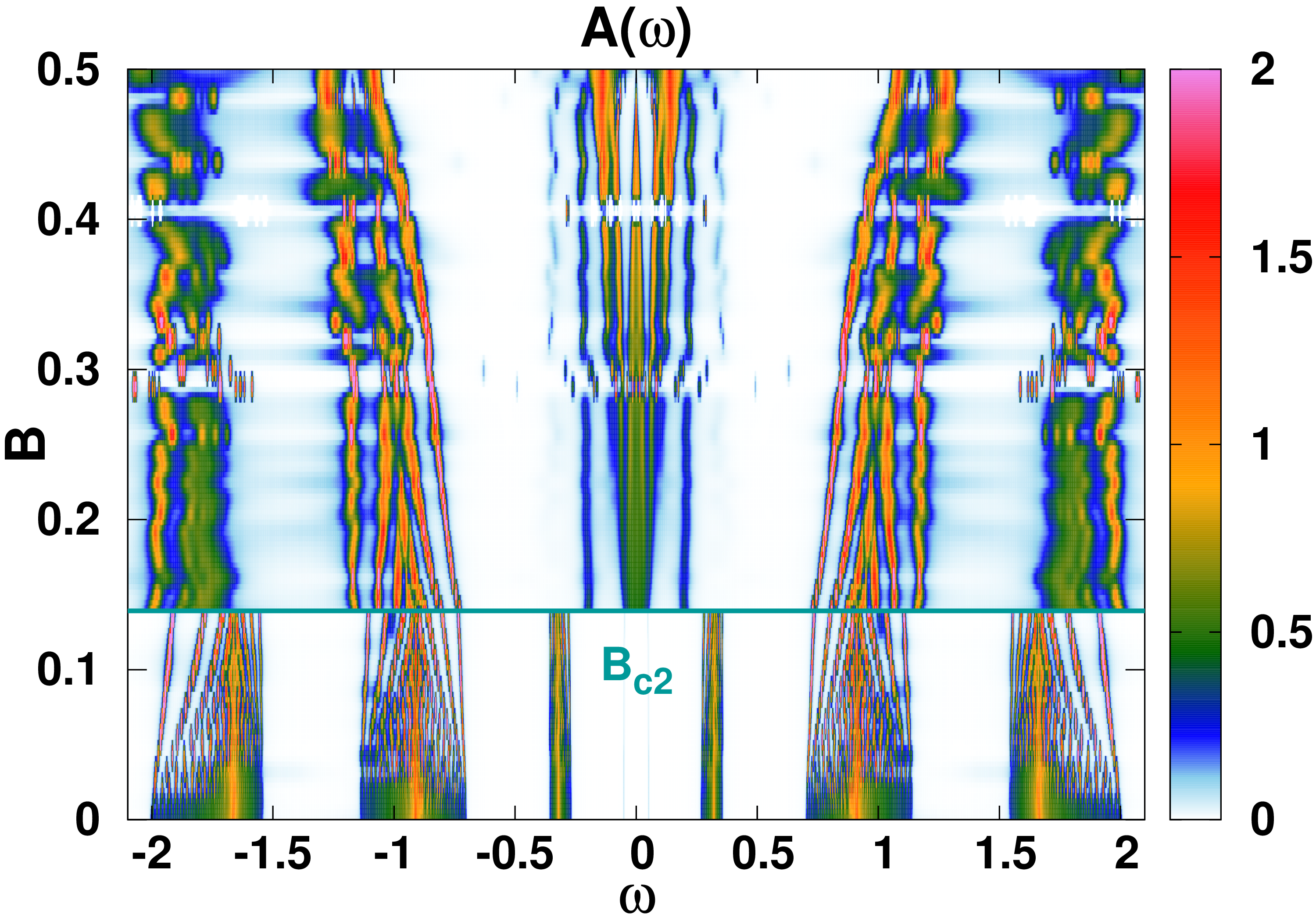}
	\includegraphics[width=0.49\textwidth]{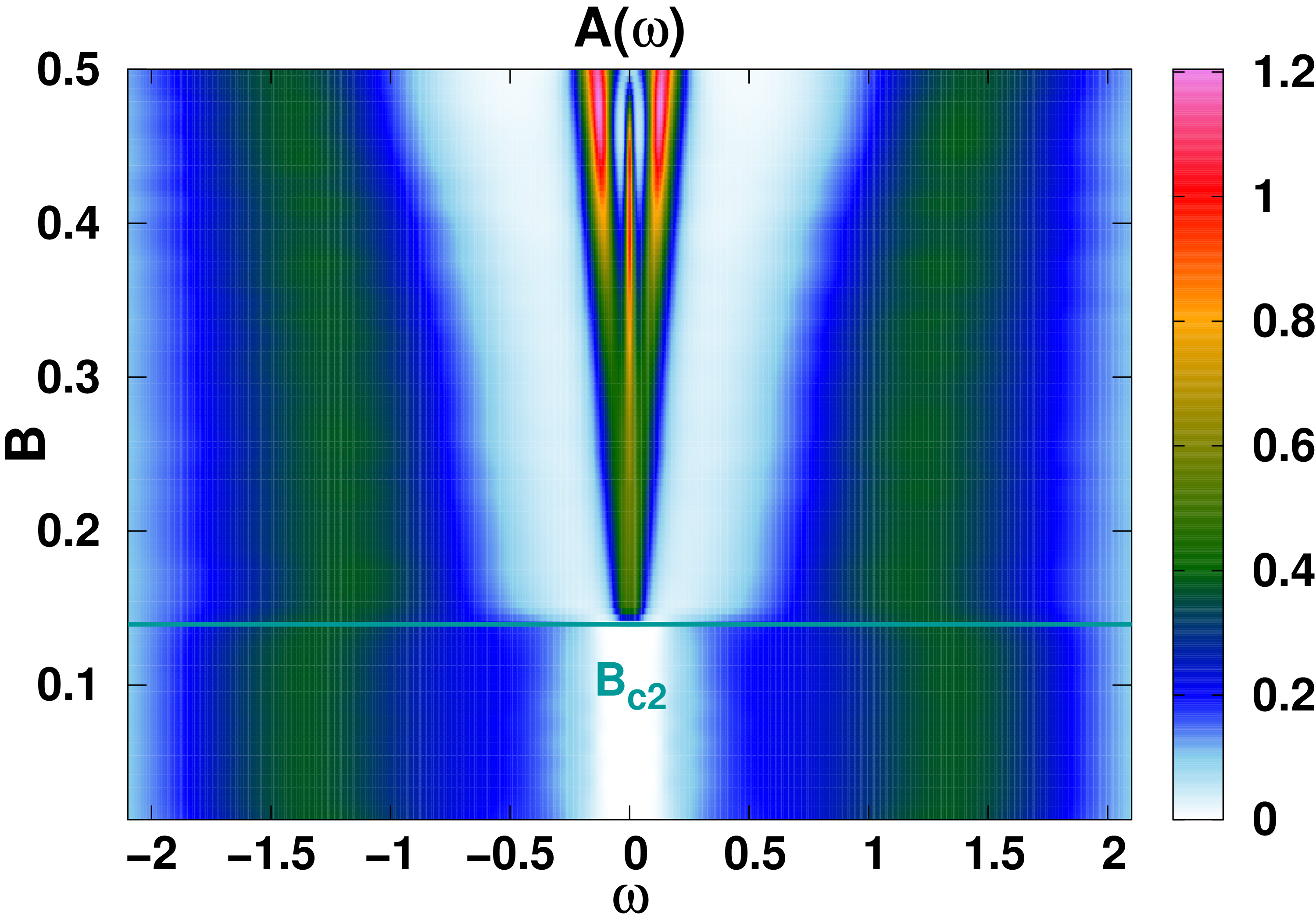}
	\caption{Spectral function $A(\omega)$ for $U\!=\!2.5$ and $T\!=\!0.01$ (corresponding to P1 in the phase diagram of Fig.~\ref{fig:PhasediagramB0} of the main text) plotted as a heat map in the $(\omega,B)$ plane. Left panel (which is just a reproduction of the right panel of Fig.~\ref{fig:InsulatorMetal} of the main text): Results obtained via the second line of Eq.~(\ref{equ:spectralfunctionDMFT}) where the ED impurity self-energy $\Sigma_\text{imp}(i\nu)$ has been analytically continued by a Nevanlinna fit. Right panel: Results obtained from the first line of Eq.~(\ref{equ:spectralfunctionDMFT}) where the QMC impurity Green's function $G_\text{imp}(i\nu)$ has been analytically continued via the Maximum Entropy approach.}
	\label{fig:spectralfunctioncompare}
\end{figure*}

Let us finally mention that the calculations above do not present a fully consistent expansion of $E_\text{kin}$ and $E_\text{pot}$ in powers of $t$.
In fact, to achieve this also higher-order contributions in $t$ to the self-energy have to be taken into account\cite{S-sen1995large,S-Pairault2000}. However, since such contributions would include nonlocal (momentum dependent) correlations which are beyond our DMFT treatment we neglect such terms here and will address this question in a future research work.


\section{Calculation of spectral function and conductivity}
\label{Sec:spectralfunctionconductivity}

In the middle and right panel of Fig.~\ref{fig:InsulatorMetal} of the main text we present data for the DC conductivity $\sigma_{xx}(\omega\!=\!0)$ and the local spectral function $A(\omega)$ of the Hubbard-Hofstadter model as a function of the magnetic field, respectively. In this part of the SM we will briefly outline how these observables have been obtained from our DMFT data. We will start with the calculation of $A(\omega)$ and then proceed to $\sigma_{xx}(\omega\!=\!0)$ which requires the (momentum dependent) spectral function as an input.

\subsection{Calculation of the local spectral function \texorpdfstring{$A(\omega)$}{A(w)}}
\label{Sec:spectralfunction}

The local spectral function $A(\omega)$ is defined through the negative imaginary part of the local (i.e., momentum-summed) retarded Green's function which depends on the real frequency $\omega$:
\begin{align}
	\label{equ:spectralfunctionfromimpurity}
	A(\omega)&=-\frac{1}{\pi}\Im G^\text{R}_\text{loc}(\omega)=-\frac{1}{\pi}\Im\frac{1}{q}\Tr\sum_\mathbf{k}G^\text{R}(\omega,\mathbf{k})\nonumber\\&=-\frac{1}{\pi}\Im\frac{1}{q}\Tr\sum_\mathbf{k}\frac{1}{(\omega+\mu)\mathds{1}_{q\times q}-\varepsilon_\mathbf{k}-\Sigma^\text{R}(\omega,\mathbf{k})},
\end{align}
where $G^\text{R}(\omega,\mathbf{k})$ is the retarded Green's function, $\varepsilon_\mathbf{k}$ is the non-interacting dispersion of the Hofstadter model defined in Eq.~(\ref{equ:HarperMatrix}) and $\Sigma^\text{R}(\omega,\mathbf{k})$ is the retarded self-energy. 
Let us recall that all these objects are $q\!\times\!q$ matrices in orbital space as detailed in Sec.~\ref{Sec:DMFTHofstadter}. 
Moreover, in this section it has been discussed that, within the framework of DMFT, the self-energy is purely local and can be, hence, represented as a scalar self-energy $\Sigma^\text{R}(\omega)$ times the $q\!\times\!q$ unit matrix, i.e., $\Sigma^\text{R}(\omega,\mathbf{k})\!=\!\Sigma^\text{R}(\omega)\mathds{1}_{q\times q}$. 
The diagonal elements of the local Green's function are all equivalent which implies that $\frac{1}{q}\Tr\sum_\mathbf{k}G^\text{R}(\omega,\mathbf{k})\!=\!\sum_\mathbf{k}\left[G^\text{R}(\omega,\mathbf{k})\right]_{ll}$, $l\!=\!1,\ldots,q$,  i.e., the trace (normalized by $q$) is equal to each of the diagonal elements of this local correlation function. 
The DMFT self-consistency condition in Eq.~(\ref{equ:DMFTselfconsistency}) implies that this local Green's function is equivalent to the impurity Green's function $G^\text{R}_\text{imp}(\omega)$. 
Hence, within the framework of DMFT the local spectral function can be expressed as
\begin{align}
	\label{equ:spectralfunctionDMFT}
	A(\omega)&=-\frac{1}{\pi}\Im G_\text{imp}^\text{R}(\omega)\nonumber\\&=-\frac{1}{\pi}\sum_\mathbf{k}\left[\frac{1}{[\omega+\mu-\Sigma^\text{R}_\text{imp}(\omega)+i\delta]\mathds{1}_{q\times q}-\varepsilon_\mathbf{k}}\right]_{ll}.
\end{align}
The evaluation of $A(\omega)$ then requires an analytical continuation $i\nu\!\rightarrow\!\omega\!+\!i\delta$ of the DMFT impurity Green's function $G_\text{imp}(i\nu)$ or self-energy $\Sigma_\text{imp}(i\nu)$ to the real frequency axis as they are typically calculated for imaginary (Matsubara) frequencies $i\nu$ in DMFT at finite temperatures. 
Within ED the analytical continuation of $G_\text{imp}(i\nu)$ is, in principle, straightforward as the Lehmann representation allows for a direct evaluation of the Green's function on the real frequency axis. 
However, due to the finite Hilbert space of the auxiliary AIM in this approach, the resulting spectral function would consist of a (large) number of $\delta$-peaks which have no direct physical interpretation as they strongly depend on the exact values of the auxiliary bath parameters $\epsilon_{n}$ and $V_n$ [see Eq.~(\ref{equ:AIM})].

We have, hence, analytically continued the self-energy to the real frequency axis $\Sigma_\text{imp}(i\nu)\!\-\rightarrow\!\Sigma_\text{imp}^\text{R}(\omega+i\delta)$ and exploited the second line of Eq.~(\ref{equ:spectralfunctionDMFT}) to determine the local spectral function $A(\omega)$. 
As for the actual continuation $i\nu\!\rightarrow\!\omega+i\delta$ we have used the recently developed Nevanlinna technique\cite{S-Fei2021,S-Nogaki2023,S-Nogaki2023a} which guarantees that the spectral function is analytic in the upper complex frequency plane and positive on the real frequency axis. 
The results for $A(\omega)$ as a function of $\omega$ and $B$ for solution S2 (i.e., where the magnetic field is decreased from $B\!=\!0.5$ to $0$) at point P1 ($U\!=\!2.5$, $T\!=\!0.01$) in the phase diagram of Fig.~\ref{fig:PhasediagramB0} in the main text are shown as a heat map in the $(\omega,B)$ plane in the right panel of Fig.~\ref{fig:InsulatorMetal} in the main text and are replotted for convenience in the left panel of Fig.~\ref{fig:spectralfunctioncompare} in this SM.
Our numerical DMFT data include the most important spectral features and capture well the overall distribution of spectral weight.
In particular, we observe (i) a gap for $B\!<\!B_{c2}$ and final spectral weight for $B\!>\!B_{c2}$ around $\omega\!=\!0$ which is a clear hallmark of the metal-insulator transition at the magnetic field $B_{c2}$, (ii) a reduction of spectral weight around $\omega\!=\!0$ for $B\!\rightarrow\!0.5$ at $\omega\!=\!0$ due to the emergence of a Dirac cone in the non-interacting density of states, (iii) a size of the spectral gap around $\omega\!=\!0$ in the insulating phase $B\!<\!B_{c2}$ which is given by $E_g\!\approx\!U-W\!\approx\!0.5$ consistent with the literature\cite{S-Sangiovanni2006,S-Wang2009} and (iv) Hubbard bands which are centered around $\omega\!\approx\!\pm 1.125$ which is in reasonably quantitative agreement with the expected value $\frac{U}{2}$.

\begin{figure}
	\centering
	\includegraphics[width=1.0\linewidth]{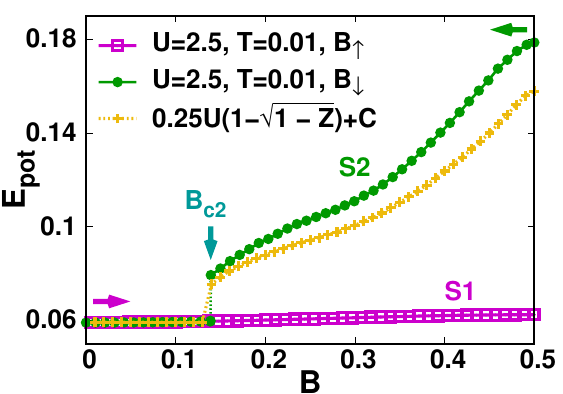}
	\caption{Potential energy $E_\text{pot}\!=\!U\langle n_{i\uparrow}n_{i\downarrow}\rangle$ for point P1 in Fig.~\ref{fig:PhasediagramB0} of the main text corresponding to $U\!=\!2.5$ and $T\!=\!0.01$. Data are shown for scans where the magnetic field has been decreased from $B\!=\!0.5$ to $B\!=\!0$ (filled green circles, solution S2) and where the magnetic field has been increased from $B\!=\!0$ to $B\!=\!0.5$ (empty violet squares, solution S1), respectively. Yellow crosses depict the potential energy estimated from the Gutzwiller approximation where $Z$ is the quasiparticle renormalization factor of DMFT and $C\!=\!E_\text{pot}^\text{ins}$ corresponds to the potential energy of DMFT in the insulating state.}
	\label{fig:potentialenergyU2.5beta100metaltoinsulator}
\end{figure}

To benchmark our ED/Nevanlinna results we have compared them to CTQMC calculations with the \texttt{w2dyanmics} software package\cite{S-Wallerberger2019} which are in excellent agreement for all thermodynamic observables and Matsubara data as it is demonstrated in Sec.~\ref{Sec:compareQMC}.
Within the QMC method the local impurity Green's function $G_\text{imp}(i\nu)$ can be directly analytically continued to obtain $A(\omega)$ via the first line of Eq.~(\ref{equ:spectralfunctionDMFT}) as discretization errors due to a finite bath are absent in this approach.
However, due to the intrinsic statistical noise a Nevanlinna fit is unreliable and, therefore, we have exploited a standard Maximum Entropy (MaxEnt) method\cite{S-Jarrell1996} as provided by the \texttt{ana\_cont} package\cite{S-Kaufmann2023}.
The corresponding results are presented in the right panel of Fig.~\ref{fig:spectralfunctioncompare} and show an excellent agreement with the ED findings for the main spectral features around $\omega\!=\!0$ and the general distribution of spectral weight.
In particular, they reproduce (i) the gap around $\omega\!=\!0$ for $B\!<\!B_{c2}$, (ii) the reduction of spectral weight at $\omega\!=\!0$ for $B\!\rightarrow\!0.5$, (iii) a size of the spectral gap in the insulating state of $E_g\!\approx\!0.5\!=\!U-W$ and (iv) Hubbard bands which are centered around $\omega\!\approx\! 1.125$.

Let us, however, note that the two methods (ED/Nevanlinna vs. CTQMC/MaxEnt) discussed here differ in specific features of the spectral function $A(\omega)$ at high frequencies (i.e., in the Hubbard bands).
While the Nevanlinna fit resolves well the fine spectral details (magnetic mini bands) within the Hubbard bands which originate from the non-interacting Hofstadter spectrum, it also introduces a splitting of these bands into six parts (three for the upper and three for the lower Hubbard band) which is likely an artifact originating from the choice of six bath sites in our ED algorithm.
On the contrary, such spurious features are absent in the QMC spectral function as no bath discretization is required.
However, the MaxEnt technique washes out {\em all} features within the Hubbard bands and just leaves a structureless bunch of spectral weight at high energies without any remainders of the underlying spectral structure of the Hofstadter model.
In this respect, ED/Nevanlinna and CTQMC/MaxEnt can be viewed as complimentary approaches:
while the first method is able to resolve very fine details of the spectrum at high energies but also introduces spurious features due to the finite ED bath the second approach avoids such artifacts but is unable to resolve {\em any} particularities of the Hubbard bands.
In this respect, let us stress that an exact reproduction of the spectral function at all energy scales by means of analytical continuation of Matsubara data is an infeasible task as this problem is intrinsically ill-defined from a mathematical perspective.



\subsection{Calculation of the DC conductivity \texorpdfstring{$\sigma_{xx}(\omega\!=\!0)$}{sigma(w=0)}}
\label{Sec:conductivity}

\begin{figure*}[t!]
	\centering
	\includegraphics[width=1.0\textwidth]{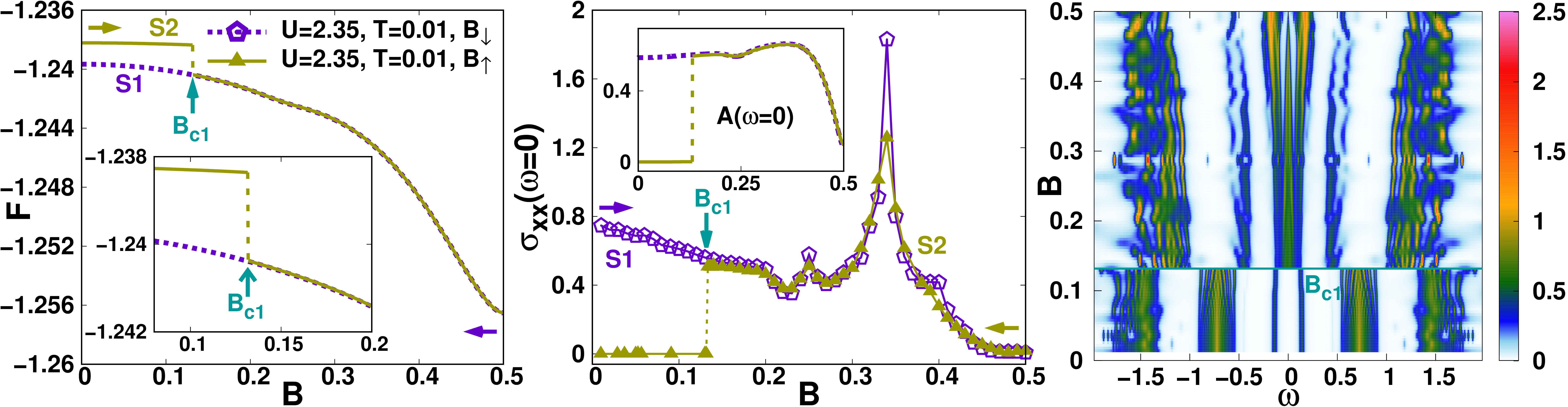}
	\caption{Same as in Fig.~\ref{fig:InsulatorMetal} of the main text but for the point P2 in the phase diagram of Fig.~\ref{fig:PhasediagramB0} of the main text corresponding to the parameters $U\!=\!2.35$ and $T\!=\!0.01$.}
	\label{fig:MetalInsulator}
\end{figure*}

\begin{figure}[t!]
	\centering
	\includegraphics[width=0.5\textwidth]{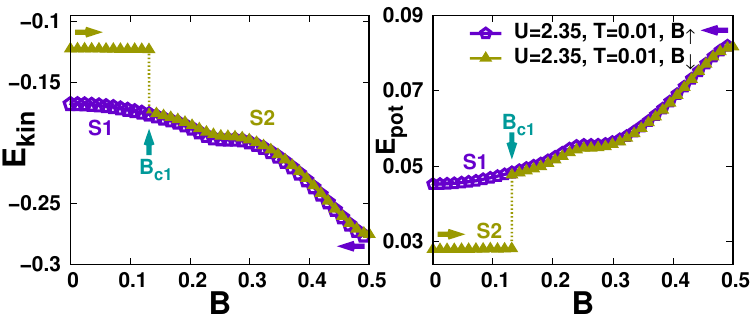}
	\caption{Kinetic (left panel) and potential (right panel) energies for $U\!=\!2.35$ and $T\!=\!0.01$ (corresponding to the point P2 in the phase diagram in Fig.~\ref{fig:PhasediagramB0} of the main text) as a function of the orbital magnetic field $B$ obtained by the two sets of calculations $B_\uparrow$ and $B_\downarrow$.}
	\label{fig:PotentialKineticEnergyU2.5Beta100}
\end{figure}

To calculate the DC conductivity we exploit the linear response Kubo formalism\cite{S-Kubo1957}. Within this theory the electrical conductivity can be expressed as\cite{S-Markov2019}:
\begin{align}
	\label{equ:Kuboconductivity}
	\sigma_{xx}&(\omega=0)=\nonumber\\&-\frac{2\pi}{q}\int_{-\infty}^\infty d\omega_1 f'(\omega_1)\sum_\mathbf{k}\Tr[v_\mathbf{k}^xA(\omega_1,\mathbf{k})v_\mathbf{k}^xA(\omega_1,\mathbf{k})],
\end{align}
where $f(\omega)\!=\!\frac{1}{1+e^{\beta\omega}}$ denotes the Fermi function [and $f'(\omega)$ its first derivative with respect to $\omega$], $A(\omega,\mathbf{k})\!=\!-\frac{1}{\pi}G^\text{R}(\omega,\mathbf{k})\!=\!-\frac{1}{\pi}\Im\frac{1}{[\omega+\mu-\Sigma^\text{R}_\text{imp}(\omega)]\mathds{1}_{q\times q}-\varepsilon_\mathbf{k}}$ is the momentum dependent spectral function  [which corresponds to a $q\!\times\!q$ matrix in orbital space, see Eq.~(\ref{equ:spectralfunctionfromimpurity})], and $v_\mathbf{k}^x\!=\!\frac{\partial \varepsilon_\mathbf{k}}{\partial k_x}$ is the current operator (which is also a $q\!\times\!q$ matrix in orbital space). 
The retarded impurity self-energy $\Sigma^\text{R}_\text{imp}(\omega)$ can be readily obtained from our DMFT data for the Matsubara self-energy $\Sigma_\text{imp}(i\nu)$ by means of analytic continuation.
Let us note that while for point P1 in the phase diagram of Fig.~\ref{fig:PhasediagramB0} of the main text the Nevanlinna technique has been used for this task (see Sec.~\ref{Sec:spectralfunction}), for the spectral functions for points P2-P4 presented in Secs.~\ref{Sec:InsulatorToMetal} and \ref{Sec:additionaldata} a simpler Pad\'{e} fit has been exploited.
The validity of this approach has been verified by comparing Pad\'{e} and Nevanlinna data for point P1.

\section{Potential Energy}
\label{sec:potentialenergyU2.5beta100}

For completeness, we present in Fig.~\ref{fig:potentialenergyU2.5beta100metaltoinsulator} the magnetic field dependence of the potential energy $E_\text{pot}\!=\!U\langle n_{i\uparrow} n_{i\downarrow}\rangle$ per lattice site (which is just the average double occupancy $D\!=\!\langle n_{i\uparrow} n_{i\downarrow}\rangle$ multiplied with the interaction strength $U$).
We observe an analogous behavior as for the free energy, DC conductivity, spectral function at the Fermi level and, in particular, the kinetic energy in Figs.~\ref{fig:InsulatorMetal} and \ref{fig:KineticEnergyBeta400} of the main text.
If we begin with our calculations for $E_\text{pot}$ (and $\left \lvert E_\text{kin}\right\rvert$ in the main text) at $B\!=\!0$ and gradually increase the magnetic field (empty violet squares, S1) the system remains in the insulating state from which it started at zero field for $U\!=\!2.5$ and $T\!=\!0.01$ (corresponding to P1 in the phase diagram in Fig.~\ref{fig:PhasediagramB0} of the main text) which is indicated by the rather low values of the potential energy and the continuous evolution with increasing $B$.
At the largest magnetic field $B\!=\!0.5$ we can, however, stabilize a second solution S2 with a substantially higher potential energy (filled green circles) indicative of a metal.
Upon decreasing the magnetic field, the potential energy (or the absolute value of the kinetic energy in Fig.~\ref{fig:KineticEnergyBeta400} in the main text) rapidly decreases. 
At the critical magnetic field $B_{c2}$ it features a discontinuity and collapses to the insulating state S1 which provides further evidence for the occurrence of a first-order metal-to-insulator transition at this magnetic field which has been observed for other observables in the main text.

As for the kinetic energy in the main text, we also compare our results for $E_\text{pot}$ obtained from DMFT with the Gutzwiller estimate provided in Refs.~\cite{S-Brinkman1970,S-Vollhardt1984,S-Fabrizio2017} (yellow crosses). 
We observe a good agreement with our numerical DMFT data (albeit not as perfect as for $E_\text{kin}$ in Fig.~\ref{fig:KineticEnergyBeta400} of the main text).
The constant $C$ corresponds to the potential energy in the insulating state $E_\text{pot}^\text{ins}\!=\!0.06$ which originates from the incoherent part of the spectrum and also agrees reasonably well with the value $0.05$ predicted by the improved Gutzwiller treatment of Ref.~\cite{S-Fabrizio2017}.

\section{Insulator-to-metal transition}
\label{Sec:InsulatorToMetal}

\begin{figure*}[t!]
	\centering
	\includegraphics[width=0.7\linewidth]{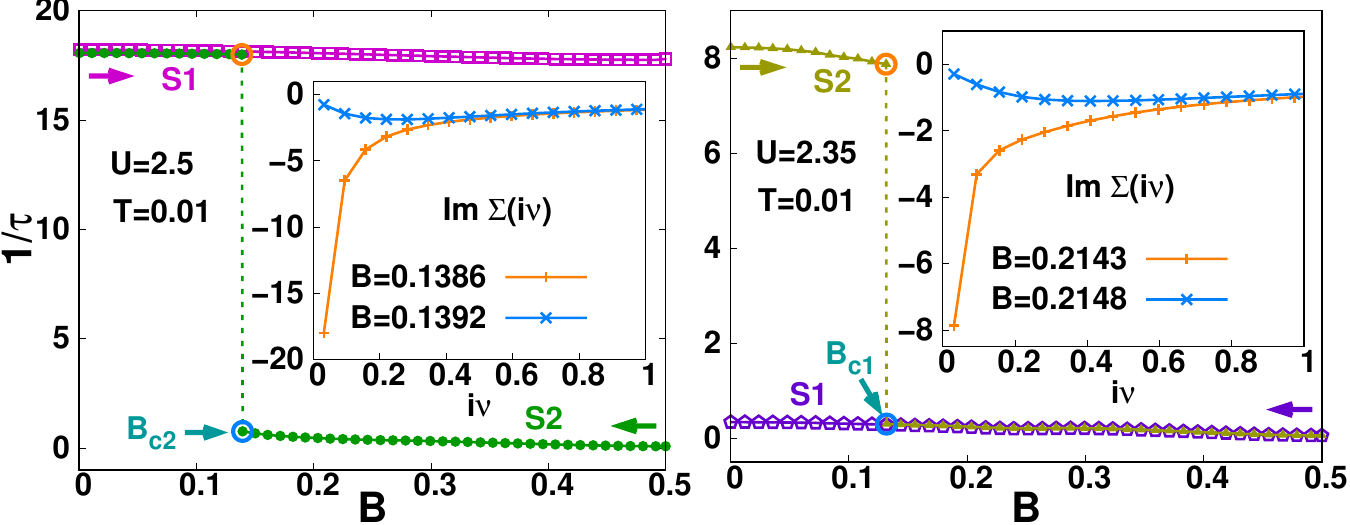}
	\caption{Inverse quasiparticle life time $\frac{1}{\tau}\!=\!-\Im \Sigma(i\nu_1)$, where $\nu_1\!=\!\pi T$ is the first (positive) Matsubara frequency, vs. orbital magnetic field $B$ at $T\!=\!0.01$ for $U\!=\!2.5$ (left panel) and $U\!=\!2.35$ (right panel) corresponding to the points P1 and P2 in the phase diagram of Fig.~\ref{fig:PhasediagramB0} in the main text, respectively. The insets depict the frequency dependence of the imaginary part of the self-energy $\Im \Sigma(i\nu)$ for $B$ values right before and after the phase transition which are marked by small colored circles in the respective main panel.}
	\label{fig:InverseLifeTimeBeta400}
\end{figure*}

In the main text we have demonstrated the emergence of a metal-to-insulator transition driven by a decrease of the orbital magnetic field (c.f. state S2 in Fig.~\ref{fig:InsulatorMetal} and the related discussion). 
However, the opposite transition from an insulator to a metal triggered by an increase of the orbital magnetic field could not be observed since the corresponding boundary $B_{c1}$ between the coexistence region and the metallic state has not been reached for the largest magnetic field $B\!=\!0.5$.
While both $B_{c1}$ {\em and} $B_{c2}$ are indeed located within the interval $B\!\in\![0,0.5]$ for a geometrically frustrated lattice as discussed in Sec.~\ref{sec:FrustratedLattice} (see Fig.~\ref{fig:motttransitionfrustratedt1.0}), we explore here the possibility for finding an insulator-to-metal transition and the related $B_{c1}$ in the bipartite case.
This can be achieved by considering specific values of $U$ and/or $T$ such as the point P2 in Fig.~\ref{fig:PhasediagramB0} of the main text which is located inside the metallic part of the coexistence regime (for the system without magnetic field) and characterized by the interaction $U\!=\!2.35$ and the temperature $T\!=\!0.01$.
The free energy for this parameter set is depicted in the left panel of Fig.~\ref{fig:MetalInsulator}. Starting the calculations at $B\!=\!0.5$ and gradually decreasing the magnetic field ($B_\downarrow$, dashed violet line) $F$ is a continuous function of $B$ indicating that the system remains in the state $S1$ for all $B\!\in\![0,0.5]$.
However, at $B\!=\!0$ another state $S2$ can be stabilized in our DMFT calculations (solid olive line).
Upon increasing $B$ this state remains stable up to a magnetic field  $B_{c1}\!=\!0.1317$ where the free energy of state S2 features a discontinuity and collapses onto state S1.
This clearly indicates a first-order phase transition between the states S2 and S1 where $B_{c1}$ marks the boundary between the coexistence region between S1 and S2 (for $B\!<\!B_{c1})$ and the region where only one phase (S1=S2) exists.

\begin{figure*}[t!]
	\centering
	\includegraphics[width=\textwidth]{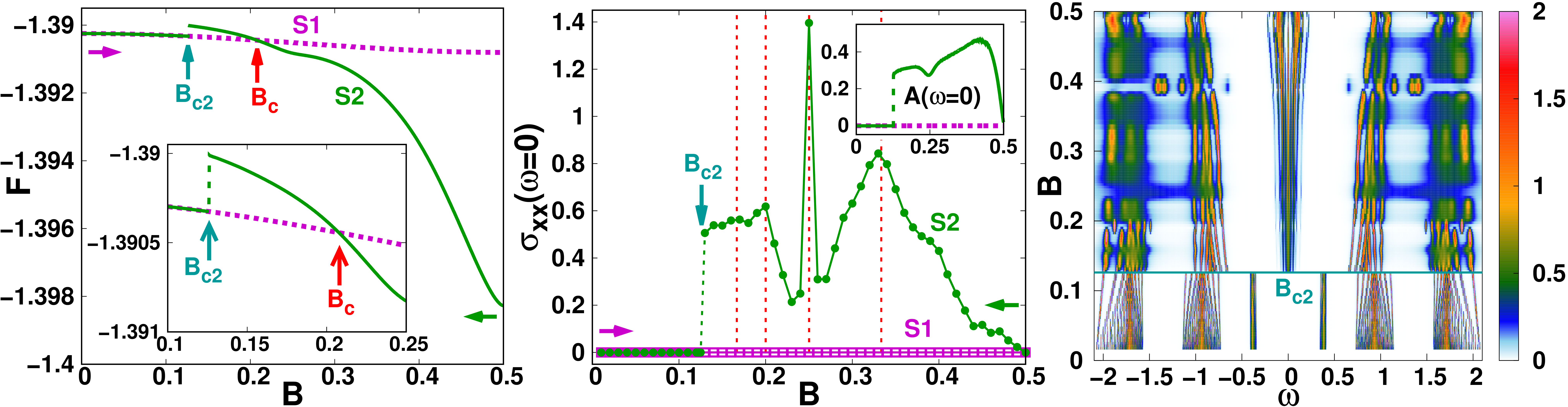}
	\caption{Same as in Fig.~\ref{fig:InsulatorMetal} of the main text but for the point P3 in the phase diagram of Fig.~\ref{fig:PhasediagramB0} of the main text corresponding to the parameters $U\!=\!2.65$ and $T\!=\!0.0025$.}
	\label{fig:InsulatorMetalBeta100}
\end{figure*}

\begin{figure*}[t!]
	\centering
	\includegraphics[width=\textwidth]{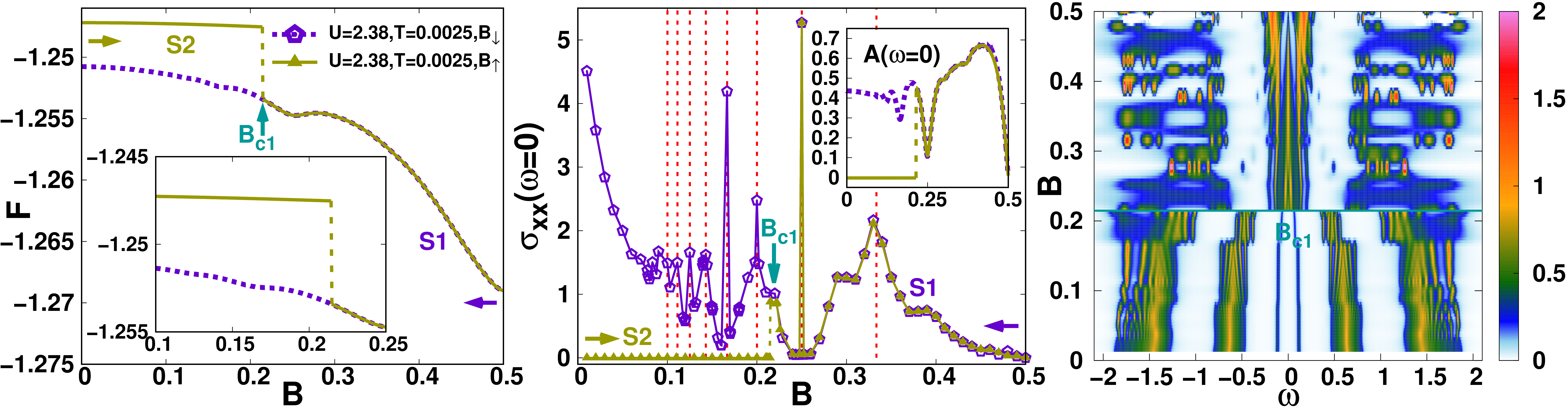}
	\caption{Same as in Fig.~\ref{fig:MetalInsulator} but for point P4 in the phase diagram of Fig.~\ref{fig:PhasediagramB0} of the main text corresponding to the parameters $U\!=\!2.38$ and $T\!=\!0.0025$.}
	\label{fig:MetalInsulatorBeta100}
\end{figure*}

\begin{figure}
	\centering
	\includegraphics[width=0.5\textwidth]{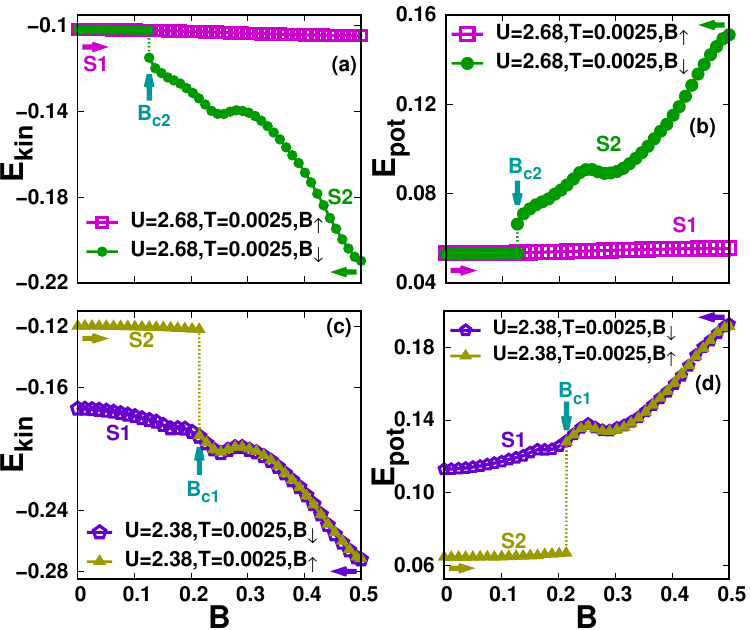}
	\caption{Kinetic (left) and potential (right) energies for $T\!=\!0.0025$ and $U\!=\!2.68$ (upper panel) and $U\!=\!2.38$ (lower panel).}
	\label{fig:EkinBeta100}
\end{figure}

\begin{figure}
	\centering
	\includegraphics[width=0.5\textwidth]{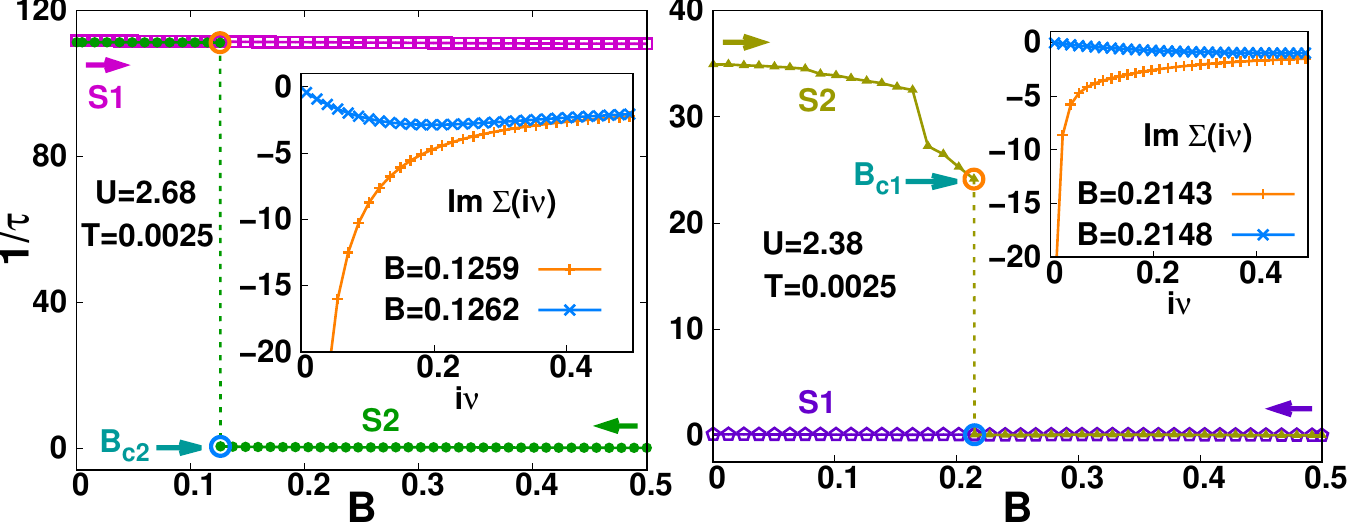}
	\caption{Same as in Fig~\ref{fig:InverseLifeTimeBeta400} but for $T\!=\!0.0025$ and $U\!=\!2.68$ (left panel) and $U\!=\!2.38$ (right panel).}
	\label{fig:InverseLiftimeBeta400}
\end{figure}

\begin{figure}[t!]
	\centering
	\includegraphics[width=0.5\textwidth]{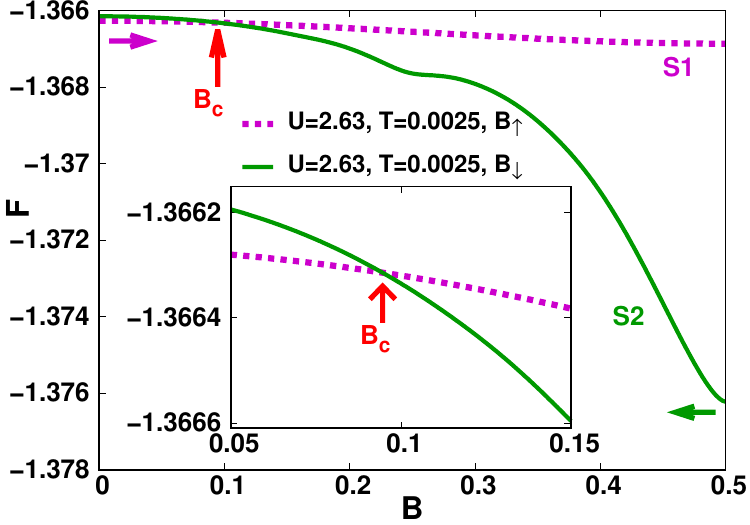}
	\caption{Same as left panel in Fig.~\ref{fig:InsulatorMetal} of the main text but for the point P5 in the phase diagram of Fig.~\ref{fig:PhasediagramB0} of the main text corresponding to the parameters $U\!=\!2.63$ and $T\!=\!0.0025$.}
	\label{fig:FreeEnergyU2.63}
\end{figure}

To identify the nature of the two states S1 and S2 we have analyzed the DC conductivity $\sigma_{xx}(\omega\!=\!0)$ and the local spectral function $A(\omega)$ in the middle and left panels of Fig.~\ref{fig:MetalInsulator}, respectively.
We observe that the conductivity is finite for the state S1 (empty violet pentagons) indicative of a conductor. 
On the contrary, $\sigma_{xx}(\omega\!=\!0)$ is zero for $B\!<\!B_{c1}$ in state S2 (olive triangles) implying that the system is an insulator.
At $B\!=\!B_{c1}$ the conductivity of state S2 jumps to a finite value equivalent to the one obtained for S1.
This proves that the observed change in phase is indeed a transition from an insulator S2 to a metal S1 driven by an increase of the orbital magnetic field.
This picture is confirmed by the evolution of the spectral function at the Fermi level $A(\omega\!=\!0)$ with the orbital magnetic field shown in the inset of the middle panel of Fig.~\ref{fig:MetalInsulator} which we have estimated from our Matsubara time Green's function as $\frac{\beta}{\pi}G\left(\tau\!=\!\frac{\beta}{2}\right)$.
For S1 (dashed violet line) we observe finite spectral weight in the entire $B$ interval typical for a metal while $A(\omega\!=\!0)$ is zero for solution S2 (solid olive line) in the region $B\!<\!B_{c1}$ which is a clear hallmark of an insulator.
In this respect, let us mention that both $\sigma_{xx}(\omega\!=\!0)$ and $A(\omega\!=\!0)$ also approach zero for the largest magnetic fields close to $B\!=\!0.5$ as a Dirac cone gradually emerges in the non-interacting density of states for $B\!\rightarrow\!0.5$.

To provide a comprehensive picture of the spectral function in the entire frequency regime we present $A(\omega)$ for all magnetic field strengths $B$ as a heat map in the $(\omega,B)$ plane for solution S2 (right panel of Fig.~\ref{fig:MetalInsulator}).
We observe the very same spectral features which have been discussed for the right panel of Fig.~\ref{fig:InsulatorMetal} in the main text and in Sec.~\ref{Sec:spectralfunction} in this SM: We can identify (i) a gap around $\omega\!=\!0$ for $B\!<\!B_{c1}$ and final spectral weight for $B\!>\!B_{c1}$ indicating a (Mott) metal-insulator transition, (ii) a reduction of spectral weight at $\omega\!=\!0$ for $B\!\rightarrow\!0.5$ due to the formation of a Dirac cone in the non-interacting density of states, (iii) a size $E_g\!\approx\!U-W$ of the spectral gap in the insulating phase which is consistent with previous results\cite{S-Sangiovanni2006,S-Wang2009} and (iv) Hubbard bands which are roughly centered around $\pm\frac{U}{2}$.


\begin{figure*}
	\centering
	\includegraphics[width=1.0\linewidth]{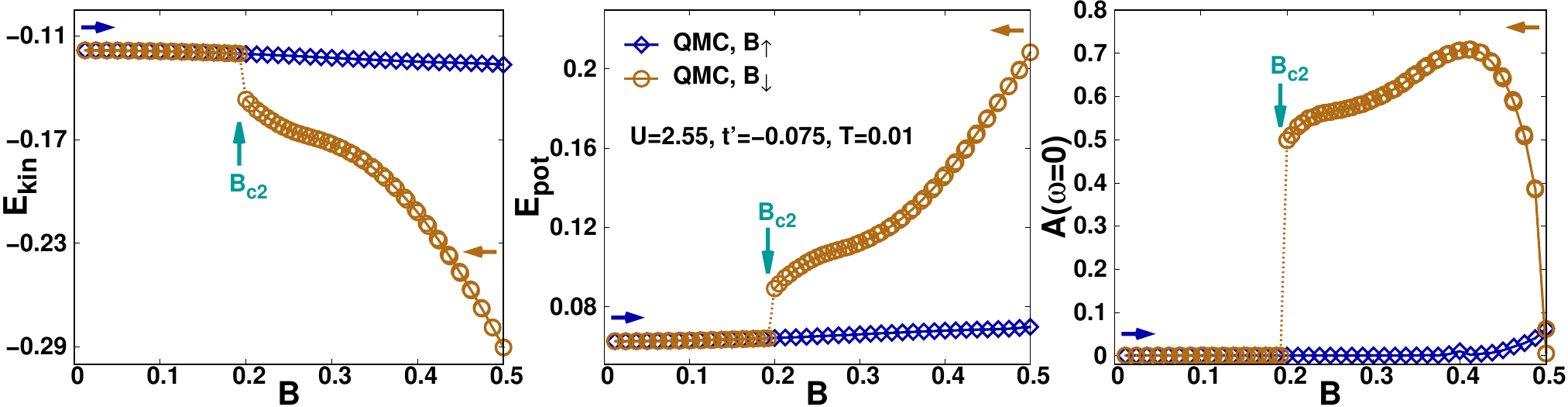}
	\caption{Kinetic energy $E_\text{kin}$ (left panel), potential energy $E_\text{pot}$ (middle panel) and spectral function at the Fermi level $A(\omega\!=\!0)$ (right panel) as a function of the magnetic field $B$ for a square lattice with geometrical frustration induced by the next-nearest neighbor hopping parameter $t_{i(i+2)}\!=\!t'\!=\!-0.3t$ for $U\!=\!2.55$ and $T\!=\!0.01$. Brown circles indicate data where the magnetic field has been gradually decreased from $B\!=\!0.5$ to $B\!=\!0$ while violet diamonds correspond to data where the magnetic field has been increased from $B\!=\!0$ to $B\!=\!0.5$.}
	\label{fig:motttransitonfrustrated}
\end{figure*}

\begin{figure*}
	\centering
	\includegraphics[width=1.0\linewidth]{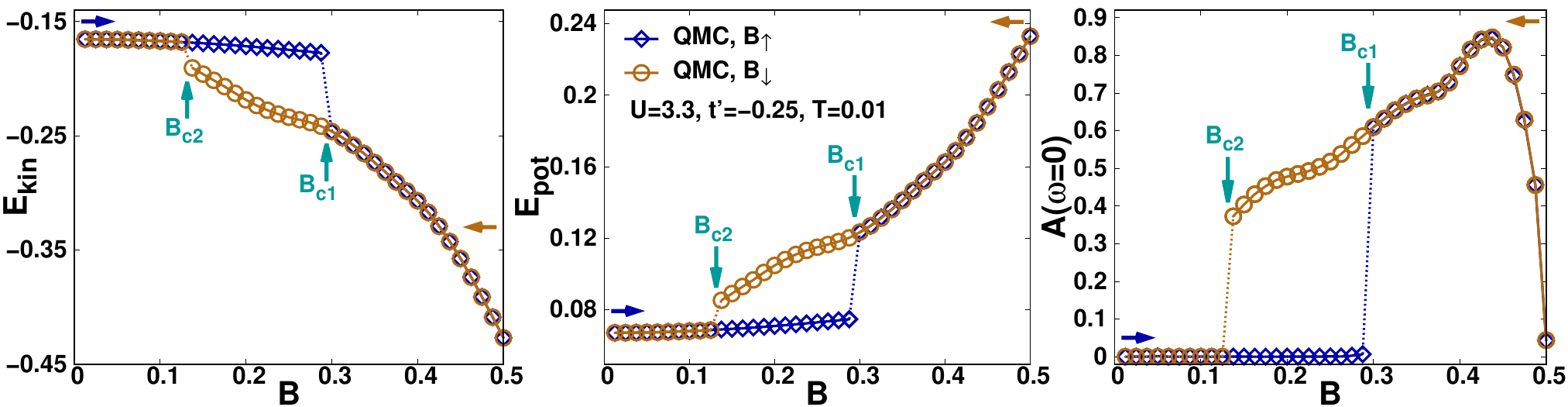}
	\caption{Same as in Fig.~\ref{fig:motttransitonfrustrated} but for $t'\!=\!-1.0t$, $U\!=\!3.3$ and $T\!=\!0.01$.}
	\label{fig:motttransitionfrustratedt1.0}
\end{figure*}


We complete our analysis of P2 by discussing the kinetic and potential energies of the system as a function of the magnetic field strength in the left and right panel of Fig.~\ref{fig:PotentialKineticEnergyU2.5Beta100}, respectively.
As for the kinetic and potential energies at P1 in Fig.~\ref{fig:KineticEnergyBeta400} of the main text and Fig.~\ref{fig:potentialenergyU2.5beta100metaltoinsulator} in Sec.~\ref{sec:potentialenergyU2.5beta100} of this SM, respectively, these observables are very small (in absolute value) and weakly $B$ dependent in the insulating state S2 for $B\!<\!B_{c1}$ (olive triangles in Fig.~\ref{fig:PotentialKineticEnergyU2.5Beta100}).
For the metallic solution S1 (empty violet pentagons) on the other hand, to which S2 collapses for $B\!>\!B_{c1}$, the energies are much larger and features a substantial variation with the magnetic field strength.
The finite jump which we observe at $B_{c1}$ for solution S2 is a clear hallmark of a first order phase transition between the metallic and the insulating state of the system.

Let us finally remark that for the point P2 in the phase diagram of Fig.~\ref{fig:PhasediagramB0} the free energy of the metallic solution is {\em always} lower than the one of the insulator (see left panel of Fig.~\ref{fig:MetalInsulator}).
Hence, for this set of parameters only a coexistence between the two phases but no real phase transition in the thermodynamics sense is observed.

\section{Quasiparticle lifetime and self-energy}
\label{Sec:Quasiparticle}

To further substantiate our findings, we have also analyzed the inverse quasiparticle life time $\frac{1}{\tau}$ which can be approximated by the (negative) imaginary part of the electronic self-energy $-\Im \Sigma(i\nu_1\!=\!i\pi T)$ at the first Matsubara frequency $\nu_1\!=\!\pi T$.
For the bipartite lattice, a low value of this quantity indicates a metal while a large value corresponds to an insulator (which is not straightforward for the non-bipartite case, e.g., when the system is geometrically frustrated by a next-nearest neighbor hopping $t'$, see Sec.~\ref{sec:FrustratedLattice}).
The results for this quantity are shown in Fig.~\ref{fig:InverseLifeTimeBeta400} at $T\!=\!0.01$ for $U\!=\!2.5$ (left panel) and $U\!=\!2.35$ (right panel).
In both cases the Mott transition is reflected by a drastic change of $\frac{1}{\tau}$ at the magnetic fields $B_{c2}$ and $B_{c1}$, respectively.

More specifically, for $U\!=\!2.5$ (left panel) the state S1 (empty purple squares) features a very large value of $\frac{1}{\tau}$ corresponding to an insulating state.
On the contrary, the state S2 (filled green circles) is a metal for $B\!>\!B_{c2}$ indicated by the very low values of $\frac{1}{\tau}$.
At $B_{c2}$ it undergoes a metal-to-insulator transition and collapses on state S1 as it has been observed for the DC conductivity $\sigma_{xx}(\omega\!=\!0)$ and the kinetic energy in Figs.~\ref{fig:InsulatorMetal} and \ref{fig:KineticEnergyBeta400} of the main text as well as for the potential energy in Fig.\ref{fig:potentialenergyU2.5beta100metaltoinsulator} of Sec.~\ref{sec:potentialenergyU2.5beta100} in this SM, respectively.

On the contrary, for $U\!=\!2.35$ (right panel of Fig.~\ref{fig:InverseLifeTimeBeta400}) the state S1 (empty violet pentagons) is metallic for all values of $B$.
For $B\!<\!B_{c1}$ the state S2 features large values of $\frac{1}{\tau}$ indicating that the system is in an insulating phase.
At $B\!=B_{c1}$ S2 eventually collapses on S1 marking the first order phase transition to the metallic state.

The above picture is also confirmed by comparing the imaginary part of the self-energy as a function of (positive) Matsubara frequencies right before and after the respective phase transition (see insets in Fig.~\ref{fig:InverseLifeTimeBeta400}).
On the metallic side (blue crosses), the self-energy shows a non-monotonous behavior with an upturn upon approaching small frequencies.
On the contrary, the insulating self-energy (orange crosses) diverges for $\nu\!\rightarrow\!0$ which is associated with the strong suppression of spectral weight at the Fermi level in the bipartite system.

\section{Additional data for the bipartite case}
\label{Sec:additionaldata}

In this section, we present results for the points P3, P4 and P5 in the phase diagram of Fig.~\ref{fig:PhasediagramB0} of the main text corresponding to $T\!=\!0.0025$ and $U\!=\!2.68$, $2.38$ and $2.63$, respectively.
The results at P3 and P4 for the free energy, optical conductivity, local spectral function, kinetic and potential energy as well as inverse quasi-particle life-time are presented in Figs.~\ref{fig:InsulatorMetalBeta100}, \ref{fig:MetalInsulatorBeta100}, \ref{fig:EkinBeta100} and \ref{fig:InverseLiftimeBeta400}.
The results for P3 and P4 are completely analogous to the ones for P1 and P2 and, hence, we refer the reader to the discussion of the latter for a detailed description.
Let us mention that in particular the calculations for P3 ($U\!=\!2.68$, $T\!=\!0.0025$) are aimed at describing the magnetic field driven metal-insulator transition in VO$_2$\cite{S-Matsuda2020,S-Matsuda2022}.
Estimating the critical temperature $T_c$ in the phase diagram in Fig.~\ref{fig:PhasediagramB0} as $T_c\!\sim\!0.0166$ then the ratio $\frac{T}{T_c}\!\sim\!0.15$ for P3 is close to the experimental value of $0.14$ in Ref.~\cite{S-Matsuda2020}.
To estimate the critical magnetic field $B_\text{crit}\!=\!B_c\frac{\Phi_0}{a^2}$ we assume a lattice constant of $\!=\!0.6${\AA} which is compatible with both theoretical\cite{S-Lu2019} and experimental\cite{S-Choi2020} predictions.
This gives $B_\text{crit}\!=2400T$ which is roughly five times the experimental value of $B_\text{exp}\!=\!500T$.
Given the very simple modelization of the complex multi-orbital material VO$_2$ by a single band Hubbard model the experimental and theoretical results are in quite good qualitative agreement.
In this respect, let us also mention that the actual magnitude of the critical field can be also controlled by the choice of the Hubbard interaction $U$ and frustration via a next-nearest neighbor hopping parameter $t'$.

In Fig.~\ref{fig:FreeEnergyU2.63} we depict the free energy at point P5 of Fig.~\ref{fig:PhasediagramB0} in the main text which corresponds to a slightly smaller value of $U\!=\!2.63$ than the one for P3 ($U\!=\!2.68$).
Since the system is more metallic in this case, the critical magnetic field $B_c\!=\!0.0946$, where the free energy of the insulator becomes lower than the one of the metal, is smaller for $U\!=\!2.63$ with respect to $U\!=\!2.68$ (where we find $B_c\!=\!0.2077$).
Moreover, P3 is located in the (insulating part of the) coexistence region of the field-free model and, hence, the metallic solution can be stabilized down to $B\!=\!0$. 
This implies that the system is in the coexistence region for all values of $B$ and, hence, the coexistence region boundaries $B_{c1}$ and $B_{c2}$, which would be indicated by a discontinuity in the free energy, cannot be observed in this case.

\section{Mott transition in the frustrated lattice}
\label{sec:FrustratedLattice}

\begin{figure*}[t!]
	\centering
	\includegraphics[width=\textwidth]{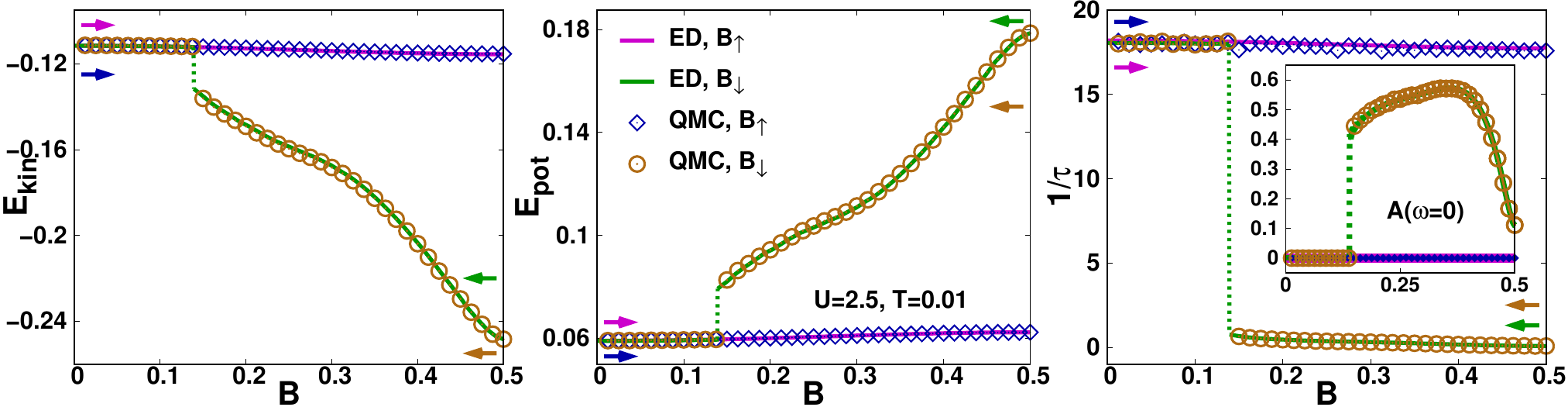}
	\caption{Kinetic energy $E_\text{kin}$ (left panel), potential energy $E_\text{pot}$ (middle panel), inverse quasi-particle lifetime $1/\tau\!=\!-\text{Im}\Sigma(i\nu_1), \nu_1\!=\!\frac{\pi}{\beta}$ (right panel) and local spectral function at zero frequency $A(\omega\!=\!0)$ (inset in the right panel), for $U\!=\!2.5$ and $T\!=\!0.01$ as a function of the orbital magnetic field $B$ obtained from ED and QMC calculations. $B_\uparrow$ and $B_\downarrow$ indicate that the calculations have been started at $B\!=\!0$ and $B\!=\!0.5$, respectively.}
	\label{fig:CompareEDvsQMC}
\end{figure*}

In this section, we present result for the orbital magnetic field driven Mott metal-insulator transition for a Hubbard model with geometric frustration which is introduced by a next-nearest neighbor hopping parameter $t'$.
This demonstrates that the observed transition is independent of the details of the underlying lattice structure and, in particular, of the strength of (antiferro)magnetic fluctuations.
In fact, while the bipartite system considered in the main text and Secs.~\ref{Sec:InsulatorToMetal}--\ref{Sec:additionaldata} of this SM features a strong tendency to antiferromagnetism, the latter is reduced by the introduction of frustration via the next-nearest neighbor hopping parameter $t'$.
This justifies the neglect of such antiferromagnetic fluctuations (or the related antiferromagnetic order) as the latter can be controlled by the parameter $t'$ without (qualitatively) changing the results for the orbital magnetic field driven metal-insulator transition in the main text.
Let us note that the corresponding calculations have been carried out with the CTQMC solver provided in the \texttt{w2dynamics} package as ED calculations are more difficult to converge in the frustrated (non-bipartite) case.

The results for the kinetic energy $E_\text{kin}$, the potential energy $E_\text{pot}$ and the spectral function at the Fermi level $A(\omega\!=\!0)$ for a moderate next-nearest neighbor hopping parameter $t'\!=\!-0.3t$ are shown in the left, middle and right panels of Fig.~\ref{fig:motttransitonfrustrated}, respectively.
As for the bipartite case discussed in the main text and Secs.~\ref{Sec:InsulatorToMetal}-\ref{Sec:additionaldata} we clearly observe a transition from a metal to an insulator at a critical magnetic field strength $B\!=\!B_{c2}$ upon decreasing the magnetic field (brown circles).
On the contrary, for increasing magnetic field the system remains in its insulating state throughout the entire $B$ range (violet diamonds).
More specifically, the transition from a metal to an insulator is indicated by a drastic drop of the spectral function at the Fermi level $A(\omega\!=\!0)$ at $B_{c2}$ in the right panel of Fig.~\ref{fig:motttransitonfrustrated} [which has been estimated from the CTQMC imaginary times Green's function $G(\tau)$ via the standard formula $A(\omega\!=\!0)\!=\!\frac{\beta}{\pi}G(\tau\!=\!\frac{\beta}{2})$ which is quite accurate at the low temperatures considered here]. 
This picture is confirmed by the abrupt decrease of the potential energy $E_\text{pot}$ and the (absolute value of the) kinetic energy $E_\text{kin}$ in the middle and left panels.

\begin{figure}[t!]
	\centering
	\includegraphics[width=0.5\textwidth]{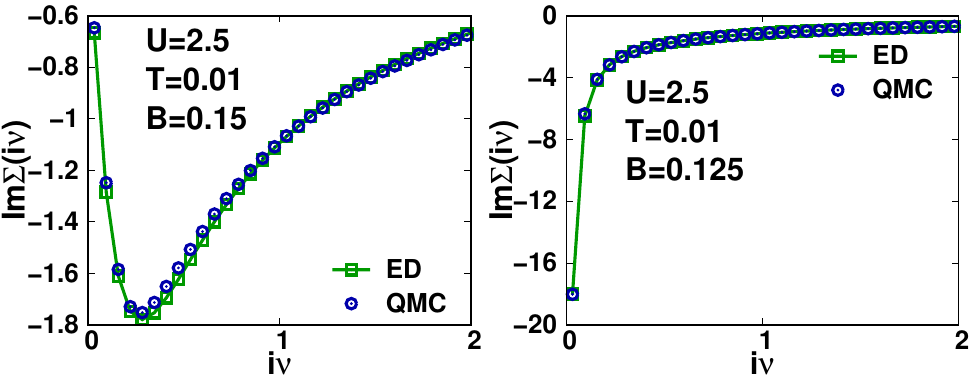}
	\caption{ED vs. QMC self-energies for $U\!=\!2.5$ and $T\!=\!0.01$ for two magnetic fields slightly above (left panel) and below (right panel) the phase transition between the insulating and the metallic phase in Fig.~\ref{fig:InsulatorMetalBeta100}.}
	\label{fig:selfenergies}
\end{figure}

To provide further evidence for the independence of the emergence of an orbital magnetic field driven Mott transition on the geometric structure of the underlying lattice we have also performed analogous calculations for the extreme case of a fully frustrated system where $t'\!=\!-t$, i.e., the diagonal next-nearest and the nearest neighbor hopping amplitudes are equal (see Fig.~\ref{fig:motttransitionfrustratedt1.0}).
Again, we observe a scenario analogous to $t'\!=\!-0.3t$ with only one difference: For $t'\!=\!-t$ {\em both} the transition from the metal to the insulator at $B_{c2}$ upon decreasing the orbital magnetic field (brown circles) as well as the transition from the insulator to the metal at $B_{c1}$ upon increasing the orbital magnetic field (violet diamonds) are located in the interval $B\!\in\![0,0.5]$.
This demonstrates that the size of the coexistence region as a function of $B$ can be tuned by the choice of the hopping matrix $t_{ij}$.

Summing up, the results above clearly show that the orbital magnetic field driven metal-insulator Mott transition is a generic strong-coupling feature and is indeed independent of the amount of geometric frustration in the system which controls the strength of antiferromagnetic fluctuations.
Moreover, the size of the critical region $[B_{c2},B_{c1}]$ can be controlled by the hopping matrix $t_{ij}$ where both boundaries are located in the interval $B\!\in\![0,0.5]$ only for the fully frustrated case $t'\!=\!-t$.

\section{Comparison with QMC calculations}
\label{Sec:compareQMC}

In this section, we present a benchmark of our ED calculations against corresponding QMC results obtained from a CTQMC solver in its hybridization expansion implementation as provided by the \texttt{w2dynamics} package\cite{S-Wallerberger2019}.
We focus on the point P1 in the phase diagram of Fig.~\ref{fig:PhasediagramB0} in the main text corresponding to the parameters $U\!=\!2.5$ and $T\!=\!0.01$.
We find an excellent agreement between ED (green and pink solid lines for solutions S2 and S1, respectively) and QMC (brown circles and blue diamonds for solutions S2 and S1, respectively) data for the kinetic and potential energies $E_\text{kin}$ and $E_\text{pot}$, the inverse quasi-particle lifetime $1/\tau$ and the local spectral function at zero frequency $A(\omega\!=\!0)$ in the left, middle and right panels as well as in the inset of the right panel of Fig.~\ref{fig:CompareEDvsQMC}, respectively.
In particular, the magnetic field $B_{c2}$ defining the boundary between the coexistence region and the insulating state is the same for ED and QMC which confirms the robustness of our ED calculations.
Let us mention, that the fluctuations in the QMC results for the inverse quasi-particle life time $1/\tau$ in the insulating solution (brown circles for $B\!<\!B_{c2}$ and blue diamonds) originate from statistical noise which can be systematically improved by increasing the number of Monte Carlo samples in the simulation.

Finally, in Fig.~\ref{fig:selfenergies} we show a comparison between ED (green empty squares) and QMC (blue empty circles) self-energies for point P1 in the phase diagram Fig.~\ref{fig:PhasediagramB0} in the main text.
The magnetic fields correspond to the metallic regime right above $B_{c2}$ (left panel) and the insulating regime right below $B_{c2}$ (right panel) in Fig.~\ref{fig:InsulatorMetal} of the main text, respectively.
The agreement between ED and QMC is indeed excellent for the insulating case while for the metallic state small deviations at low frequencies can be observed.
This tiny discrepancy originates from the fact that in the metallic part of the phase diagram a finite number of bath sites cannot fully capture the Kondo screening of the local moment.
However, such small deviations in $\Sigma_\text{imp}(i\nu)$ do not play a role for the observed phase transition and the related thermodynamic quantities as we have seen in Fig.~\ref{fig:CompareEDvsQMC}.

\FloatBarrier

\makeatletter
\def\@bibitem#1{\item\Hy@raisedlink{\hyper@anchorstart{SM-#1}\hyper@anchorend}}
\def\Hy@BibItemShut{\hyper@anchorend} 
\def\bibitem{\@ifnextchar[\@lbibitem\@bibitem}
\makeatother


\begin{thebibliography}{74}%
	\makeatletter
	\providecommand \@ifxundefined [1]{%
		\@ifx{#1\undefined}
	}%
	\providecommand \@ifnum [1]{%
		\ifnum #1\expandafter \@firstoftwo
		\else \expandafter \@secondoftwo
		\fi
	}%
	\providecommand \@ifx [1]{%
		\ifx #1\expandafter \@firstoftwo
		\else \expandafter \@secondoftwo
		\fi
	}%
	\providecommand \natexlab [1]{#1}%
	\providecommand \enquote  [1]{``#1''}%
	\providecommand \bibnamefont  [1]{#1}%
	\providecommand \bibfnamefont [1]{#1}%
	\providecommand \citenamefont [1]{#1}%
	\providecommand \href@noop [0]{\@secondoftwo}%
	\providecommand \href [0]{\begingroup \@sanitize@url \@href}%
	\providecommand \@href[1]{\@@startlink{#1}\@@href}%
	\providecommand \@@href[1]{\endgroup#1\@@endlink}%
	\providecommand \@sanitize@url [0]{\catcode `\\12\catcode `\$12\catcode
		`\&12\catcode `\#12\catcode `\^12\catcode `\_12\catcode `\%12\relax}%
	\providecommand \@@startlink[1]{}%
	\providecommand \@@endlink[0]{}%
	\providecommand \url  [0]{\begingroup\@sanitize@url \@url }%
	\providecommand \@url [1]{\endgroup\@href {#1}{\urlprefix }}%
	\providecommand \urlprefix  [0]{URL }%
	\providecommand \Eprint [0]{\href }%
	\providecommand \doibase [0]{https://doi.org/}%
	\providecommand \selectlanguage [0]{\@gobble}%
	\providecommand \bibinfo  [0]{\@secondoftwo}%
	\providecommand \bibfield  [0]{\@secondoftwo}%
	\providecommand \translation [1]{[#1]}%
	\providecommand \BibitemOpen [0]{}%
	\providecommand \bibitemStop [0]{}%
	\providecommand \bibitemNoStop [0]{.\EOS\space}%
	\providecommand \EOS [0]{\spacefactor3000\relax}%
	\providecommand \BibitemShut  [1]{\csname bibitem#1\endcsname}%
	\let\auto@bib@innerbib\@empty
	\bibitem [{\citenamefont {Ashcroft}\ and\ \citenamefont
		{Mermin}(1976)}]{Ashcroft1976}%
	\BibitemOpen
	\bibfield  {author} {\bibinfo {author} {\bibfnamefont {N.}~\bibnamefont
			{Ashcroft}}\ and\ \bibinfo {author} {\bibfnamefont {N.}~\bibnamefont
			{Mermin}},\ }\href@noop {} {\emph {\bibinfo {title} {Solid State Physics}}}\
	(\bibinfo  {publisher} {Hartcourt College Publishers, New York},\ \bibinfo
	{year} {1976})\BibitemShut {NoStop}%
	\bibitem [{\citenamefont {Sch\"afer}\ \emph {et~al.}(2015)\citenamefont
		{Sch\"afer}, \citenamefont {Geles}, \citenamefont {Rost}, \citenamefont
		{Rohringer}, \citenamefont {Arrigoni}, \citenamefont {Held}, \citenamefont
		{Bl\"umer}, \citenamefont {Aichhorn},\ and\ \citenamefont
		{Toschi}}]{Schafer2015}%
	\BibitemOpen
	\bibfield  {author} {\bibinfo {author} {\bibfnamefont {T.}~\bibnamefont
			{Sch\"afer}}, \bibinfo {author} {\bibfnamefont {F.}~\bibnamefont {Geles}},
		\bibinfo {author} {\bibfnamefont {D.}~\bibnamefont {Rost}}, \bibinfo {author}
		{\bibfnamefont {G.}~\bibnamefont {Rohringer}}, \bibinfo {author}
		{\bibfnamefont {E.}~\bibnamefont {Arrigoni}}, \bibinfo {author}
		{\bibfnamefont {K.}~\bibnamefont {Held}}, \bibinfo {author} {\bibfnamefont
			{N.}~\bibnamefont {Bl\"umer}}, \bibinfo {author} {\bibfnamefont
			{M.}~\bibnamefont {Aichhorn}},\ and\ \bibinfo {author} {\bibfnamefont
			{A.}~\bibnamefont {Toschi}},\ }\href
	{https://doi.org/10.1103/PhysRevB.91.125109} {\bibfield  {journal} {\bibinfo
			{journal} {Phys. Rev. B}\ }\textbf {\bibinfo {volume} {91}},\ \bibinfo
		{pages} {125109} (\bibinfo {year} {2015})}\BibitemShut {NoStop}%
	\bibitem [{\citenamefont {Anderson}(1958)}]{Anderson1958}%
	\BibitemOpen
	\bibfield  {author} {\bibinfo {author} {\bibfnamefont {P.~W.}\ \bibnamefont
			{Anderson}},\ }\href {https://doi.org/10.1103/PhysRev.109.1492} {\bibfield
		{journal} {\bibinfo  {journal} {Phys. Rev.}\ }\textbf {\bibinfo {volume}
			{109}},\ \bibinfo {pages} {1492} (\bibinfo {year} {1958})}\BibitemShut
	{NoStop}%
	\bibitem [{\citenamefont {Evers}\ and\ \citenamefont
		{Mirlin}(2008)}]{Evers2008}%
	\BibitemOpen
	\bibfield  {author} {\bibinfo {author} {\bibfnamefont {F.}~\bibnamefont
			{Evers}}\ and\ \bibinfo {author} {\bibfnamefont {A.~D.}\ \bibnamefont
			{Mirlin}},\ }\href {https://doi.org/10.1103/RevModPhys.80.1355} {\bibfield
		{journal} {\bibinfo  {journal} {Rev. Mod. Phys.}\ }\textbf {\bibinfo {volume}
			{80}},\ \bibinfo {pages} {1355} (\bibinfo {year} {2008})}\BibitemShut
	{NoStop}%
	\bibitem [{\citenamefont {Peierls}(1996)}]{Peierls1996}%
	\BibitemOpen
	\bibfield  {author} {\bibinfo {author} {\bibfnamefont {R.~E.}\ \bibnamefont
			{Peierls}},\ }\href@noop {} {\emph {\bibinfo {title} {Quantum theory of
				solids}}}\ (\bibinfo  {publisher} {Clarendon Press},\ \bibinfo {address} {New
		York},\ \bibinfo {year} {1996})\BibitemShut {NoStop}%
	\bibitem [{\citenamefont {Zaanen}\ \emph {et~al.}(1985)\citenamefont {Zaanen},
		\citenamefont {Sawatzky},\ and\ \citenamefont {Allen}}]{Zaanen1985}%
	\BibitemOpen
	\bibfield  {author} {\bibinfo {author} {\bibfnamefont {J.}~\bibnamefont
			{Zaanen}}, \bibinfo {author} {\bibfnamefont {G.~A.}\ \bibnamefont
			{Sawatzky}},\ and\ \bibinfo {author} {\bibfnamefont {J.~W.}\ \bibnamefont
			{Allen}},\ }\href {https://doi.org/10.1103/PhysRevLett.55.418} {\bibfield
		{journal} {\bibinfo  {journal} {Phys. Rev. Lett.}\ }\textbf {\bibinfo
			{volume} {55}},\ \bibinfo {pages} {418} (\bibinfo {year} {1985})}\BibitemShut
	{NoStop}%
	\bibitem [{\citenamefont {Mott}(1968)}]{Mott1968}%
	\BibitemOpen
	\bibfield  {author} {\bibinfo {author} {\bibfnamefont {N.~F.}\ \bibnamefont
			{Mott}},\ }\href {https://doi.org/10.1103/RevModPhys.40.677} {\bibfield
		{journal} {\bibinfo  {journal} {Rev. Mod. Phys.}\ }\textbf {\bibinfo {volume}
			{40}},\ \bibinfo {pages} {677} (\bibinfo {year} {1968})}\BibitemShut
	{NoStop}%
	\bibitem [{\citenamefont {Manzeli}\ \emph {et~al.}(2017)\citenamefont
		{Manzeli}, \citenamefont {Ovchinnikov}, \citenamefont {Pasquier},
		\citenamefont {Yazyev},\ and\ \citenamefont {Kis}}]{Manzeli2017}%
	\BibitemOpen
	\bibfield  {author} {\bibinfo {author} {\bibfnamefont {S.}~\bibnamefont
			{Manzeli}}, \bibinfo {author} {\bibfnamefont {D.}~\bibnamefont
			{Ovchinnikov}}, \bibinfo {author} {\bibfnamefont {D.}~\bibnamefont
			{Pasquier}}, \bibinfo {author} {\bibfnamefont {O.~V.}\ \bibnamefont
			{Yazyev}},\ and\ \bibinfo {author} {\bibfnamefont {A.}~\bibnamefont {Kis}},\
	}\href {https://doi.org/10.1038/natrevmats.2017.33} {\bibfield  {journal}
		{\bibinfo  {journal} {Nature Reviews Materials}\ }\textbf {\bibinfo {volume}
			{2}},\ \bibinfo {pages} {17033} (\bibinfo {year} {2017})}\BibitemShut
	{NoStop}%
	\bibitem [{\citenamefont {Moon}(2021)}]{Moon2021}%
	\BibitemOpen
	\bibfield  {author} {\bibinfo {author} {\bibfnamefont {B.~H.}\ \bibnamefont
			{Moon}},\ }\href {https://doi.org/10.1007/s42247-021-00202-9} {\bibfield
		{journal} {\bibinfo  {journal} {Emergent Materials}\ }\textbf {\bibinfo
			{volume} {4}},\ \bibinfo {pages} {989} (\bibinfo {year} {2021})}\BibitemShut
	{NoStop}%
	\bibitem [{\citenamefont {Haule}\ and\ \citenamefont
		{Pascut}(2017)}]{Haule2017}%
	\BibitemOpen
	\bibfield  {author} {\bibinfo {author} {\bibfnamefont {K.}~\bibnamefont
			{Haule}}\ and\ \bibinfo {author} {\bibfnamefont {G.~L.}\ \bibnamefont
			{Pascut}},\ }\href {https://doi.org/10.1038/s41598-017-10374-2} {\bibfield
		{journal} {\bibinfo  {journal} {Scientific Reports}\ }\textbf {\bibinfo
			{volume} {7}},\ \bibinfo {pages} {10375} (\bibinfo {year}
		{2017})}\BibitemShut {NoStop}%
	\bibitem [{\citenamefont {Fei}\ \emph {et~al.}(2022)\citenamefont {Fei},
		\citenamefont {Wu}, \citenamefont {Zhang},\ and\ \citenamefont
		{Yin}}]{Fei2022}%
	\BibitemOpen
	\bibfield  {author} {\bibinfo {author} {\bibfnamefont {Y.}~\bibnamefont
			{Fei}}, \bibinfo {author} {\bibfnamefont {Z.}~\bibnamefont {Wu}}, \bibinfo
		{author} {\bibfnamefont {W.}~\bibnamefont {Zhang}},\ and\ \bibinfo {author}
		{\bibfnamefont {Y.}~\bibnamefont {Yin}},\ }\href
	{https://doi.org/10.1007/s43673-022-00049-0} {\bibfield  {journal} {\bibinfo
			{journal} {AAPPS Bulletin}\ }\textbf {\bibinfo {volume} {32}},\ \bibinfo
		{pages} {20} (\bibinfo {year} {2022})}\BibitemShut {NoStop}%
	\bibitem [{\citenamefont {Gurunatha}\ \emph {et~al.}(2020)\citenamefont
		{Gurunatha}, \citenamefont {Sathasivam}, \citenamefont {Li}, \citenamefont
		{Portnoi}, \citenamefont {Parkin},\ and\ \citenamefont {I.}}]{Gurunatha2020}%
	\BibitemOpen
	\bibfield  {author} {\bibinfo {author} {\bibfnamefont {K.~L.}\ \bibnamefont
			{Gurunatha}}, \bibinfo {author} {\bibfnamefont {S.}~\bibnamefont
			{Sathasivam}}, \bibinfo {author} {\bibfnamefont {J.}~\bibnamefont {Li}},
		\bibinfo {author} {\bibfnamefont {M.}~\bibnamefont {Portnoi}}, \bibinfo
		{author} {\bibfnamefont {I.~P.}\ \bibnamefont {Parkin}},\ and\ \bibinfo
		{author} {\bibfnamefont {P.}~\bibnamefont {I.}},\ }\href@noop {} {\bibfield
		{journal} {\bibinfo  {journal} {Adv. Funct. Mater.}\ }\textbf {\bibinfo
			{volume} {30}},\ \bibinfo {pages} {2005311} (\bibinfo {year}
		{2020})}\BibitemShut {NoStop}%
	\bibitem [{\citenamefont {Ling}\ \emph {et~al.}(2019)\citenamefont {Ling},
		\citenamefont {Zhao}, \citenamefont {Hu}, \citenamefont {Li}, \citenamefont
		{Zhao}, \citenamefont {Wang}, \citenamefont {Zhao},\ and\ \citenamefont
		{Jin}}]{Ling2019}%
	\BibitemOpen
	\bibfield  {author} {\bibinfo {author} {\bibfnamefont {C.}~\bibnamefont
			{Ling}}, \bibinfo {author} {\bibfnamefont {Z.}~\bibnamefont {Zhao}}, \bibinfo
		{author} {\bibfnamefont {X.}~\bibnamefont {Hu}}, \bibinfo {author}
		{\bibfnamefont {J.}~\bibnamefont {Li}}, \bibinfo {author} {\bibfnamefont
			{X.}~\bibnamefont {Zhao}}, \bibinfo {author} {\bibfnamefont {Z.}~\bibnamefont
			{Wang}}, \bibinfo {author} {\bibfnamefont {Y.}~\bibnamefont {Zhao}},\ and\
		\bibinfo {author} {\bibfnamefont {H.}~\bibnamefont {Jin}},\ }\href
	{https://doi.org/10.1021/acsanm.9b01640} {\bibfield  {journal} {\bibinfo
			{journal} {ACS Appl. Nano Mater.}\ }\textbf {\bibinfo {volume} {2}},\
		\bibinfo {pages} {6738} (\bibinfo {year} {2019})}\BibitemShut {NoStop}%
	\bibitem [{\citenamefont {Mazza}\ \emph {et~al.}(2016)\citenamefont {Mazza},
		\citenamefont {Amaricci}, \citenamefont {Capone},\ and\ \citenamefont
		{Fabrizio}}]{Mazza2016}%
	\BibitemOpen
	\bibfield  {author} {\bibinfo {author} {\bibfnamefont {G.}~\bibnamefont
			{Mazza}}, \bibinfo {author} {\bibfnamefont {A.}~\bibnamefont {Amaricci}},
		\bibinfo {author} {\bibfnamefont {M.}~\bibnamefont {Capone}},\ and\ \bibinfo
		{author} {\bibfnamefont {M.}~\bibnamefont {Fabrizio}},\ }\href
	{https://doi.org/10.1103/PhysRevLett.117.176401} {\bibfield  {journal}
		{\bibinfo  {journal} {Phys. Rev. Lett.}\ }\textbf {\bibinfo {volume} {117}},\
		\bibinfo {pages} {176401} (\bibinfo {year} {2016})}\BibitemShut {NoStop}%
	\bibitem [{\citenamefont {Kagawa}\ \emph {et~al.}(2004)\citenamefont {Kagawa},
		\citenamefont {Itou}, \citenamefont {Miyagawa},\ and\ \citenamefont
		{Kanoda}}]{Kagawa2004}%
	\BibitemOpen
	\bibfield  {author} {\bibinfo {author} {\bibfnamefont {F.}~\bibnamefont
			{Kagawa}}, \bibinfo {author} {\bibfnamefont {T.}~\bibnamefont {Itou}},
		\bibinfo {author} {\bibfnamefont {K.}~\bibnamefont {Miyagawa}},\ and\
		\bibinfo {author} {\bibfnamefont {K.}~\bibnamefont {Kanoda}},\ }\href
	{https://doi.org/10.1103/PhysRevLett.93.127001} {\bibfield  {journal}
		{\bibinfo  {journal} {Phys. Rev. Lett.}\ }\textbf {\bibinfo {volume} {93}},\
		\bibinfo {pages} {127001} (\bibinfo {year} {2004})}\BibitemShut {NoStop}%
	\bibitem [{\citenamefont {Sun}\ \emph {et~al.}(2021)\citenamefont {Sun},
		\citenamefont {Wang}, \citenamefont {Mu}, \citenamefont {Wang}, \citenamefont
		{Wang}, \citenamefont {Wu}, \citenamefont {Wang}, \citenamefont {Zhou},\ and\
		\citenamefont {Chen}}]{Sun2021}%
	\BibitemOpen
	\bibfield  {author} {\bibinfo {author} {\bibfnamefont {Z.~L.}\ \bibnamefont
			{Sun}}, \bibinfo {author} {\bibfnamefont {A.~F.}\ \bibnamefont {Wang}},
		\bibinfo {author} {\bibfnamefont {H.~M.}\ \bibnamefont {Mu}}, \bibinfo
		{author} {\bibfnamefont {H.~H.}\ \bibnamefont {Wang}}, \bibinfo {author}
		{\bibfnamefont {Z.~F.}\ \bibnamefont {Wang}}, \bibinfo {author}
		{\bibfnamefont {T.}~\bibnamefont {Wu}}, \bibinfo {author} {\bibfnamefont
			{Z.~Y.}\ \bibnamefont {Wang}}, \bibinfo {author} {\bibfnamefont {X.~Y.}\
			\bibnamefont {Zhou}},\ and\ \bibinfo {author} {\bibfnamefont {X.~H.}\
			\bibnamefont {Chen}},\ }\href {https://doi.org/10.1038/s41535-021-00397-4}
	{\bibfield  {journal} {\bibinfo  {journal} {npj Quantum Materials}\ }\textbf
		{\bibinfo {volume} {6}},\ \bibinfo {pages} {94} (\bibinfo {year}
		{2021})}\BibitemShut {NoStop}%
	\bibitem [{\citenamefont {Ramirez}(1997)}]{Ramirez1997}%
	\BibitemOpen
	\bibfield  {author} {\bibinfo {author} {\bibfnamefont {A.~P.}\ \bibnamefont
			{Ramirez}},\ }\href {https://doi.org/10.1088/0953-8984/9/39/005} {\bibfield
		{journal} {\bibinfo  {journal} {Journal of Physics: Condensed Matter}\
		}\textbf {\bibinfo {volume} {9}},\ \bibinfo {pages} {8171} (\bibinfo {year}
		{1997})}\BibitemShut {NoStop}%
	\bibitem [{\citenamefont {Dagotto}\ \emph {et~al.}(2001)\citenamefont
		{Dagotto}, \citenamefont {Hotta},\ and\ \citenamefont {Moreo}}]{Dagotto2001}%
	\BibitemOpen
	\bibfield  {author} {\bibinfo {author} {\bibfnamefont {E.}~\bibnamefont
			{Dagotto}}, \bibinfo {author} {\bibfnamefont {T.}~\bibnamefont {Hotta}},\
		and\ \bibinfo {author} {\bibfnamefont {A.}~\bibnamefont {Moreo}},\ }\href
	{https://doi.org/https://doi.org/10.1016/S0370-1573(00)00121-6} {\bibfield
		{journal} {\bibinfo  {journal} {Physics Reports}\ }\textbf {\bibinfo {volume}
			{344}},\ \bibinfo {pages} {1} (\bibinfo {year} {2001})}\BibitemShut {NoStop}%
	\bibitem [{\citenamefont {Salamon}\ and\ \citenamefont
		{Jaime}(2001)}]{Salamon2001}%
	\BibitemOpen
	\bibfield  {author} {\bibinfo {author} {\bibfnamefont {M.~B.}\ \bibnamefont
			{Salamon}}\ and\ \bibinfo {author} {\bibfnamefont {M.}~\bibnamefont
			{Jaime}},\ }\href {https://doi.org/10.1103/RevModPhys.73.583} {\bibfield
		{journal} {\bibinfo  {journal} {Rev. Mod. Phys.}\ }\textbf {\bibinfo {volume}
			{73}},\ \bibinfo {pages} {583} (\bibinfo {year} {2001})}\BibitemShut
	{NoStop}%
	\bibitem [{\citenamefont {Matsuda}\ \emph {et~al.}(2020)\citenamefont
		{Matsuda}, \citenamefont {Nakamura}, \citenamefont {Ikeda}, \citenamefont
		{Takeyama}, \citenamefont {Suga}, \citenamefont {Nakahara},\ and\
		\citenamefont {Muraoka}}]{Matsuda2020}%
	\BibitemOpen
	\bibfield  {author} {\bibinfo {author} {\bibfnamefont {Y.~H.}\ \bibnamefont
			{Matsuda}}, \bibinfo {author} {\bibfnamefont {D.}~\bibnamefont {Nakamura}},
		\bibinfo {author} {\bibfnamefont {A.}~\bibnamefont {Ikeda}}, \bibinfo
		{author} {\bibfnamefont {S.}~\bibnamefont {Takeyama}}, \bibinfo {author}
		{\bibfnamefont {Y.}~\bibnamefont {Suga}}, \bibinfo {author} {\bibfnamefont
			{H.}~\bibnamefont {Nakahara}},\ and\ \bibinfo {author} {\bibfnamefont
			{Y.}~\bibnamefont {Muraoka}},\ }\href
	{https://doi.org/10.1038/s41467-020-17416-w} {\bibfield  {journal} {\bibinfo
			{journal} {Nature Communications}\ }\textbf {\bibinfo {volume} {11}},\
		\bibinfo {pages} {3591} (\bibinfo {year} {2020})}\BibitemShut {NoStop}%
	\bibitem [{\citenamefont {Matsuda}\ \emph {et~al.}(2022)\citenamefont
		{Matsuda}, \citenamefont {Muraoka}, \citenamefont {Nakamura}, \citenamefont
		{Ikeda}, \citenamefont {Ishii}, \citenamefont {Zhou}, \citenamefont
		{Sawabe},\ and\ \citenamefont {Takeyama}}]{Matsuda2022}%
	\BibitemOpen
	\bibfield  {author} {\bibinfo {author} {\bibfnamefont {Y.~H.}\ \bibnamefont
			{Matsuda}}, \bibinfo {author} {\bibfnamefont {Y.}~\bibnamefont {Muraoka}},
		\bibinfo {author} {\bibfnamefont {D.}~\bibnamefont {Nakamura}}, \bibinfo
		{author} {\bibfnamefont {A.}~\bibnamefont {Ikeda}}, \bibinfo {author}
		{\bibfnamefont {Y.}~\bibnamefont {Ishii}}, \bibinfo {author} {\bibfnamefont
			{X.-G.}\ \bibnamefont {Zhou}}, \bibinfo {author} {\bibfnamefont
			{H.}~\bibnamefont {Sawabe}},\ and\ \bibinfo {author} {\bibfnamefont
			{S.}~\bibnamefont {Takeyama}},\ }\href
	{https://doi.org/https://doi.org/10.7566/JPSJ.91.101008} {\bibfield
		{journal} {\bibinfo  {journal} {Journal of the Physical Society of Japan}\
		}\textbf {\bibinfo {volume} {91}},\ \bibinfo {pages} {101008} (\bibinfo
		{year} {2022})}\BibitemShut {NoStop}%
	\bibitem [{\citenamefont {Fukuoka}\ \emph {et~al.}(2024)\citenamefont
		{Fukuoka}, \citenamefont {Oka}, \citenamefont {Ihara}, \citenamefont
		{Kawamoto}, \citenamefont {Imajo},\ and\ \citenamefont {Kindo}}]{Fukoka2024}%
	\BibitemOpen
	\bibfield  {author} {\bibinfo {author} {\bibfnamefont {S.}~\bibnamefont
			{Fukuoka}}, \bibinfo {author} {\bibfnamefont {T.}~\bibnamefont {Oka}},
		\bibinfo {author} {\bibfnamefont {Y.}~\bibnamefont {Ihara}}, \bibinfo
		{author} {\bibfnamefont {A.}~\bibnamefont {Kawamoto}}, \bibinfo {author}
		{\bibfnamefont {S.}~\bibnamefont {Imajo}},\ and\ \bibinfo {author}
		{\bibfnamefont {K.}~\bibnamefont {Kindo}},\ }\href
	{https://doi.org/10.1103/PhysRevB.109.195142} {\bibfield  {journal} {\bibinfo
			{journal} {Phys. Rev. B}\ }\textbf {\bibinfo {volume} {109}},\ \bibinfo
		{pages} {195142} (\bibinfo {year} {2024})}\BibitemShut {NoStop}%
	\bibitem [{\citenamefont {Cao}\ \emph {et~al.}(2018)\citenamefont {Cao},
		\citenamefont {Fatemi}, \citenamefont {Demir}, \citenamefont {Fang},
		\citenamefont {Tomarken}, \citenamefont {Luo}, \citenamefont
		{Sanchez-Yamagishi}, \citenamefont {Watanabe}, \citenamefont {Taniguchi},
		\citenamefont {Kaxiras}, \citenamefont {Ashoori},\ and\ \citenamefont
		{Jarillo-Herrero}}]{Cao2018}%
	\BibitemOpen
	\bibfield  {author} {\bibinfo {author} {\bibfnamefont {Y.}~\bibnamefont
			{Cao}}, \bibinfo {author} {\bibfnamefont {V.}~\bibnamefont {Fatemi}},
		\bibinfo {author} {\bibfnamefont {A.}~\bibnamefont {Demir}}, \bibinfo
		{author} {\bibfnamefont {S.}~\bibnamefont {Fang}}, \bibinfo {author}
		{\bibfnamefont {S.~L.}\ \bibnamefont {Tomarken}}, \bibinfo {author}
		{\bibfnamefont {J.~Y.}\ \bibnamefont {Luo}}, \bibinfo {author} {\bibfnamefont
			{J.~D.}\ \bibnamefont {Sanchez-Yamagishi}}, \bibinfo {author} {\bibfnamefont
			{K.}~\bibnamefont {Watanabe}}, \bibinfo {author} {\bibfnamefont
			{T.}~\bibnamefont {Taniguchi}}, \bibinfo {author} {\bibfnamefont
			{E.}~\bibnamefont {Kaxiras}}, \bibinfo {author} {\bibfnamefont {R.~C.}\
			\bibnamefont {Ashoori}},\ and\ \bibinfo {author} {\bibfnamefont
			{P.}~\bibnamefont {Jarillo-Herrero}},\ }\href
	{https://doi.org/10.1038/nature26154} {\bibfield  {journal} {\bibinfo
			{journal} {Nature}\ }\textbf {\bibinfo {volume} {556}},\ \bibinfo {pages}
		{80} (\bibinfo {year} {2018})}\BibitemShut {NoStop}%
	\bibitem [{\citenamefont {Pizarro}\ \emph {et~al.}(2019)\citenamefont
		{Pizarro}, \citenamefont {Calderón},\ and\ \citenamefont
		{Bascones}}]{Pizarro2019}%
	\BibitemOpen
	\bibfield  {author} {\bibinfo {author} {\bibfnamefont {J.~M.}\ \bibnamefont
			{Pizarro}}, \bibinfo {author} {\bibfnamefont {M.~J.}\ \bibnamefont
			{Calderón}},\ and\ \bibinfo {author} {\bibfnamefont {E.}~\bibnamefont
			{Bascones}},\ }\href {https://doi.org/10.1088/2399-6528/ab0fa9} {\bibfield
		{journal} {\bibinfo  {journal} {Journal of Physics Communications}\ }\textbf
		{\bibinfo {volume} {3}},\ \bibinfo {pages} {035024} (\bibinfo {year}
		{2019})}\BibitemShut {NoStop}%
	\bibitem [{\citenamefont {Kwan}\ \emph {et~al.}(2020)\citenamefont {Kwan},
		\citenamefont {Parameswaran},\ and\ \citenamefont {Sondhi}}]{Kwan2020}%
	\BibitemOpen
	\bibfield  {author} {\bibinfo {author} {\bibfnamefont {Y.~H.}\ \bibnamefont
			{Kwan}}, \bibinfo {author} {\bibfnamefont {S.~A.}\ \bibnamefont
			{Parameswaran}},\ and\ \bibinfo {author} {\bibfnamefont {S.~L.}\ \bibnamefont
			{Sondhi}},\ }\href {https://doi.org/10.1103/PhysRevB.101.205116} {\bibfield
		{journal} {\bibinfo  {journal} {Phys. Rev. B}\ }\textbf {\bibinfo {volume}
			{101}},\ \bibinfo {pages} {205116} (\bibinfo {year} {2020})}\BibitemShut
	{NoStop}%
	\bibitem [{\citenamefont {Czajka}\ \emph {et~al.}(2006)\citenamefont {Czajka},
		\citenamefont {Gorczyca}, \citenamefont {Ma\ifmmode~\acute{s}\else
			\'{s}\fi{}ka},\ and\ \citenamefont {Mierzejewski}}]{Czajka2006}%
	\BibitemOpen
	\bibfield  {author} {\bibinfo {author} {\bibfnamefont {K.}~\bibnamefont
			{Czajka}}, \bibinfo {author} {\bibfnamefont {A.}~\bibnamefont {Gorczyca}},
		\bibinfo {author} {\bibfnamefont {M.~M.}\ \bibnamefont
			{Ma\ifmmode~\acute{s}\else \'{s}\fi{}ka}},\ and\ \bibinfo {author}
		{\bibfnamefont {M.}~\bibnamefont {Mierzejewski}},\ }\href
	{https://doi.org/10.1103/PhysRevB.74.125116} {\bibfield  {journal} {\bibinfo
			{journal} {Phys. Rev. B}\ }\textbf {\bibinfo {volume} {74}},\ \bibinfo
		{pages} {125116} (\bibinfo {year} {2006})}\BibitemShut {NoStop}%
	\bibitem [{\citenamefont {Acheche}\ \emph {et~al.}(2017)\citenamefont
		{Acheche}, \citenamefont {Arsenault},\ and\ \citenamefont
		{Tremblay}}]{Acheche2017}%
	\BibitemOpen
	\bibfield  {author} {\bibinfo {author} {\bibfnamefont {S.}~\bibnamefont
			{Acheche}}, \bibinfo {author} {\bibfnamefont {L.-F.}\ \bibnamefont
			{Arsenault}},\ and\ \bibinfo {author} {\bibfnamefont {A.-M.~S.}\ \bibnamefont
			{Tremblay}},\ }\href {https://doi.org/10.1103/PhysRevB.96.235135} {\bibfield
		{journal} {\bibinfo  {journal} {Phys. Rev. B}\ }\textbf {\bibinfo {volume}
			{96}},\ \bibinfo {pages} {235135} (\bibinfo {year} {2017})}\BibitemShut
	{NoStop}%
	\bibitem [{\citenamefont {Markov}\ \emph {et~al.}(2019)\citenamefont {Markov},
		\citenamefont {Rohringer},\ and\ \citenamefont {Rubtsov}}]{Markov2019}%
	\BibitemOpen
	\bibfield  {author} {\bibinfo {author} {\bibfnamefont {A.~A.}\ \bibnamefont
			{Markov}}, \bibinfo {author} {\bibfnamefont {G.}~\bibnamefont {Rohringer}},\
		and\ \bibinfo {author} {\bibfnamefont {A.~N.}\ \bibnamefont {Rubtsov}},\
	}\href {https://doi.org/10.1103/PhysRevB.100.115102} {\bibfield  {journal}
		{\bibinfo  {journal} {Phys. Rev. B}\ }\textbf {\bibinfo {volume} {100}},\
		\bibinfo {pages} {115102} (\bibinfo {year} {2019})}\BibitemShut {NoStop}%
	\bibitem [{\citenamefont {Vu\ifmmode \check{c}\else \v{c}\fi{}i\ifmmode
			\check{c}\else \v{c}\fi{}evi\ifmmode~\acute{c}\else \'{c}\fi{}}\ and\
		\citenamefont {\ifmmode~\check{Z}\else
			\v{Z}\fi{}itko}(2021{\natexlab{a}})}]{Vucicevic2021}%
	\BibitemOpen
	\bibfield  {author} {\bibinfo {author} {\bibfnamefont {J.~c.~v.}\
			\bibnamefont {Vu\ifmmode \check{c}\else \v{c}\fi{}i\ifmmode \check{c}\else
				\v{c}\fi{}evi\ifmmode~\acute{c}\else \'{c}\fi{}}}\ and\ \bibinfo {author}
		{\bibfnamefont {R.}~\bibnamefont {\ifmmode~\check{Z}\else \v{Z}\fi{}itko}},\
	}\href {https://doi.org/10.1103/PhysRevLett.127.196601} {\bibfield  {journal}
		{\bibinfo  {journal} {Phys. Rev. Lett.}\ }\textbf {\bibinfo {volume} {127}},\
		\bibinfo {pages} {196601} (\bibinfo {year} {2021}{\natexlab{a}})}\BibitemShut
	{NoStop}%
	\bibitem [{\citenamefont {Vu\ifmmode \check{c}\else \v{c}\fi{}i\ifmmode
			\check{c}\else \v{c}\fi{}evi\ifmmode~\acute{c}\else \'{c}\fi{}}\ and\
		\citenamefont {\ifmmode~\check{Z}\else
			\v{Z}\fi{}itko}(2021{\natexlab{b}})}]{Vucicevic2021a}%
	\BibitemOpen
	\bibfield  {author} {\bibinfo {author} {\bibfnamefont {J.}~\bibnamefont
			{Vu\ifmmode \check{c}\else \v{c}\fi{}i\ifmmode \check{c}\else
				\v{c}\fi{}evi\ifmmode~\acute{c}\else \'{c}\fi{}}}\ and\ \bibinfo {author}
		{\bibfnamefont {R.}~\bibnamefont {\ifmmode~\check{Z}\else \v{Z}\fi{}itko}},\
	}\href {https://doi.org/10.1103/PhysRevB.104.205101} {\bibfield  {journal}
		{\bibinfo  {journal} {Phys. Rev. B}\ }\textbf {\bibinfo {volume} {104}},\
		\bibinfo {pages} {205101} (\bibinfo {year} {2021}{\natexlab{b}})}\BibitemShut
	{NoStop}%
	\bibitem [{\citenamefont {Ding}\ \emph {et~al.}(2022)\citenamefont {Ding},
		\citenamefont {Wang}, \citenamefont {Moritz}, \citenamefont {Schattner},
		\citenamefont {Huang},\ and\ \citenamefont {Devereaux}}]{Ding2022}%
	\BibitemOpen
	\bibfield  {author} {\bibinfo {author} {\bibfnamefont {J.~K.}\ \bibnamefont
			{Ding}}, \bibinfo {author} {\bibfnamefont {W.~O.}\ \bibnamefont {Wang}},
		\bibinfo {author} {\bibfnamefont {B.}~\bibnamefont {Moritz}}, \bibinfo
		{author} {\bibfnamefont {Y.}~\bibnamefont {Schattner}}, \bibinfo {author}
		{\bibfnamefont {E.~W.}\ \bibnamefont {Huang}},\ and\ \bibinfo {author}
		{\bibfnamefont {T.~P.}\ \bibnamefont {Devereaux}},\ }\href
	{https://doi.org/10.1038/s42005-022-00968-2} {\bibfield  {journal} {\bibinfo
			{journal} {Communications Physics}\ }\textbf {\bibinfo {volume} {5}},\
		\bibinfo {pages} {204} (\bibinfo {year} {2022})}\BibitemShut {NoStop}%
	\bibitem [{Roh()}]{Rohringer2024Supp}%
	\BibitemOpen
	\href@noop {} {\bibinfo {title} {See supplemental material at
			http://link.aps.org/supplemental/10.1103/bw7f-p8mp for technical details of
			the numerical calculations and additional numerical results.}}\BibitemShut
	{Stop}%
	\bibitem [{\citenamefont {Wallerberger}\ \emph {et~al.}(2019)\citenamefont
		{Wallerberger}, \citenamefont {Hausoel}, \citenamefont {Gunacker},
		\citenamefont {Kowalski}, \citenamefont {Parragh}, \citenamefont {Goth},
		\citenamefont {Held},\ and\ \citenamefont {Sangiovanni}}]{Wallerberger2019}%
	\BibitemOpen
	\bibfield  {author} {\bibinfo {author} {\bibfnamefont {M.}~\bibnamefont
			{Wallerberger}}, \bibinfo {author} {\bibfnamefont {A.}~\bibnamefont
			{Hausoel}}, \bibinfo {author} {\bibfnamefont {P.}~\bibnamefont {Gunacker}},
		\bibinfo {author} {\bibfnamefont {A.}~\bibnamefont {Kowalski}}, \bibinfo
		{author} {\bibfnamefont {N.}~\bibnamefont {Parragh}}, \bibinfo {author}
		{\bibfnamefont {F.}~\bibnamefont {Goth}}, \bibinfo {author} {\bibfnamefont
			{K.}~\bibnamefont {Held}},\ and\ \bibinfo {author} {\bibfnamefont
			{G.}~\bibnamefont {Sangiovanni}},\ }\href
	{https://doi.org/https://doi.org/10.1016/j.cpc.2018.09.007} {\bibfield
		{journal} {\bibinfo  {journal} {Computer Physics Communications}\ }\textbf
		{\bibinfo {volume} {235}},\ \bibinfo {pages} {388 } (\bibinfo {year}
		{2019})}\BibitemShut {NoStop}%
	\bibitem [{\citenamefont {Wang}\ and\ \citenamefont {Landau}(2001)}]{Wang2001}%
	\BibitemOpen
	\bibfield  {author} {\bibinfo {author} {\bibfnamefont {F.}~\bibnamefont
			{Wang}}\ and\ \bibinfo {author} {\bibfnamefont {D.~P.}\ \bibnamefont
			{Landau}},\ }\href {https://doi.org/10.1103/PhysRevLett.86.2050} {\bibfield
		{journal} {\bibinfo  {journal} {Phys. Rev. Lett.}\ }\textbf {\bibinfo
			{volume} {86}},\ \bibinfo {pages} {2050} (\bibinfo {year}
		{2001})}\BibitemShut {NoStop}%
	\bibitem [{\citenamefont {Sherman}\ and\ \citenamefont
		{Morrison}(1950)}]{Sherman1950}%
	\BibitemOpen
	\bibfield  {author} {\bibinfo {author} {\bibfnamefont {J.}~\bibnamefont
			{Sherman}}\ and\ \bibinfo {author} {\bibfnamefont {W.~J.}\ \bibnamefont
			{Morrison}},\ }\href {https://doi.org/10.1214/aoms/1177729893} {\bibfield
		{journal} {\bibinfo  {journal} {The Annals of Mathematical Statistics}\
		}\textbf {\bibinfo {volume} {21}},\ \bibinfo {pages} {124 } (\bibinfo {year}
		{1950})}\BibitemShut {NoStop}%
	\bibitem [{\citenamefont {Hager}(1989)}]{Hager1989}%
	\BibitemOpen
	\bibfield  {author} {\bibinfo {author} {\bibfnamefont {W.~W.}\ \bibnamefont
			{Hager}},\ }\href {https://doi.org/10.1137/1031049} {\bibfield  {journal}
		{\bibinfo  {journal} {SIAM Review}\ }\textbf {\bibinfo {volume} {31}},\
		\bibinfo {pages} {221} (\bibinfo {year} {1989})}\BibitemShut {NoStop}%
	\bibitem [{\citenamefont {Harville}(1997)}]{Harville1997}%
	\BibitemOpen
	\bibfield  {author} {\bibinfo {author} {\bibfnamefont {D.~A.}\ \bibnamefont
			{Harville}},\ }\href {https://doi.org/https://doi.org/10.1007/b98818} {\emph
		{\bibinfo {title} {Matrix Algebra From a Statistician's Perspective}}}\
	(\bibinfo  {publisher} {Springer-Verlag},\ \bibinfo {address} {New York},\
	\bibinfo {year} {1997})\BibitemShut {NoStop}%
	\bibitem [{\citenamefont {Niven}\ \emph {et~al.}(1991)\citenamefont {Niven},
		\citenamefont {Zuckerman},\ and\ \citenamefont {Montgomery}}]{Niven1991}%
	\BibitemOpen
	\bibfield  {author} {\bibinfo {author} {\bibfnamefont {I.~M.}\ \bibnamefont
			{Niven}}, \bibinfo {author} {\bibfnamefont {H.~S.}\ \bibnamefont
			{Zuckerman}},\ and\ \bibinfo {author} {\bibfnamefont {H.~L.}\ \bibnamefont
			{Montgomery}},\ }\href@noop {} {\emph {\bibinfo {title} {An Introduction to
				the Theory of Numbers}}},\ \bibinfo {edition} {5th}\ ed.\ (\bibinfo
	{publisher} {John Wiley and Sons},\ \bibinfo {address} {New York},\ \bibinfo
	{year} {1991})\BibitemShut {NoStop}%
	\bibitem [{\citenamefont {Inc.}(2025)}]{Wolfram2024}%
	\BibitemOpen
	\bibfield  {author} {\bibinfo {author} {\bibfnamefont {W.~R.}\ \bibnamefont
			{Inc.}},\ }\href {https://www.wolfram.com/mathematica} {\bibinfo {title}
		{Mathematica, {V}ersion 14.1}} (\bibinfo {year} {1988--2025}),\ \bibinfo
	{note} {{C}hampaign, IL, 2024}\BibitemShut {NoStop}%
	\bibitem [{\citenamefont {Vollhardt}\ and\ \citenamefont
		{W\"olfe}(1992)}]{Vollhardt1992}%
	\BibitemOpen
	\bibfield  {author} {\bibinfo {author} {\bibfnamefont {D.}~\bibnamefont
			{Vollhardt}}\ and\ \bibinfo {author} {\bibfnamefont {P.}~\bibnamefont
			{W\"olfe}},\ }\bibinfo {title} {Electronic phase transitions}\ (\bibinfo
	{publisher} {Elsevier Science Publishers B.V.},\ \bibinfo {address} {New
		York},\ \bibinfo {year} {1992})\ Chap.~\bibinfo {chapter} {1}, pp.\ \bibinfo
	{pages} {1--78}\BibitemShut {NoStop}%
	\bibitem [{\citenamefont {Georges}\ and\ \citenamefont
		{Krauth}(1992)}]{Georges1992}%
	\BibitemOpen
	\bibfield  {author} {\bibinfo {author} {\bibfnamefont {A.}~\bibnamefont
			{Georges}}\ and\ \bibinfo {author} {\bibfnamefont {W.}~\bibnamefont
			{Krauth}},\ }\href {https://doi.org/10.1103/PhysRevLett.69.1240} {\bibfield
		{journal} {\bibinfo  {journal} {Phys. Rev. Lett.}\ }\textbf {\bibinfo
			{volume} {69}},\ \bibinfo {pages} {1240} (\bibinfo {year}
		{1992})}\BibitemShut {NoStop}%
	\bibitem [{\citenamefont {Georges}\ and\ \citenamefont
		{Kotliar}(1992)}]{Georges1992a}%
	\BibitemOpen
	\bibfield  {author} {\bibinfo {author} {\bibfnamefont {A.}~\bibnamefont
			{Georges}}\ and\ \bibinfo {author} {\bibfnamefont {G.}~\bibnamefont
			{Kotliar}},\ }\href {https://doi.org/10.1103/PhysRevB.45.6479} {\bibfield
		{journal} {\bibinfo  {journal} {Phys. Rev. B}\ }\textbf {\bibinfo {volume}
			{45}},\ \bibinfo {pages} {6479} (\bibinfo {year} {1992})}\BibitemShut
	{NoStop}%
	\bibitem [{\citenamefont {Georges}\ \emph {et~al.}(1996)\citenamefont
		{Georges}, \citenamefont {Kotliar}, \citenamefont {Krauth},\ and\
		\citenamefont {Rozenberg}}]{Georges1996}%
	\BibitemOpen
	\bibfield  {author} {\bibinfo {author} {\bibfnamefont {A.}~\bibnamefont
			{Georges}}, \bibinfo {author} {\bibfnamefont {G.}~\bibnamefont {Kotliar}},
		\bibinfo {author} {\bibfnamefont {W.}~\bibnamefont {Krauth}},\ and\ \bibinfo
		{author} {\bibfnamefont {M.~J.}\ \bibnamefont {Rozenberg}},\ }\href
	{https://doi.org/10.1103/RevModPhys.68.13} {\bibfield  {journal} {\bibinfo
			{journal} {Rev. Mod. Phys.}\ }\textbf {\bibinfo {volume} {68}},\ \bibinfo
		{pages} {13} (\bibinfo {year} {1996})}\BibitemShut {NoStop}%
	\bibitem [{\citenamefont {Laloux}\ \emph {et~al.}(1994)\citenamefont {Laloux},
		\citenamefont {Georges},\ and\ \citenamefont {Krauth}}]{laloux1994effect}%
	\BibitemOpen
	\bibfield  {author} {\bibinfo {author} {\bibfnamefont {L.}~\bibnamefont
			{Laloux}}, \bibinfo {author} {\bibfnamefont {A.}~\bibnamefont {Georges}},\
		and\ \bibinfo {author} {\bibfnamefont {W.}~\bibnamefont {Krauth}},\
	}\href@noop {} {\bibfield  {journal} {\bibinfo  {journal} {Phys. Rev. B}\
		}\textbf {\bibinfo {volume} {50}},\ \bibinfo {pages} {3092} (\bibinfo {year}
		{1994})}\BibitemShut {NoStop}%
	\bibitem [{\citenamefont {Bauer}\ and\ \citenamefont
		{Hewson}(2007)}]{Bauer2007}%
	\BibitemOpen
	\bibfield  {author} {\bibinfo {author} {\bibfnamefont {J.}~\bibnamefont
			{Bauer}}\ and\ \bibinfo {author} {\bibfnamefont {A.~C.}\ \bibnamefont
			{Hewson}},\ }\href {https://doi.org/10.1103/PhysRevB.76.035118} {\bibfield
		{journal} {\bibinfo  {journal} {Phys. Rev. B}\ }\textbf {\bibinfo {volume}
			{76}},\ \bibinfo {pages} {035118} (\bibinfo {year} {2007})}\BibitemShut
	{NoStop}%
	\bibitem [{\citenamefont {van Dongen}\ and\ \citenamefont
		{Leinung}(1997)}]{van1997mott}%
	\BibitemOpen
	\bibfield  {author} {\bibinfo {author} {\bibfnamefont {P.~G.}\ \bibnamefont
			{van Dongen}}\ and\ \bibinfo {author} {\bibfnamefont {C.}~\bibnamefont
			{Leinung}},\ }\href@noop {} {\bibfield  {journal} {\bibinfo  {journal}
			{Annalen der Physik}\ }\textbf {\bibinfo {volume} {509}},\ \bibinfo {pages}
		{45} (\bibinfo {year} {1997})}\BibitemShut {NoStop}%
	\bibitem [{\citenamefont
		{Vollhardt}(1984{\natexlab{a}})}]{vollhardt1984normal}%
	\BibitemOpen
	\bibfield  {author} {\bibinfo {author} {\bibfnamefont {D.}~\bibnamefont
			{Vollhardt}},\ }\href@noop {} {\bibfield  {journal} {\bibinfo  {journal}
			{Reviews of modern physics}\ }\textbf {\bibinfo {volume} {56}},\ \bibinfo
		{pages} {99} (\bibinfo {year} {1984}{\natexlab{a}})}\BibitemShut {NoStop}%
	\bibitem [{\citenamefont {Kubo}(1957)}]{Kubo1957}%
	\BibitemOpen
	\bibfield  {author} {\bibinfo {author} {\bibfnamefont {R.}~\bibnamefont
			{Kubo}},\ }\href@noop {} {\bibfield  {journal} {\bibinfo  {journal} {Journal
				of the Physical Society of Japan}\ }\textbf {\bibinfo {volume} {12}},\
		\bibinfo {pages} {570} (\bibinfo {year} {1957})}\BibitemShut {NoStop}%
	\bibitem [{\citenamefont {Sangiovanni}\ \emph {et~al.}(2006)\citenamefont
		{Sangiovanni}, \citenamefont {Toschi}, \citenamefont {Koch}, \citenamefont
		{Held}, \citenamefont {Capone}, \citenamefont {Castellani}, \citenamefont
		{Gunnarsson}, \citenamefont {Mo}, \citenamefont {Allen}, \citenamefont {Kim},
		\citenamefont {Sekiyama}, \citenamefont {Yamasaki}, \citenamefont {Suga},\
		and\ \citenamefont {Metcalf}}]{Sangiovanni2006}%
	\BibitemOpen
	\bibfield  {author} {\bibinfo {author} {\bibfnamefont {G.}~\bibnamefont
			{Sangiovanni}}, \bibinfo {author} {\bibfnamefont {A.}~\bibnamefont {Toschi}},
		\bibinfo {author} {\bibfnamefont {E.}~\bibnamefont {Koch}}, \bibinfo {author}
		{\bibfnamefont {K.}~\bibnamefont {Held}}, \bibinfo {author} {\bibfnamefont
			{M.}~\bibnamefont {Capone}}, \bibinfo {author} {\bibfnamefont
			{C.}~\bibnamefont {Castellani}}, \bibinfo {author} {\bibfnamefont
			{O.}~\bibnamefont {Gunnarsson}}, \bibinfo {author} {\bibfnamefont {S.-K.}\
			\bibnamefont {Mo}}, \bibinfo {author} {\bibfnamefont {J.~W.}\ \bibnamefont
			{Allen}}, \bibinfo {author} {\bibfnamefont {H.-D.}\ \bibnamefont {Kim}},
		\bibinfo {author} {\bibfnamefont {A.}~\bibnamefont {Sekiyama}}, \bibinfo
		{author} {\bibfnamefont {A.}~\bibnamefont {Yamasaki}}, \bibinfo {author}
		{\bibfnamefont {S.}~\bibnamefont {Suga}},\ and\ \bibinfo {author}
		{\bibfnamefont {P.}~\bibnamefont {Metcalf}},\ }\href
	{https://doi.org/10.1103/PhysRevB.73.205121} {\bibfield  {journal} {\bibinfo
			{journal} {Phys. Rev. B}\ }\textbf {\bibinfo {volume} {73}},\ \bibinfo
		{pages} {205121} (\bibinfo {year} {2006})}\BibitemShut {NoStop}%
	\bibitem [{\citenamefont {Wang}\ \emph {et~al.}(2009)\citenamefont {Wang},
		\citenamefont {Gull}, \citenamefont {de' Medici}, \citenamefont {Capone},\
		and\ \citenamefont {Millis}}]{Wang2009}%
	\BibitemOpen
	\bibfield  {author} {\bibinfo {author} {\bibfnamefont {X.}~\bibnamefont
			{Wang}}, \bibinfo {author} {\bibfnamefont {E.}~\bibnamefont {Gull}}, \bibinfo
		{author} {\bibfnamefont {L.}~\bibnamefont {de' Medici}}, \bibinfo {author}
		{\bibfnamefont {M.}~\bibnamefont {Capone}},\ and\ \bibinfo {author}
		{\bibfnamefont {A.~J.}\ \bibnamefont {Millis}},\ }\href
	{https://doi.org/10.1103/PhysRevB.80.045101} {\bibfield  {journal} {\bibinfo
			{journal} {Phys. Rev. B}\ }\textbf {\bibinfo {volume} {80}},\ \bibinfo
		{pages} {045101} (\bibinfo {year} {2009})}\BibitemShut {NoStop}%
	\bibitem [{\citenamefont {Fei}\ \emph {et~al.}(2021)\citenamefont {Fei},
		\citenamefont {Yeh},\ and\ \citenamefont {Gull}}]{Fei2021}%
	\BibitemOpen
	\bibfield  {author} {\bibinfo {author} {\bibfnamefont {J.}~\bibnamefont
			{Fei}}, \bibinfo {author} {\bibfnamefont {C.-N.}\ \bibnamefont {Yeh}},\ and\
		\bibinfo {author} {\bibfnamefont {E.}~\bibnamefont {Gull}},\ }\href
	{https://doi.org/10.1103/PhysRevLett.126.056402} {\bibfield  {journal}
		{\bibinfo  {journal} {Phys. Rev. Lett.}\ }\textbf {\bibinfo {volume} {126}},\
		\bibinfo {pages} {056402} (\bibinfo {year} {2021})}\BibitemShut {NoStop}%
	\bibitem [{\citenamefont {Nogaki}\ \emph
		{et~al.}(2023{\natexlab{a}})\citenamefont {Nogaki}, \citenamefont {Fei},
		\citenamefont {Gull},\ and\ \citenamefont {Shinaoka}}]{Nogaki2023}%
	\BibitemOpen
	\bibfield  {author} {\bibinfo {author} {\bibfnamefont {K.}~\bibnamefont
			{Nogaki}}, \bibinfo {author} {\bibfnamefont {J.}~\bibnamefont {Fei}},
		\bibinfo {author} {\bibfnamefont {E.}~\bibnamefont {Gull}},\ and\ \bibinfo
		{author} {\bibfnamefont {H.}~\bibnamefont {Shinaoka}},\ }\href
	{https://doi.org/10.21468/SciPostPhysCodeb.19} {\bibfield  {journal}
		{\bibinfo  {journal} {SciPost Phys. Codebases}\ ,\ \bibinfo {pages} {19}}
		(\bibinfo {year} {2023}{\natexlab{a}})}\BibitemShut {NoStop}%
	\bibitem [{\citenamefont {Nogaki}\ \emph
		{et~al.}(2023{\natexlab{b}})\citenamefont {Nogaki}, \citenamefont {Fei},
		\citenamefont {Gull},\ and\ \citenamefont {Shinaoka}}]{Nogaki2023a}%
	\BibitemOpen
	\bibfield  {author} {\bibinfo {author} {\bibfnamefont {K.}~\bibnamefont
			{Nogaki}}, \bibinfo {author} {\bibfnamefont {J.}~\bibnamefont {Fei}},
		\bibinfo {author} {\bibfnamefont {E.}~\bibnamefont {Gull}},\ and\ \bibinfo
		{author} {\bibfnamefont {H.}~\bibnamefont {Shinaoka}},\ }\href
	{https://doi.org/10.21468/SciPostPhysCodeb.19-r1.0} {\bibfield  {journal}
		{\bibinfo  {journal} {SciPost Phys. Codebases}\ ,\ \bibinfo {pages} {19}}
		(\bibinfo {year} {2023}{\natexlab{b}})}\BibitemShut {NoStop}%
	\bibitem [{\citenamefont {Jarrell}\ and\ \citenamefont
		{Gubernatis}(1996)}]{Jarrell1996}%
	\BibitemOpen
	\bibfield  {author} {\bibinfo {author} {\bibfnamefont {M.}~\bibnamefont
			{Jarrell}}\ and\ \bibinfo {author} {\bibfnamefont {J.~E.}\ \bibnamefont
			{Gubernatis}},\ }\href {https://doi.org/DOI: 10.1016/0370-1573(95)00074-7}
	{\bibfield  {journal} {\bibinfo  {journal} {Physics Reports}\ }\textbf
		{\bibinfo {volume} {269}},\ \bibinfo {pages} {133} (\bibinfo {year}
		{1996})}\BibitemShut {NoStop}%
	\bibitem [{\citenamefont {Kaufmann}\ and\ \citenamefont
		{Held}(2023)}]{Kaufmann2023}%
	\BibitemOpen
	\bibfield  {author} {\bibinfo {author} {\bibfnamefont {J.}~\bibnamefont
			{Kaufmann}}\ and\ \bibinfo {author} {\bibfnamefont {K.}~\bibnamefont
			{Held}},\ }\href {https://doi.org/https://doi.org/10.1016/j.cpc.2022.108519}
	{\bibfield  {journal} {\bibinfo  {journal} {Computer Physics Communications}\
		}\textbf {\bibinfo {volume} {282}},\ \bibinfo {pages} {108519} (\bibinfo
		{year} {2023})}\BibitemShut {NoStop}%
	\bibitem [{Note1()}]{Note1}%
	\BibitemOpen
	\bibinfo {note} {Larger magnetic fields ($B > 0.5$) are not meaningful in our
		model, as it is symmetric under the transformation $B\rightarrow
		1-B$.}\BibitemShut {Stop}%
	\bibitem [{\citenamefont {Hatsugai}\ and\ \citenamefont
		{Kohmoto}(1990)}]{Hatsugai1990}%
	\BibitemOpen
	\bibfield  {author} {\bibinfo {author} {\bibfnamefont {Y.}~\bibnamefont
			{Hatsugai}}\ and\ \bibinfo {author} {\bibfnamefont {M.}~\bibnamefont
			{Kohmoto}},\ }\href {https://doi.org/10.1103/PhysRevB.42.8282} {\bibfield
		{journal} {\bibinfo  {journal} {Phys. Rev. B}\ }\textbf {\bibinfo {volume}
			{42}},\ \bibinfo {pages} {8282} (\bibinfo {year} {1990})}\BibitemShut
	{NoStop}%
	\bibitem [{\citenamefont {Hyttrek}(2024)}]{Hyttrek2024}%
	\BibitemOpen
	\bibfield  {author} {\bibinfo {author} {\bibfnamefont {N.}~\bibnamefont
			{Hyttrek}},\ }\emph {\bibinfo {title} {Topological quantization of the Hall
			conductivity in superconducting cuprates – a Hubbard-Hofstadter model
			study}},\ \href@noop {} {Master's thesis},\ \bibinfo  {school} {University of
		Hamburg} (\bibinfo {year} {2024})\BibitemShut {NoStop}%
	\bibitem [{\citenamefont {Bulla}\ and\ \citenamefont
		{Potthoff}(2000)}]{Bulla2000}%
	\BibitemOpen
	\bibfield  {author} {\bibinfo {author} {\bibfnamefont {R.}~\bibnamefont
			{Bulla}}\ and\ \bibinfo {author} {\bibfnamefont {M.}~\bibnamefont
			{Potthoff}},\ }\href {https://doi.org/10.1007/s100510050030} {\bibfield
		{journal} {\bibinfo  {journal} {The European Physical Journal B - Condensed
				Matter and Complex Systems}\ }\textbf {\bibinfo {volume} {13}},\ \bibinfo
		{pages} {257} (\bibinfo {year} {2000})}\BibitemShut {NoStop}%
	\bibitem [{\citenamefont {Gutzwiller}(1965)}]{gutzwiller1965correlation}%
	\BibitemOpen
	\bibfield  {author} {\bibinfo {author} {\bibfnamefont {M.~C.}\ \bibnamefont
			{Gutzwiller}},\ }\href@noop {} {\bibfield  {journal} {\bibinfo  {journal}
			{Physical Review}\ }\textbf {\bibinfo {volume} {137}},\ \bibinfo {pages}
		{A1726} (\bibinfo {year} {1965})}\BibitemShut {NoStop}%
	\bibitem [{\citenamefont {Brinkman}\ and\ \citenamefont
		{Rice}(1970)}]{Brinkman1970}%
	\BibitemOpen
	\bibfield  {author} {\bibinfo {author} {\bibfnamefont {W.~F.}\ \bibnamefont
			{Brinkman}}\ and\ \bibinfo {author} {\bibfnamefont {T.~M.}\ \bibnamefont
			{Rice}},\ }\href {https://doi.org/10.1103/PhysRevB.2.4302} {\bibfield
		{journal} {\bibinfo  {journal} {Phys. Rev. B}\ }\textbf {\bibinfo {volume}
			{2}},\ \bibinfo {pages} {4302} (\bibinfo {year} {1970})}\BibitemShut
	{NoStop}%
	\bibitem [{\citenamefont {Vollhardt}(1984{\natexlab{b}})}]{Vollhardt1984}%
	\BibitemOpen
	\bibfield  {author} {\bibinfo {author} {\bibfnamefont {D.}~\bibnamefont
			{Vollhardt}},\ }\href {https://doi.org/10.1103/RevModPhys.56.99} {\bibfield
		{journal} {\bibinfo  {journal} {Rev. Mod. Phys.}\ }\textbf {\bibinfo {volume}
			{56}},\ \bibinfo {pages} {99} (\bibinfo {year}
		{1984}{\natexlab{b}})}\BibitemShut {NoStop}%
	\bibitem [{\citenamefont {Wysoki\ifmmode~\acute{n}\else \'{n}\fi{}ski}\ and\
		\citenamefont {Fabrizio}(2017)}]{Fabrizio2017}%
	\BibitemOpen
	\bibfield  {author} {\bibinfo {author} {\bibfnamefont {M.~M.}\ \bibnamefont
			{Wysoki\ifmmode~\acute{n}\else \'{n}\fi{}ski}}\ and\ \bibinfo {author}
		{\bibfnamefont {M.}~\bibnamefont {Fabrizio}},\ }\href
	{https://doi.org/10.1103/PhysRevB.95.161106} {\bibfield  {journal} {\bibinfo
			{journal} {Phys. Rev. B}\ }\textbf {\bibinfo {volume} {95}},\ \bibinfo
		{pages} {161106} (\bibinfo {year} {2017})}\BibitemShut {NoStop}%
	\bibitem [{Note2()}]{Note2}%
	\BibitemOpen
	\bibinfo {note} {The improved treatment of fluctuations in Ref.~\cite
		{Fabrizio2017} reduces $U_c$ by a factor of $\sim 0.822$. Moreover, according
		to our phase diagram the critical $U_c\protect \!=\protect \!2.71$ at
		$T\protect \!=\protect \!0$ is reduced to $U_c\protect \!=\protect \!2.45$ at
		$T\protect \!=\protect \!0.01$. Hence, we have $U_c=(2.45/2.71)\times
		0.822\times 8|E_\protect \text {kin}^0|=5.945|E_\protect \text {kin}^0| \sim
		6|E_\protect \text {kin}^0|$.}\BibitemShut {Stop}%
	\bibitem [{\citenamefont {Naumis}(2016)}]{naumis2016topological}%
	\BibitemOpen
	\bibfield  {author} {\bibinfo {author} {\bibfnamefont {G.~G.}\ \bibnamefont
			{Naumis}},\ }\href@noop {} {\bibfield  {journal} {\bibinfo  {journal}
			{Physics Letters A}\ }\textbf {\bibinfo {volume} {380}},\ \bibinfo {pages}
		{1772} (\bibinfo {year} {2016})}\BibitemShut {NoStop}%
	\bibitem [{\citenamefont {Pairault}\ \emph {et~al.}(1998)\citenamefont
		{Pairault}, \citenamefont {S\'en\'echal},\ and\ \citenamefont
		{Tremblay}}]{Pairault1998}%
	\BibitemOpen
	\bibfield  {author} {\bibinfo {author} {\bibfnamefont {S.}~\bibnamefont
			{Pairault}}, \bibinfo {author} {\bibfnamefont {D.}~\bibnamefont
			{S\'en\'echal}},\ and\ \bibinfo {author} {\bibfnamefont {A.-M.~S.}\
			\bibnamefont {Tremblay}},\ }\href
	{https://doi.org/10.1103/PhysRevLett.80.5389} {\bibfield  {journal} {\bibinfo
			{journal} {Phys. Rev. Lett.}\ }\textbf {\bibinfo {volume} {80}},\ \bibinfo
		{pages} {5389} (\bibinfo {year} {1998})}\BibitemShut {NoStop}%
	\bibitem [{\citenamefont {Pairault}\ \emph {et~al.}(2000)\citenamefont
		{Pairault}, \citenamefont {S\'en\'echal},\ and\ \citenamefont
		{Tremblay}}]{Pairault2000}%
	\BibitemOpen
	\bibfield  {author} {\bibinfo {author} {\bibfnamefont {S.}~\bibnamefont
			{Pairault}}, \bibinfo {author} {\bibfnamefont {D.}~\bibnamefont
			{S\'en\'echal}},\ and\ \bibinfo {author} {\bibfnamefont {A.-M.~S.}\
			\bibnamefont {Tremblay}},\ }\href {https://doi.org/10.1007/s100510070253}
	{\bibfield  {journal} {\bibinfo  {journal} {Eur. Phys. J. B}\ }\textbf
		{\bibinfo {volume} {16}},\ \bibinfo {pages} {85} (\bibinfo {year}
		{2000})}\BibitemShut {NoStop}%
	\bibitem [{\citenamefont {Sen}\ and\ \citenamefont
		{Chitra}(1995)}]{sen1995large}%
	\BibitemOpen
	\bibfield  {author} {\bibinfo {author} {\bibfnamefont {D.}~\bibnamefont
			{Sen}}\ and\ \bibinfo {author} {\bibfnamefont {R.}~\bibnamefont {Chitra}},\
	}\href@noop {} {\bibfield  {journal} {\bibinfo  {journal} {Phys. Rev. B}\
		}\textbf {\bibinfo {volume} {51}},\ \bibinfo {pages} {1922} (\bibinfo {year}
		{1995})}\BibitemShut {NoStop}%
	\bibitem [{\citenamefont {Aharonov}\ and\ \citenamefont
		{Bohm}(1959)}]{aharonov1959significance}%
	\BibitemOpen
	\bibfield  {author} {\bibinfo {author} {\bibfnamefont {Y.}~\bibnamefont
			{Aharonov}}\ and\ \bibinfo {author} {\bibfnamefont {D.}~\bibnamefont
			{Bohm}},\ }\href@noop {} {\bibfield  {journal} {\bibinfo  {journal} {Physical
				review}\ }\textbf {\bibinfo {volume} {115}},\ \bibinfo {pages} {485}
		(\bibinfo {year} {1959})}\BibitemShut {NoStop}%
	\bibitem [{\citenamefont {Choi}\ \emph {et~al.}(2020)\citenamefont {Choi},
		\citenamefont {Ahn}, \citenamefont {Moon},\ and\ \citenamefont
		{Lee}}]{Choi2020}%
	\BibitemOpen
	\bibfield  {author} {\bibinfo {author} {\bibfnamefont {S.}~\bibnamefont
			{Choi}}, \bibinfo {author} {\bibfnamefont {G.}~\bibnamefont {Ahn}}, \bibinfo
		{author} {\bibfnamefont {S.~J.}\ \bibnamefont {Moon}},\ and\ \bibinfo
		{author} {\bibfnamefont {S.}~\bibnamefont {Lee}},\ }\href
	{https://doi.org/10.1038/s41598-020-66439-2} {\bibfield  {journal} {\bibinfo
			{journal} {Scientific Reports}\ }\textbf {\bibinfo {volume} {10}},\ \bibinfo
		{pages} {9721} (\bibinfo {year} {2020})}\BibitemShut {NoStop}%
	\bibitem [{\citenamefont {Lu}\ and\ \citenamefont {Robertson}(2019)}]{Lu2019}%
	\BibitemOpen
	\bibfield  {author} {\bibinfo {author} {\bibfnamefont {H.}~\bibnamefont
			{Lu}}\ and\ \bibinfo {author} {\bibfnamefont {J.}~\bibnamefont {Robertson}},\
	}\href {https://doi.org/https://doi.org/10.1002/pssb.201900210} {\bibfield
		{journal} {\bibinfo  {journal} {Physica Status Solidi (b)}\ }\textbf
		{\bibinfo {volume} {256}},\ \bibinfo {pages} {1900210} (\bibinfo {year}
		{2019})}\BibitemShut {NoStop}%
	\bibitem [{\citenamefont {Dalibard}(2015)}]{Dalibard2015}%
	\BibitemOpen
	\bibfield  {author} {\bibinfo {author} {\bibfnamefont {J.}~\bibnamefont
			{Dalibard}},\ }\href {https://arxiv.org/abs/1504.05520} {\bibinfo {title}
		{Introduction to the physics of artificial gauge fields}} (\bibinfo {year}
	{2015}),\ \Eprint {https://arxiv.org/abs/1504.05520} {arXiv:1504.05520
		[cond-mat.quant-gas]} \BibitemShut {NoStop}%
	\bibitem [{\citenamefont {Kang}\ \emph {et~al.}(2021)\citenamefont {Kang},
		\citenamefont {Bernevig},\ and\ \citenamefont {Vafek}}]{kang2021cascades}%
	\BibitemOpen
	\bibfield  {author} {\bibinfo {author} {\bibfnamefont {J.}~\bibnamefont
			{Kang}}, \bibinfo {author} {\bibfnamefont {B.~A.}\ \bibnamefont {Bernevig}},\
		and\ \bibinfo {author} {\bibfnamefont {O.}~\bibnamefont {Vafek}},\
	}\href@noop {} {\bibfield  {journal} {\bibinfo  {journal} {Phys. Rev. Lett.}\
		}\textbf {\bibinfo {volume} {127}},\ \bibinfo {pages} {266402} (\bibinfo
		{year} {2021})}\BibitemShut {NoStop}%
	\bibitem [{\citenamefont {Kometter}\ \emph {et~al.}(2023)\citenamefont
		{Kometter}, \citenamefont {Yu}, \citenamefont {Devakul}, \citenamefont
		{Reddy}, \citenamefont {Zhang}, \citenamefont {Foutty}, \citenamefont
		{Watanabe}, \citenamefont {Taniguchi}, \citenamefont {Fu},\ and\
		\citenamefont {Feldman}}]{kometter2023hofstadter}%
	\BibitemOpen
	\bibfield  {author} {\bibinfo {author} {\bibfnamefont {C.~R.}\ \bibnamefont
			{Kometter}}, \bibinfo {author} {\bibfnamefont {J.}~\bibnamefont {Yu}},
		\bibinfo {author} {\bibfnamefont {T.}~\bibnamefont {Devakul}}, \bibinfo
		{author} {\bibfnamefont {A.~P.}\ \bibnamefont {Reddy}}, \bibinfo {author}
		{\bibfnamefont {Y.}~\bibnamefont {Zhang}}, \bibinfo {author} {\bibfnamefont
			{B.~A.}\ \bibnamefont {Foutty}}, \bibinfo {author} {\bibfnamefont
			{K.}~\bibnamefont {Watanabe}}, \bibinfo {author} {\bibfnamefont
			{T.}~\bibnamefont {Taniguchi}}, \bibinfo {author} {\bibfnamefont
			{L.}~\bibnamefont {Fu}},\ and\ \bibinfo {author} {\bibfnamefont {B.~E.}\
			\bibnamefont {Feldman}},\ }\href@noop {} {\bibfield  {journal} {\bibinfo
			{journal} {Nature Physics}\ }\textbf {\bibinfo {volume} {19}},\ \bibinfo
		{pages} {1861} (\bibinfo {year} {2023})}\BibitemShut {NoStop}%
\end{thebibliography}

\begin{thebibliography}{33}%
	\makeatletter
	\providecommand \@ifxundefined [1]{%
		\@ifx{#1\undefined}
	}%
	\providecommand \@ifnum [1]{%
		\ifnum #1\expandafter \@firstoftwo
		\else \expandafter \@secondoftwo
		\fi
	}%
	\providecommand \@ifx [1]{%
		\ifx #1\expandafter \@firstoftwo
		\else \expandafter \@secondoftwo
		\fi
	}%
	\providecommand \natexlab [1]{#1}%
	\providecommand \enquote  [1]{``#1''}%
	\providecommand \bibnamefont  [1]{#1}%
	\providecommand \bibfnamefont [1]{#1}%
	\providecommand \citenamefont [1]{#1}%
	\providecommand \href@noop [0]{\@secondoftwo}%
	\providecommand \href [0]{\begingroup \@sanitize@url \@href}%
	\providecommand \@href[1]{\@@startlink{#1}\@@href}%
	\providecommand \@@href[1]{\endgroup#1\@@endlink}%
	\providecommand \@sanitize@url [0]{\catcode `\\12\catcode `\$12\catcode
		`\&12\catcode `\#12\catcode `\^12\catcode `\_12\catcode `\%12\relax}%
	\providecommand \@@startlink[1]{}%
	\providecommand \@@endlink[0]{}%
	\providecommand \url  [0]{\begingroup\@sanitize@url \@url }%
	\providecommand \@url [1]{\endgroup\@href {#1}{\urlprefix }}%
	\providecommand \urlprefix  [0]{URL }%
	\providecommand \Eprint [0]{\href }%
	\providecommand \doibase [0]{https://doi.org/}%
	\providecommand \selectlanguage [0]{\@gobble}%
	\providecommand \bibinfo  [0]{\@secondoftwo}%
	\providecommand \bibfield  [0]{\@secondoftwo}%
	\providecommand \translation [1]{[#1]}%
	\providecommand \BibitemOpen [0]{}%
	\providecommand \bibitemStop [0]{}%
	\providecommand \bibitemNoStop [0]{.\EOS\space}%
	\providecommand \EOS [0]{\spacefactor3000\relax}%
	\providecommand \BibitemShut  [1]{\csname bibitem#1\endcsname}%
	\let\auto@bib@innerbib\@empty
	\bibitem [{\citenamefont {Markov}\ \emph {et~al.}(2019)\citenamefont {Markov},
		\citenamefont {Rohringer},\ and\ \citenamefont {Rubtsov}}]{S-Markov2019}%
	\BibitemOpen
	\bibfield  {author} {\bibinfo {author} {\bibfnamefont {A.~A.}\ \bibnamefont
			{Markov}}, \bibinfo {author} {\bibfnamefont {G.}~\bibnamefont {Rohringer}},\
		and\ \bibinfo {author} {\bibfnamefont {A.~N.}\ \bibnamefont {Rubtsov}},\
	}\href {https://doi.org/10.1103/PhysRevB.100.115102} {\bibfield  {journal}
		{\bibinfo  {journal} {Phys. Rev. B}\ }\textbf {\bibinfo {volume} {100}},\
		\bibinfo {pages} {115102} (\bibinfo {year} {2019})}\BibitemShut {NoStop}%
	\bibitem [{\citenamefont {Vu\ifmmode \check{c}\else \v{c}\fi{}i\ifmmode
			\check{c}\else \v{c}\fi{}evi\ifmmode~\acute{c}\else \'{c}\fi{}}\ and\
		\citenamefont {\ifmmode~\check{Z}\else
			\v{Z}\fi{}itko}(2021{\natexlab{a}})}]{S-Vucicevic2021a}%
	\BibitemOpen
	\bibfield  {author} {\bibinfo {author} {\bibfnamefont {J.}~\bibnamefont
			{Vu\ifmmode \check{c}\else \v{c}\fi{}i\ifmmode \check{c}\else
				\v{c}\fi{}evi\ifmmode~\acute{c}\else \'{c}\fi{}}}\ and\ \bibinfo {author}
		{\bibfnamefont {R.}~\bibnamefont {\ifmmode~\check{Z}\else \v{Z}\fi{}itko}},\
	}\href {https://doi.org/10.1103/PhysRevB.104.205101} {\bibfield  {journal}
		{\bibinfo  {journal} {Phys. Rev. B}\ }\textbf {\bibinfo {volume} {104}},\
		\bibinfo {pages} {205101} (\bibinfo {year} {2021}{\natexlab{a}})}\BibitemShut
	{NoStop}%
	\bibitem [{\citenamefont {Hatsugai}\ and\ \citenamefont
		{Kohmoto}(1990)}]{S-Hatsugai1990}%
	\BibitemOpen
	\bibfield  {author} {\bibinfo {author} {\bibfnamefont {Y.}~\bibnamefont
			{Hatsugai}}\ and\ \bibinfo {author} {\bibfnamefont {M.}~\bibnamefont
			{Kohmoto}},\ }\href {https://doi.org/10.1103/PhysRevB.42.8282} {\bibfield
		{journal} {\bibinfo  {journal} {Phys. Rev. B}\ }\textbf {\bibinfo {volume}
			{42}},\ \bibinfo {pages} {8282} (\bibinfo {year} {1990})}\BibitemShut
	{NoStop}%
	\bibitem [{\citenamefont {Hyttrek}(2024)}]{S-Hyttrek2024}%
	\BibitemOpen
	\bibfield  {author} {\bibinfo {author} {\bibfnamefont {N.}~\bibnamefont
			{Hyttrek}},\ }\emph {\bibinfo {title} {Topological quantization of the Hall
			conductivity in superconducting cuprates – a Hubbard-Hofstadter model
			study}},\ \href@noop {} {Master's thesis},\ \bibinfo  {school} {University of
		Hamburg} (\bibinfo {year} {2024})\BibitemShut {NoStop}%
	\bibitem [{\citenamefont {Acheche}\ \emph {et~al.}(2017)\citenamefont
		{Acheche}, \citenamefont {Arsenault},\ and\ \citenamefont
		{Tremblay}}]{S-Acheche2017}%
	\BibitemOpen
	\bibfield  {author} {\bibinfo {author} {\bibfnamefont {S.}~\bibnamefont
			{Acheche}}, \bibinfo {author} {\bibfnamefont {L.-F.}\ \bibnamefont
			{Arsenault}},\ and\ \bibinfo {author} {\bibfnamefont {A.-M.~S.}\ \bibnamefont
			{Tremblay}},\ }\href {https://doi.org/10.1103/PhysRevB.96.235135} {\bibfield
		{journal} {\bibinfo  {journal} {Phys. Rev. B}\ }\textbf {\bibinfo {volume}
			{96}},\ \bibinfo {pages} {235135} (\bibinfo {year} {2017})}\BibitemShut
	{NoStop}%
	\bibitem [{\citenamefont {Vu\ifmmode \check{c}\else \v{c}\fi{}i\ifmmode
			\check{c}\else \v{c}\fi{}evi\ifmmode~\acute{c}\else \'{c}\fi{}}\ and\
		\citenamefont {\ifmmode~\check{Z}\else
			\v{Z}\fi{}itko}(2021{\natexlab{b}})}]{S-Vucicevic2021}%
	\BibitemOpen
	\bibfield  {author} {\bibinfo {author} {\bibfnamefont {J.~c.~v.}\
			\bibnamefont {Vu\ifmmode \check{c}\else \v{c}\fi{}i\ifmmode \check{c}\else
				\v{c}\fi{}evi\ifmmode~\acute{c}\else \'{c}\fi{}}}\ and\ \bibinfo {author}
		{\bibfnamefont {R.}~\bibnamefont {\ifmmode~\check{Z}\else \v{Z}\fi{}itko}},\
	}\href {https://doi.org/10.1103/PhysRevLett.127.196601} {\bibfield  {journal}
		{\bibinfo  {journal} {Phys. Rev. Lett.}\ }\textbf {\bibinfo {volume} {127}},\
		\bibinfo {pages} {196601} (\bibinfo {year} {2021}{\natexlab{b}})}\BibitemShut
	{NoStop}%
	\bibitem [{\citenamefont {Wallerberger}\ \emph {et~al.}(2019)\citenamefont
		{Wallerberger}, \citenamefont {Hausoel}, \citenamefont {Gunacker},
		\citenamefont {Kowalski}, \citenamefont {Parragh}, \citenamefont {Goth},
		\citenamefont {Held},\ and\ \citenamefont
		{Sangiovanni}}]{S-Wallerberger2019}%
	\BibitemOpen
	\bibfield  {author} {\bibinfo {author} {\bibfnamefont {M.}~\bibnamefont
			{Wallerberger}}, \bibinfo {author} {\bibfnamefont {A.}~\bibnamefont
			{Hausoel}}, \bibinfo {author} {\bibfnamefont {P.}~\bibnamefont {Gunacker}},
		\bibinfo {author} {\bibfnamefont {A.}~\bibnamefont {Kowalski}}, \bibinfo
		{author} {\bibfnamefont {N.}~\bibnamefont {Parragh}}, \bibinfo {author}
		{\bibfnamefont {F.}~\bibnamefont {Goth}}, \bibinfo {author} {\bibfnamefont
			{K.}~\bibnamefont {Held}},\ and\ \bibinfo {author} {\bibfnamefont
			{G.}~\bibnamefont {Sangiovanni}},\ }\href
	{https://doi.org/https://doi.org/10.1016/j.cpc.2018.09.007} {\bibfield
		{journal} {\bibinfo  {journal} {Computer Physics Communications}\ }\textbf
		{\bibinfo {volume} {235}},\ \bibinfo {pages} {388 } (\bibinfo {year}
		{2019})}\BibitemShut {NoStop}%
	\bibitem [{\citenamefont {Georges}\ \emph {et~al.}(1996)\citenamefont
		{Georges}, \citenamefont {Kotliar}, \citenamefont {Krauth},\ and\
		\citenamefont {Rozenberg}}]{S-Georges1996}%
	\BibitemOpen
	\bibfield  {author} {\bibinfo {author} {\bibfnamefont {A.}~\bibnamefont
			{Georges}}, \bibinfo {author} {\bibfnamefont {G.}~\bibnamefont {Kotliar}},
		\bibinfo {author} {\bibfnamefont {W.}~\bibnamefont {Krauth}},\ and\ \bibinfo
		{author} {\bibfnamefont {M.~J.}\ \bibnamefont {Rozenberg}},\ }\href
	{https://doi.org/10.1103/RevModPhys.68.13} {\bibfield  {journal} {\bibinfo
			{journal} {Rev. Mod. Phys.}\ }\textbf {\bibinfo {volume} {68}},\ \bibinfo
		{pages} {13} (\bibinfo {year} {1996})}\BibitemShut {NoStop}%
	\bibitem [{\citenamefont {Wang}\ and\ \citenamefont
		{Landau}(2001)}]{S-Wang2001}%
	\BibitemOpen
	\bibfield  {author} {\bibinfo {author} {\bibfnamefont {F.}~\bibnamefont
			{Wang}}\ and\ \bibinfo {author} {\bibfnamefont {D.~P.}\ \bibnamefont
			{Landau}},\ }\href {https://doi.org/10.1103/PhysRevLett.86.2050} {\bibfield
		{journal} {\bibinfo  {journal} {Phys. Rev. Lett.}\ }\textbf {\bibinfo
			{volume} {86}},\ \bibinfo {pages} {2050} (\bibinfo {year}
		{2001})}\BibitemShut {NoStop}%
	\bibitem [{\citenamefont {Sherman}\ and\ \citenamefont
		{Morrison}(1950)}]{S-Sherman1950}%
	\BibitemOpen
	\bibfield  {author} {\bibinfo {author} {\bibfnamefont {J.}~\bibnamefont
			{Sherman}}\ and\ \bibinfo {author} {\bibfnamefont {W.~J.}\ \bibnamefont
			{Morrison}},\ }\href {https://doi.org/10.1214/aoms/1177729893} {\bibfield
		{journal} {\bibinfo  {journal} {The Annals of Mathematical Statistics}\
		}\textbf {\bibinfo {volume} {21}},\ \bibinfo {pages} {124 } (\bibinfo {year}
		{1950})}\BibitemShut {NoStop}%
	\bibitem [{\citenamefont {Hager}(1989)}]{S-Hager1989}%
	\BibitemOpen
	\bibfield  {author} {\bibinfo {author} {\bibfnamefont {W.~W.}\ \bibnamefont
			{Hager}},\ }\href {https://doi.org/10.1137/1031049} {\bibfield  {journal}
		{\bibinfo  {journal} {SIAM Review}\ }\textbf {\bibinfo {volume} {31}},\
		\bibinfo {pages} {221} (\bibinfo {year} {1989})}\BibitemShut {NoStop}%
	\bibitem [{\citenamefont {Harville}(1997)}]{S-Harville1997}%
	\BibitemOpen
	\bibfield  {author} {\bibinfo {author} {\bibfnamefont {D.~A.}\ \bibnamefont
			{Harville}},\ }\href {https://doi.org/https://doi.org/10.1007/b98818} {\emph
		{\bibinfo {title} {Matrix Algebra From a Statistician's Perspective}}}\
	(\bibinfo  {publisher} {Springer-Verlag},\ \bibinfo {address} {New York},\
	\bibinfo {year} {1997})\BibitemShut {NoStop}%
	\bibitem [{\citenamefont {Niven}\ \emph {et~al.}(1991)\citenamefont {Niven},
		\citenamefont {Zuckerman},\ and\ \citenamefont {Montgomery}}]{S-Niven1991}%
	\BibitemOpen
	\bibfield  {author} {\bibinfo {author} {\bibfnamefont {I.~M.}\ \bibnamefont
			{Niven}}, \bibinfo {author} {\bibfnamefont {H.~S.}\ \bibnamefont
			{Zuckerman}},\ and\ \bibinfo {author} {\bibfnamefont {H.~L.}\ \bibnamefont
			{Montgomery}},\ }\href@noop {} {\emph {\bibinfo {title} {An Introduction to
				the Theory of Numbers}}},\ \bibinfo {edition} {5th}\ ed.\ (\bibinfo
	{publisher} {John Wiley and Sons},\ \bibinfo {address} {New York},\ \bibinfo
	{year} {1991})\BibitemShut {NoStop}%
	\bibitem [{\citenamefont {Pairault}\ \emph {et~al.}(1998)\citenamefont
		{Pairault}, \citenamefont {S\'en\'echal},\ and\ \citenamefont
		{Tremblay}}]{S-Pairault1998}%
	\BibitemOpen
	\bibfield  {author} {\bibinfo {author} {\bibfnamefont {S.}~\bibnamefont
			{Pairault}}, \bibinfo {author} {\bibfnamefont {D.}~\bibnamefont
			{S\'en\'echal}},\ and\ \bibinfo {author} {\bibfnamefont {A.-M.~S.}\
			\bibnamefont {Tremblay}},\ }\href
	{https://doi.org/10.1103/PhysRevLett.80.5389} {\bibfield  {journal} {\bibinfo
			{journal} {Phys. Rev. Lett.}\ }\textbf {\bibinfo {volume} {80}},\ \bibinfo
		{pages} {5389} (\bibinfo {year} {1998})}\BibitemShut {NoStop}%
	\bibitem [{\citenamefont {Pairault}\ \emph {et~al.}(2000)\citenamefont
		{Pairault}, \citenamefont {S\'en\'echal},\ and\ \citenamefont
		{Tremblay}}]{S-Pairault2000}%
	\BibitemOpen
	\bibfield  {author} {\bibinfo {author} {\bibfnamefont {S.}~\bibnamefont
			{Pairault}}, \bibinfo {author} {\bibfnamefont {D.}~\bibnamefont
			{S\'en\'echal}},\ and\ \bibinfo {author} {\bibfnamefont {A.-M.~S.}\
			\bibnamefont {Tremblay}},\ }\href {https://doi.org/10.1007/s100510070253}
	{\bibfield  {journal} {\bibinfo  {journal} {Eur. Phys. J. B}\ }\textbf
		{\bibinfo {volume} {16}},\ \bibinfo {pages} {85} (\bibinfo {year}
		{2000})}\BibitemShut {NoStop}%
	\bibitem [{\citenamefont {Inc.}(2025)}]{S-Wolfram2024}%
	\BibitemOpen
	\bibfield  {author} {\bibinfo {author} {\bibfnamefont {W.~R.}\ \bibnamefont
			{Inc.}},\ }\href {https://www.wolfram.com/mathematica} {\bibinfo {title}
		{Mathematica, {V}ersion 14.1}} (\bibinfo {year} {1988--2025}),\ \bibinfo
	{note} {{C}hampaign, IL, 2024}\BibitemShut {NoStop}%
	\bibitem [{\citenamefont {Aharonov}\ and\ \citenamefont
		{Bohm}(1959)}]{S-aharonov1959significance}%
	\BibitemOpen
	\bibfield  {author} {\bibinfo {author} {\bibfnamefont {Y.}~\bibnamefont
			{Aharonov}}\ and\ \bibinfo {author} {\bibfnamefont {D.}~\bibnamefont
			{Bohm}},\ }\href@noop {} {\bibfield  {journal} {\bibinfo  {journal} {Physical
				review}\ }\textbf {\bibinfo {volume} {115}},\ \bibinfo {pages} {485}
		(\bibinfo {year} {1959})}\BibitemShut {NoStop}%
	\bibitem [{\citenamefont {Sen}\ and\ \citenamefont
		{Chitra}(1995)}]{S-sen1995large}%
	\BibitemOpen
	\bibfield  {author} {\bibinfo {author} {\bibfnamefont {D.}~\bibnamefont
			{Sen}}\ and\ \bibinfo {author} {\bibfnamefont {R.}~\bibnamefont {Chitra}},\
	}\href@noop {} {\bibfield  {journal} {\bibinfo  {journal} {Phys. Rev. B}\
		}\textbf {\bibinfo {volume} {51}},\ \bibinfo {pages} {1922} (\bibinfo {year}
		{1995})}\BibitemShut {NoStop}%
	\bibitem [{\citenamefont {Fei}\ \emph {et~al.}(2021)\citenamefont {Fei},
		\citenamefont {Yeh},\ and\ \citenamefont {Gull}}]{S-Fei2021}%
	\BibitemOpen
	\bibfield  {author} {\bibinfo {author} {\bibfnamefont {J.}~\bibnamefont
			{Fei}}, \bibinfo {author} {\bibfnamefont {C.-N.}\ \bibnamefont {Yeh}},\ and\
		\bibinfo {author} {\bibfnamefont {E.}~\bibnamefont {Gull}},\ }\href
	{https://doi.org/10.1103/PhysRevLett.126.056402} {\bibfield  {journal}
		{\bibinfo  {journal} {Phys. Rev. Lett.}\ }\textbf {\bibinfo {volume} {126}},\
		\bibinfo {pages} {056402} (\bibinfo {year} {2021})}\BibitemShut {NoStop}%
	\bibitem [{\citenamefont {Nogaki}\ \emph
		{et~al.}(2023{\natexlab{a}})\citenamefont {Nogaki}, \citenamefont {Fei},
		\citenamefont {Gull},\ and\ \citenamefont {Shinaoka}}]{S-Nogaki2023}%
	\BibitemOpen
	\bibfield  {author} {\bibinfo {author} {\bibfnamefont {K.}~\bibnamefont
			{Nogaki}}, \bibinfo {author} {\bibfnamefont {J.}~\bibnamefont {Fei}},
		\bibinfo {author} {\bibfnamefont {E.}~\bibnamefont {Gull}},\ and\ \bibinfo
		{author} {\bibfnamefont {H.}~\bibnamefont {Shinaoka}},\ }\href
	{https://doi.org/10.21468/SciPostPhysCodeb.19} {\bibfield  {journal}
		{\bibinfo  {journal} {SciPost Phys. Codebases}\ ,\ \bibinfo {pages} {19}}
		(\bibinfo {year} {2023}{\natexlab{a}})}\BibitemShut {NoStop}%
	\bibitem [{\citenamefont {Nogaki}\ \emph
		{et~al.}(2023{\natexlab{b}})\citenamefont {Nogaki}, \citenamefont {Fei},
		\citenamefont {Gull},\ and\ \citenamefont {Shinaoka}}]{S-Nogaki2023a}%
	\BibitemOpen
	\bibfield  {author} {\bibinfo {author} {\bibfnamefont {K.}~\bibnamefont
			{Nogaki}}, \bibinfo {author} {\bibfnamefont {J.}~\bibnamefont {Fei}},
		\bibinfo {author} {\bibfnamefont {E.}~\bibnamefont {Gull}},\ and\ \bibinfo
		{author} {\bibfnamefont {H.}~\bibnamefont {Shinaoka}},\ }\href
	{https://doi.org/10.21468/SciPostPhysCodeb.19-r1.0} {\bibfield  {journal}
		{\bibinfo  {journal} {SciPost Phys. Codebases}\ ,\ \bibinfo {pages} {19}}
		(\bibinfo {year} {2023}{\natexlab{b}})}\BibitemShut {NoStop}%
	\bibitem [{\citenamefont {Sangiovanni}\ \emph {et~al.}(2006)\citenamefont
		{Sangiovanni}, \citenamefont {Toschi}, \citenamefont {Koch}, \citenamefont
		{Held}, \citenamefont {Capone}, \citenamefont {Castellani}, \citenamefont
		{Gunnarsson}, \citenamefont {Mo}, \citenamefont {Allen}, \citenamefont {Kim},
		\citenamefont {Sekiyama}, \citenamefont {Yamasaki}, \citenamefont {Suga},\
		and\ \citenamefont {Metcalf}}]{S-Sangiovanni2006}%
	\BibitemOpen
	\bibfield  {author} {\bibinfo {author} {\bibfnamefont {G.}~\bibnamefont
			{Sangiovanni}}, \bibinfo {author} {\bibfnamefont {A.}~\bibnamefont {Toschi}},
		\bibinfo {author} {\bibfnamefont {E.}~\bibnamefont {Koch}}, \bibinfo {author}
		{\bibfnamefont {K.}~\bibnamefont {Held}}, \bibinfo {author} {\bibfnamefont
			{M.}~\bibnamefont {Capone}}, \bibinfo {author} {\bibfnamefont
			{C.}~\bibnamefont {Castellani}}, \bibinfo {author} {\bibfnamefont
			{O.}~\bibnamefont {Gunnarsson}}, \bibinfo {author} {\bibfnamefont {S.-K.}\
			\bibnamefont {Mo}}, \bibinfo {author} {\bibfnamefont {J.~W.}\ \bibnamefont
			{Allen}}, \bibinfo {author} {\bibfnamefont {H.-D.}\ \bibnamefont {Kim}},
		\bibinfo {author} {\bibfnamefont {A.}~\bibnamefont {Sekiyama}}, \bibinfo
		{author} {\bibfnamefont {A.}~\bibnamefont {Yamasaki}}, \bibinfo {author}
		{\bibfnamefont {S.}~\bibnamefont {Suga}},\ and\ \bibinfo {author}
		{\bibfnamefont {P.}~\bibnamefont {Metcalf}},\ }\href
	{https://doi.org/10.1103/PhysRevB.73.205121} {\bibfield  {journal} {\bibinfo
			{journal} {Phys. Rev. B}\ }\textbf {\bibinfo {volume} {73}},\ \bibinfo
		{pages} {205121} (\bibinfo {year} {2006})}\BibitemShut {NoStop}%
	\bibitem [{\citenamefont {Wang}\ \emph {et~al.}(2009)\citenamefont {Wang},
		\citenamefont {Gull}, \citenamefont {de' Medici}, \citenamefont {Capone},\
		and\ \citenamefont {Millis}}]{S-Wang2009}%
	\BibitemOpen
	\bibfield  {author} {\bibinfo {author} {\bibfnamefont {X.}~\bibnamefont
			{Wang}}, \bibinfo {author} {\bibfnamefont {E.}~\bibnamefont {Gull}}, \bibinfo
		{author} {\bibfnamefont {L.}~\bibnamefont {de' Medici}}, \bibinfo {author}
		{\bibfnamefont {M.}~\bibnamefont {Capone}},\ and\ \bibinfo {author}
		{\bibfnamefont {A.~J.}\ \bibnamefont {Millis}},\ }\href
	{https://doi.org/10.1103/PhysRevB.80.045101} {\bibfield  {journal} {\bibinfo
			{journal} {Phys. Rev. B}\ }\textbf {\bibinfo {volume} {80}},\ \bibinfo
		{pages} {045101} (\bibinfo {year} {2009})}\BibitemShut {NoStop}%
	\bibitem [{\citenamefont {Jarrell}\ and\ \citenamefont
		{Gubernatis}(1996)}]{S-Jarrell1996}%
	\BibitemOpen
	\bibfield  {author} {\bibinfo {author} {\bibfnamefont {M.}~\bibnamefont
			{Jarrell}}\ and\ \bibinfo {author} {\bibfnamefont {J.~E.}\ \bibnamefont
			{Gubernatis}},\ }\href {https://doi.org/DOI: 10.1016/0370-1573(95)00074-7}
	{\bibfield  {journal} {\bibinfo  {journal} {Physics Reports}\ }\textbf
		{\bibinfo {volume} {269}},\ \bibinfo {pages} {133} (\bibinfo {year}
		{1996})}\BibitemShut {NoStop}%
	\bibitem [{\citenamefont {Kaufmann}\ and\ \citenamefont
		{Held}(2023)}]{S-Kaufmann2023}%
	\BibitemOpen
	\bibfield  {author} {\bibinfo {author} {\bibfnamefont {J.}~\bibnamefont
			{Kaufmann}}\ and\ \bibinfo {author} {\bibfnamefont {K.}~\bibnamefont
			{Held}},\ }\href {https://doi.org/https://doi.org/10.1016/j.cpc.2022.108519}
	{\bibfield  {journal} {\bibinfo  {journal} {Computer Physics Communications}\
		}\textbf {\bibinfo {volume} {282}},\ \bibinfo {pages} {108519} (\bibinfo
		{year} {2023})}\BibitemShut {NoStop}%
	\bibitem [{\citenamefont {Kubo}(1957)}]{S-Kubo1957}%
	\BibitemOpen
	\bibfield  {author} {\bibinfo {author} {\bibfnamefont {R.}~\bibnamefont
			{Kubo}},\ }\href@noop {} {\bibfield  {journal} {\bibinfo  {journal} {Journal
				of the Physical Society of Japan}\ }\textbf {\bibinfo {volume} {12}},\
		\bibinfo {pages} {570} (\bibinfo {year} {1957})}\BibitemShut {NoStop}%
	\bibitem [{\citenamefont {Brinkman}\ and\ \citenamefont
		{Rice}(1970)}]{S-Brinkman1970}%
	\BibitemOpen
	\bibfield  {author} {\bibinfo {author} {\bibfnamefont {W.~F.}\ \bibnamefont
			{Brinkman}}\ and\ \bibinfo {author} {\bibfnamefont {T.~M.}\ \bibnamefont
			{Rice}},\ }\href {https://doi.org/10.1103/PhysRevB.2.4302} {\bibfield
		{journal} {\bibinfo  {journal} {Phys. Rev. B}\ }\textbf {\bibinfo {volume}
			{2}},\ \bibinfo {pages} {4302} (\bibinfo {year} {1970})}\BibitemShut
	{NoStop}%
	\bibitem [{\citenamefont {Vollhardt}(1984)}]{S-Vollhardt1984}%
	\BibitemOpen
	\bibfield  {author} {\bibinfo {author} {\bibfnamefont {D.}~\bibnamefont
			{Vollhardt}},\ }\href {https://doi.org/10.1103/RevModPhys.56.99} {\bibfield
		{journal} {\bibinfo  {journal} {Rev. Mod. Phys.}\ }\textbf {\bibinfo {volume}
			{56}},\ \bibinfo {pages} {99} (\bibinfo {year} {1984})}\BibitemShut {NoStop}%
	\bibitem [{\citenamefont {Wysoki\ifmmode~\acute{n}\else \'{n}\fi{}ski}\ and\
		\citenamefont {Fabrizio}(2017)}]{S-Fabrizio2017}%
	\BibitemOpen
	\bibfield  {author} {\bibinfo {author} {\bibfnamefont {M.~M.}\ \bibnamefont
			{Wysoki\ifmmode~\acute{n}\else \'{n}\fi{}ski}}\ and\ \bibinfo {author}
		{\bibfnamefont {M.}~\bibnamefont {Fabrizio}},\ }\href
	{https://doi.org/10.1103/PhysRevB.95.161106} {\bibfield  {journal} {\bibinfo
			{journal} {Phys. Rev. B}\ }\textbf {\bibinfo {volume} {95}},\ \bibinfo
		{pages} {161106} (\bibinfo {year} {2017})}\BibitemShut {NoStop}%
	\bibitem [{\citenamefont {Matsuda}\ \emph {et~al.}(2020)\citenamefont
		{Matsuda}, \citenamefont {Nakamura}, \citenamefont {Ikeda}, \citenamefont
		{Takeyama}, \citenamefont {Suga}, \citenamefont {Nakahara},\ and\
		\citenamefont {Muraoka}}]{S-Matsuda2020}%
	\BibitemOpen
	\bibfield  {author} {\bibinfo {author} {\bibfnamefont {Y.~H.}\ \bibnamefont
			{Matsuda}}, \bibinfo {author} {\bibfnamefont {D.}~\bibnamefont {Nakamura}},
		\bibinfo {author} {\bibfnamefont {A.}~\bibnamefont {Ikeda}}, \bibinfo
		{author} {\bibfnamefont {S.}~\bibnamefont {Takeyama}}, \bibinfo {author}
		{\bibfnamefont {Y.}~\bibnamefont {Suga}}, \bibinfo {author} {\bibfnamefont
			{H.}~\bibnamefont {Nakahara}},\ and\ \bibinfo {author} {\bibfnamefont
			{Y.}~\bibnamefont {Muraoka}},\ }\href
	{https://doi.org/10.1038/s41467-020-17416-w} {\bibfield  {journal} {\bibinfo
			{journal} {Nature Communications}\ }\textbf {\bibinfo {volume} {11}},\
		\bibinfo {pages} {3591} (\bibinfo {year} {2020})}\BibitemShut {NoStop}%
	\bibitem [{\citenamefont {Matsuda}\ \emph {et~al.}(2022)\citenamefont
		{Matsuda}, \citenamefont {Muraoka}, \citenamefont {Nakamura}, \citenamefont
		{Ikeda}, \citenamefont {Ishii}, \citenamefont {Zhou}, \citenamefont
		{Sawabe},\ and\ \citenamefont {Takeyama}}]{S-Matsuda2022}%
	\BibitemOpen
	\bibfield  {author} {\bibinfo {author} {\bibfnamefont {Y.~H.}\ \bibnamefont
			{Matsuda}}, \bibinfo {author} {\bibfnamefont {Y.}~\bibnamefont {Muraoka}},
		\bibinfo {author} {\bibfnamefont {D.}~\bibnamefont {Nakamura}}, \bibinfo
		{author} {\bibfnamefont {A.}~\bibnamefont {Ikeda}}, \bibinfo {author}
		{\bibfnamefont {Y.}~\bibnamefont {Ishii}}, \bibinfo {author} {\bibfnamefont
			{X.-G.}\ \bibnamefont {Zhou}}, \bibinfo {author} {\bibfnamefont
			{H.}~\bibnamefont {Sawabe}},\ and\ \bibinfo {author} {\bibfnamefont
			{S.}~\bibnamefont {Takeyama}},\ }\href
	{https://doi.org/https://doi.org/10.7566/JPSJ.91.101008} {\bibfield
		{journal} {\bibinfo  {journal} {Journal of the Physical Society of Japan}\
		}\textbf {\bibinfo {volume} {91}},\ \bibinfo {pages} {101008} (\bibinfo
		{year} {2022})}\BibitemShut {NoStop}%
	\bibitem [{\citenamefont {Lu}\ and\ \citenamefont
		{Robertson}(2019)}]{S-Lu2019}%
	\BibitemOpen
	\bibfield  {author} {\bibinfo {author} {\bibfnamefont {H.}~\bibnamefont
			{Lu}}\ and\ \bibinfo {author} {\bibfnamefont {J.}~\bibnamefont {Robertson}},\
	}\href {https://doi.org/https://doi.org/10.1002/pssb.201900210} {\bibfield
		{journal} {\bibinfo  {journal} {Physica Status Solidi (b)}\ }\textbf
		{\bibinfo {volume} {256}},\ \bibinfo {pages} {1900210} (\bibinfo {year}
		{2019})}\BibitemShut {NoStop}%
	\bibitem [{\citenamefont {Choi}\ \emph {et~al.}(2020)\citenamefont {Choi},
		\citenamefont {Ahn}, \citenamefont {Moon},\ and\ \citenamefont
		{Lee}}]{S-Choi2020}%
	\BibitemOpen
	\bibfield  {author} {\bibinfo {author} {\bibfnamefont {S.}~\bibnamefont
			{Choi}}, \bibinfo {author} {\bibfnamefont {G.}~\bibnamefont {Ahn}}, \bibinfo
		{author} {\bibfnamefont {S.~J.}\ \bibnamefont {Moon}},\ and\ \bibinfo
		{author} {\bibfnamefont {S.}~\bibnamefont {Lee}},\ }\href
	{https://doi.org/10.1038/s41598-020-66439-2} {\bibfield  {journal} {\bibinfo
			{journal} {Scientific Reports}\ }\textbf {\bibinfo {volume} {10}},\ \bibinfo
		{pages} {9721} (\bibinfo {year} {2020})}\BibitemShut {NoStop}%
\end{thebibliography}

%

\end{document}